\documentclass[]{article}
\usepackage{lmodern}
\usepackage{amssymb,amsmath}
\usepackage{ifxetex,ifluatex}
\usepackage{fixltx2e} % provides \textsubscript
\ifnum 0\ifxetex 1\fi\ifluatex 1\fi=0 % if pdftex
  \usepackage[T1]{fontenc}
  \usepackage[utf8]{inputenc}
\else % if luatex or xelatex
  \ifxetex
    \usepackage{mathspec}
  \else
    \usepackage{fontspec}
  \fi
  \defaultfontfeatures{Ligatures=TeX,Scale=MatchLowercase}
\fi
\IfFileExists{upquote.sty}{\usepackage{upquote}}{}
\IfFileExists{microtype.sty}{%
\usepackage{microtype}
\UseMicrotypeSet[protrusion]{basicmath} % disable protrusion for tt fonts
}{}
\usepackage[margin=1in]{geometry}
\usepackage{hyperref}
\hypersetup{unicode=true,
            pdftitle={SSNdesign -- an R package for pseudo-Bayesian optimal and adaptive sampling designs on stream networks},
            pdfauthor={Alan R. Pearse1,2,C, James M. McGree2,3, Nicholas A. Som4,5, Catherine Leigh1,2,3, Jay M. Ver Hoef6, Paul Maxwell7, Erin E. Peterson1,2,3},
            pdfborder={0 0 0},
            breaklinks=true}
\usepackage{graphicx,grffile}
\makeatletter
\def\maxwidth{\ifdim\Gin@nat@width>\linewidth\linewidth\else\Gin@nat@width\fi}
\def\maxheight{\ifdim\Gin@nat@height>\textheight\textheight\else\Gin@nat@height\fi}
\makeatother
\setkeys{Gin}{width=\maxwidth,height=\maxheight,keepaspectratio}
\IfFileExists{parskip.sty}{%
\usepackage{parskip}
}{% else
\setlength{\parindent}{0pt}
\setlength{\parskip}{6pt plus 2pt minus 1pt}
}
\ifx\paragraph\undefined\else
\let\oldparagraph\paragraph
\renewcommand{\paragraph}[1]{\oldparagraph{#1}\mbox{}}
\fi
\ifx\subparagraph\undefined\else
\let\oldsubparagraph\subparagraph
\renewcommand{\subparagraph}[1]{\oldsubparagraph{#1}\mbox{}}
\fi

\let\rmarkdownfootnote\footnote%
\def\footnote{\protect\rmarkdownfootnote}

\usepackage{titling}

  \title{\texttt{SSNdesign} -- an R package for pseudo-Bayesian optimal and
adaptive sampling designs on stream networks}
    \pretitle{\vspace{\droptitle}\centering\huge}
  \posttitle{\par}
    \author{Alan R. Pearse\textsuperscript{1,2,C}, James M.
McGree\textsuperscript{2,3}, Nicholas A. Som\textsuperscript{4,5},
Catherine Leigh\textsuperscript{1,2,3}, Jay M. Ver
Hoef\textsuperscript{6}, Paul Maxwell\textsuperscript{7}, Erin E.
Peterson\textsuperscript{1,2,3}}
    \preauthor{\centering\large\emph}
  \postauthor{\par}
      \predate{\centering\large\emph}
  \postdate{\par}
    \date{}

\usepackage{algorithmic}
\usepackage{algorithm}
\usepackage[font=normalsize]{caption}
\usepackage{afterpage}
\usepackage{pdflscape}
\usepackage{capt-of}
\usepackage{adjustbox}
\usepackage{epstopdf}
\begin{document}
\maketitle

\textsuperscript{1} Institute for Future Environments, Queensland
University of Technology, 2 George Street, Brisbane, QLD 4000

\textsuperscript{2} Australian Research Council Centre of Excellence for
Mathematical and Statistical Frontiers, Queensland University of
Technology, 2 George Street, Brisbane, QLD 4000

\textsuperscript{3} School of Mathematical Sciences, Queensland
University of Technology, 2 George Street, Brisbane, QLD 4000

\textsuperscript{4} US Fish and Wildlife Service, 1655 Heindon Road,
Arcata, CA 95521

\textsuperscript{5} Humboldt State University, 1 Harpst St., Arcata, CA
95521

\textsuperscript{6} Healthy Land and Water, 160 Ann Street, Brisbane,
QLD 4000

\textsuperscript{7} Alaska Fisheries Science Center, NOAA Fisheries,
7600 Sand Point Way NE, Seattle, WA, 98115

\textsuperscript{C} Corresponding author:
\href{mailto:ar.pearse@qut.edu.au}{\nolinkurl{ar.pearse@qut.edu.au}}

\section{Abstract}\label{abstract}

Streams and rivers are biodiverse and provide valuable ecosystem
services. Maintaining these ecosystems is an important task, so
organisations often monitor the status and trends in stream condition
and biodiversity using field sampling and, more recently, autonomous
\textit{in-situ} sensors. However, data collection is often costly and
so effective and efficient survey designs are crucial to maximise
information while minimising costs. Geostatistics and optimal and
adaptive design theory can be used to optimise the placement of sampling
sites in freshwater studies and aquatic monitoring programs.
Geostatistical modelling and experimental design on stream networks pose
statistical challenges due to the branching structure of the network,
flow connectivity and directionality, and differences in flow volume.
Thus, unique challenges of geostatistics and experimental design on
stream networks necessitates the development of new open-source software
for implementing the theory. We present \texttt{SSNdesign}, an R package
for solving optimal and adaptive design problems on stream networks that
integrates with existing open-source software. We demonstrate the
mathematical foundations of our approach, and illustrate the
functionality of \texttt{SSNdesign} using two case studies involving
real data from Queensland, Australia. In both case studies we
demonstrate that the optimal or adaptive designs outperform random and
spatially balanced survey designs. The \texttt{SSNdesign} package has
the potential to boost the efficiency of freshwater monitoring efforts
and provide much-needed information for freshwater conservation and
management.

\section{Introduction}\label{introduction}

Streams and rivers are highly biodiverse ecosystems supporting both
aquatic and terrestrial species (David et al. 2006; Poff, Olden, and
Strayer 2012) and provide important ecosystem services including clean
water, food, and energy (Vorosmarty et al. 2010). The ecological and
economic importance of waterways has driven government and
non-government organisations worldwide to invest large amounts of time
and money into monitoring, assessing and rehabilitating them (Ver Hoef,
Peterson, and Theobald 2006). However, monitoring data remain relatively
sparse (United Nations Water 2016) because the cost of sampling makes it
impossible to gather data everywhere, on every stream, at all times.
Thus, it is crucial to select sampling locations that yield as much
information as possible about water quality and aquatic ecosystem
health, especially when the stream system is large and resources for
sampling are limited.

Geostatistical models are commonly used to analyse environmental data
collected at different locations and to make predictions, with estimates
of uncertainty, at unobserved (i.e.~unsampled) sites (Cressie 1993).
These models are a generalisation of the classic linear regression
model, which contains a deterministic mean describing the relationship
between the response (i.e.~dependent variable) and the covariates
(i.e.~independent variables). In a geostatistical model, the assumption
of independence of the errors is relaxed to allow spatial
autocorrelation, which is modelled as a function of the distance
separating any two locations (Ver Hoef et al. 2001). This provides a way
to extract additional information from the data by modelling local
deviations from the mean using the spatial autocorrelation, or
covariance, between sites. However, spatial autocorrelation may exist in
streams data that is not well described using Euclidean distance, given
the branching network structure, stream flow connectivity, direction and
volume (Peterson et al. 2013). In addition, many traditional covariance
functions are invalid if an in-stream (i.e.~hydrologic) distance measure
is substituted for Euclidean distance (Ver Hoef, Peterson, and Theobald
2006; Ver Hoef 2018). The use of covariance functions based on Euclidean
distance may produce physically implausible results; for example,
implying that two adjacent streams that do not flow into each other and
that have separate watersheds are strongly related. This led to the
development of covariance functions that are specifically designed to
describe the unique spatial relationships found in streams data (Ver
Hoef, Peterson, and Theobald 2006; Ver Hoef and Peterson 2010).
Geostatistical models fit to streams data describe a number of in-stream
relationships in a way that is scientifically consistent with the
hydrological features of natural streams and, as such, are increasingly
being used for broad-scale monitoring and modelling of stream networks;
see, for example, D. J. Isaak et al. (2017).

The theoretical properties of geostatistical models can also be
exploited in optimal and adaptive experimental designs (Mueller 1999;
Mueller et al. 2007; Som et al. 2014; Falk, McGree, and Pettitt 2014),
which are used to select sampling locations that maximize information
gain and minimize costs. However, the exact locations included in an
optimal design will depend on the objectives of the monitoring program.
Common objectives include estimating the parameters of the underlying
geostatistical model (e.g.~fixed effects estimates and/or covariance
parameters), making accurate predictions at unsampled locations, or
both. Utility functions are mathematical representations of the
objectives used to measure the suitability of a design for a specific
purpose. Depending on the objective of the sampling, the best design
might be one that includes spatially balanced sites distributed across
the study area or it could be a design that includes clusters of sites
in close proximity to one another (Som et al. 2014). A variety of
utility functions are available (Som et al. 2014; Falk, McGree, and
Pettitt 2014) and are described more specifically in Section 2.4. An
adaptive design (i.e.~sequential design) is constructed by making a
series of optimal decisions about where to put sampling sites as new
information becomes available over time (Mueller et al. 2007). For
example, the spatial location of monitoring sites may change through
time, with some sites removed due to changes in access, or additional
sites added as new funding becomes available. In these situations, the
information gained from the data collected up to that point can be used
to inform where the optimal sampling locations will be at the next time
step. Hence, adaptive designs may provide additional benefits for
long-term environmental monitoring programs because one-off optimal
designs ignore the evolving nature of environmental processes and do not
allow for adjustments as monitoring needs change (Kang et al. 2016).

Bayesian and pseudo-Baeysian methods can enhance optimal and adaptive
designs. Utility functions, which measure the suitability of an
experimental design for some purpose, often depend on the parameters of
the geostatistical model that one intends to fit over the design;
however, the utility function can only be evaluated when these parameter
values are fixed (Mueller 1999). If the values change, for example,
through random variations in field conditions, then the design may no
longer be optimal. Bayesian and pseudo-Bayesian optimal design addresses
this issue by using simulation to construct more robust designs and to
incorporate prior information about the distribution of the model
parameters when constructing the design (Atkinson, Donev, and Tobias
2007). A drawback is that these methods are computationally intensive
(Atkinson, Donev, and Tobias 2007; Royle 2002). In this paper and in
\texttt{SSNdesign}, we use the pseudo-Bayesian approach. This is
different to a fully Bayesian approach because we are committed to
performing frequentist inference on the data we collect from an
experiment. The pseudo-Bayesian approach also does not take a Bayesian
view of uncertainty, particularly with respect to model uncertainty, and
as such we do not always have access to Bayesian utilities.
Nevertheless, the pseudo-Bayesian approach still allows us to
incorporate prior information in the design process, which is not
possible for purely frequentist designs, and can be more computationally
efficient than the fully Bayesian approach.

Although numerous software packages have been developed to implement
geostatistical models on streams and to solve experimental design
problems, none have done both. The \texttt{SSN} package (Ver Hoef et al.
2014) for R statistical Software (R Core Team 2018) is currently the
only software available for implementing geostatistical models on stream
networks (Ver Hoef and Peterson 2010). However, various software
packages exist to solve experimental design problems. For example,
\texttt{acebayes} provides an implementation of the approximate
coordinate exchange algorithm for finding optimal Bayesian designs given
a user-specified utility function (Overstall, Woods, and Adamou 2017).
For spatial design problems, \texttt{spsurvey} (Kincaid and Olsen 2018)
implements a variety of sampling designs including the Generalised
Random Tesselation Sampling (GRTS) design for spatially balanced samples
(Stevens and Olsen 2004). The package \texttt{geospt} (C. Melo,
Santacruz, and Melo 2012) focuses on drawing optimal and adaptive
spatial samples in the conventional 2-D geostatistical domain, with
Euclidean distance used to describe spatial relationships between
locations. However, it does not allow for stream-specific distance
measures and covariance functions. Furthermore, it does not calculate
design criteria consistent with a Bayesian or pseudo-Bayesian approach.
This is important for constructing designs that are robust to changes in
parameter values that the utility function depends on. Som et al. (2014)
and Falk et al. (2014) made use of the geostatistical modelling
functions in the \texttt{SSN} package for solving design problems on
stream networks, but neither addressed adaptive design problems and both
used customised code that was not intended for general use.

To our knowledge, \texttt{SSNdesign} is the first software package that
allows users to optimise experimental designs on stream networks using
geostatistical models and covariance functions that account for unique
stream characteristics within a robust design framework. It combines the
statistical modelling functionality found in the \texttt{SSN} package
(Ver Hoef et al. 2014) with principles of pseudo-Bayesian design
(Mueller 1999) and custom code developed by Som et al.(2014) into a
generalised toolkit for solving optimal and adaptive design problems on
stream networks. In Section 2, we discuss the mathematical principles
underpinning these tools and outline the structure of the core functions
in \texttt{SSNdesign} package along with a summary of the package's
speed and performance. This section is extended by Supplementary
Information A, which gives a deeper treatment of the mathematics
involved. In Section 3 we present two case studies using real data from
Queensland, Australia. Supplementary Information B is the package
vignette, and gives a detailed walkthrough of the code required to
reproduce the example. We conclude with a brief discussion of the
package and future developments in Section 4. For those unfamiliar with
the language of experimental design, note that a glossary of terms is
provided in Supplementary Information C.

\section{The SSNdesign package}\label{the-ssndesign-package}

\subsection{Workflow}\label{workflow}

There are three different workflows in \texttt{SSNdesign} corresponding
to different design types, as well as for importing and manipulating
stream network data (Figs. \ref{Fig:flowchart_import},
\ref{Fig:flowchart_design}). The general process is to import a stream
dataset (see Section 2.2) and if necessary, create potential sampling
locations and simulate data at those locations. The next step is to use
the main workhorse function \texttt{optimiseSSNDesign} to find optimal
or adaptive designs, or use \texttt{drawStreamNetworkSamples} to find
probability-based designs. In the case of adaptive design problems,
there are always multiple `timesteps'. The first timestep is important
because adaptive designs cannot be constructed without a pre-existing
design. Therefore, it is not possible to go straight from data
processing to an adaptive design; as such, the first design decision
must be based on either an optimal or probability-based design before
implementing the adaptive design process in subsequent timesteps (Fig.
\ref{Fig:flowchart_design}).

\afterpage{ 
    \clearpage
    \thispagestyle{empty}
    \begin{landscape}
\begin{center}
    \includegraphics[height = 0.75\textheight]{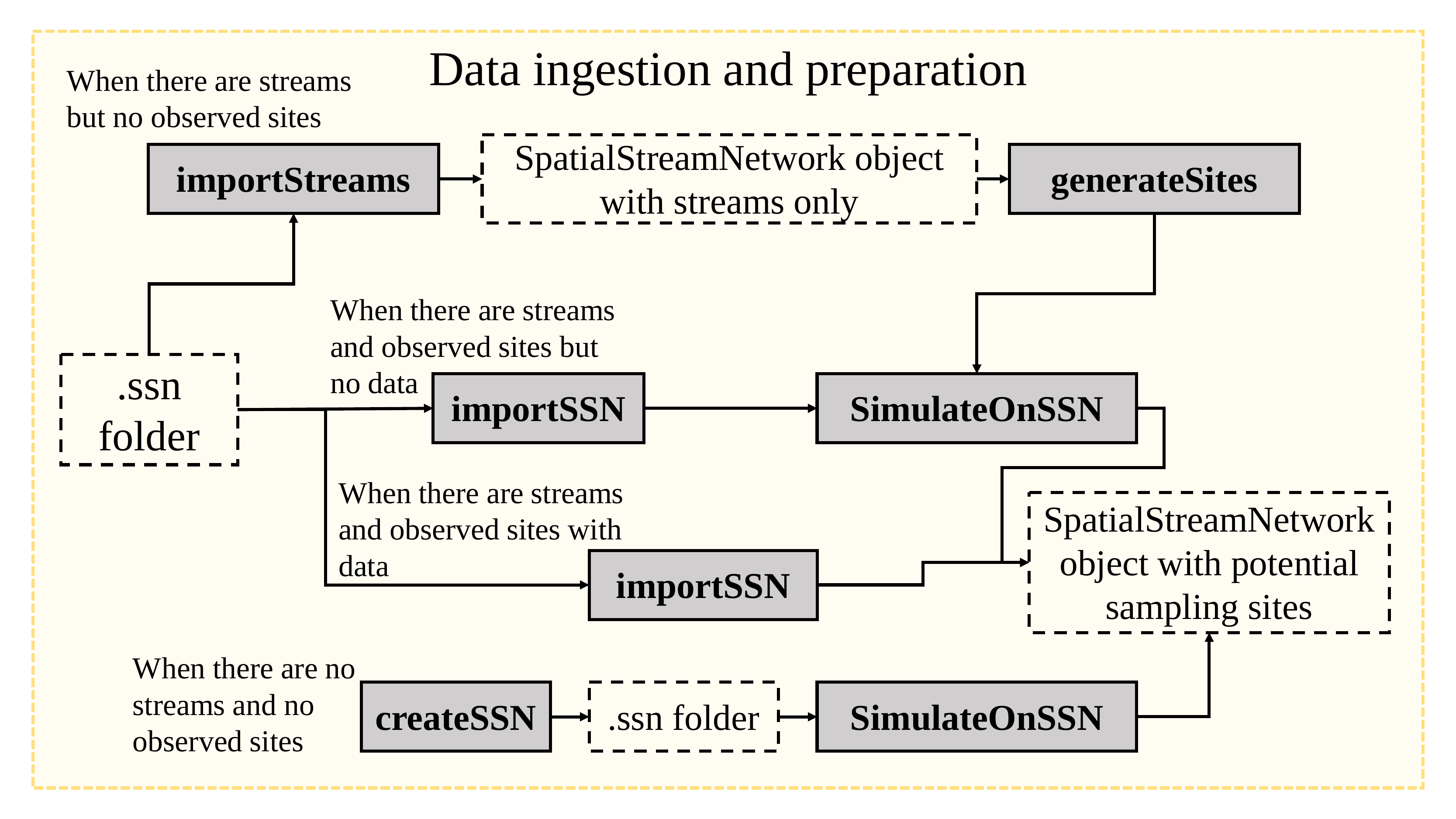}
    \captionof{figure}{A flow chart of the function calls used to import and prepare streams data for use in \texttt{SSNdesign}. Grey boxes with solid outlines indicate a call to a function (N.B. \texttt{importSSN} and \texttt{SimulateOnSSN} belong to \texttt{SSN}, not \texttt{SSNdesign}). Clear boxes with dashed outlines indicate a file, folder or R object that is created as a result of a function call.}
    \label{Fig:flowchart_import}
\end{center}
    \end{landscape}
    \clearpage
}

\afterpage{ 
    \clearpage
    \thispagestyle{empty}
    \begin{landscape}
\begin{center}
    \includegraphics[height = 0.75\textheight]{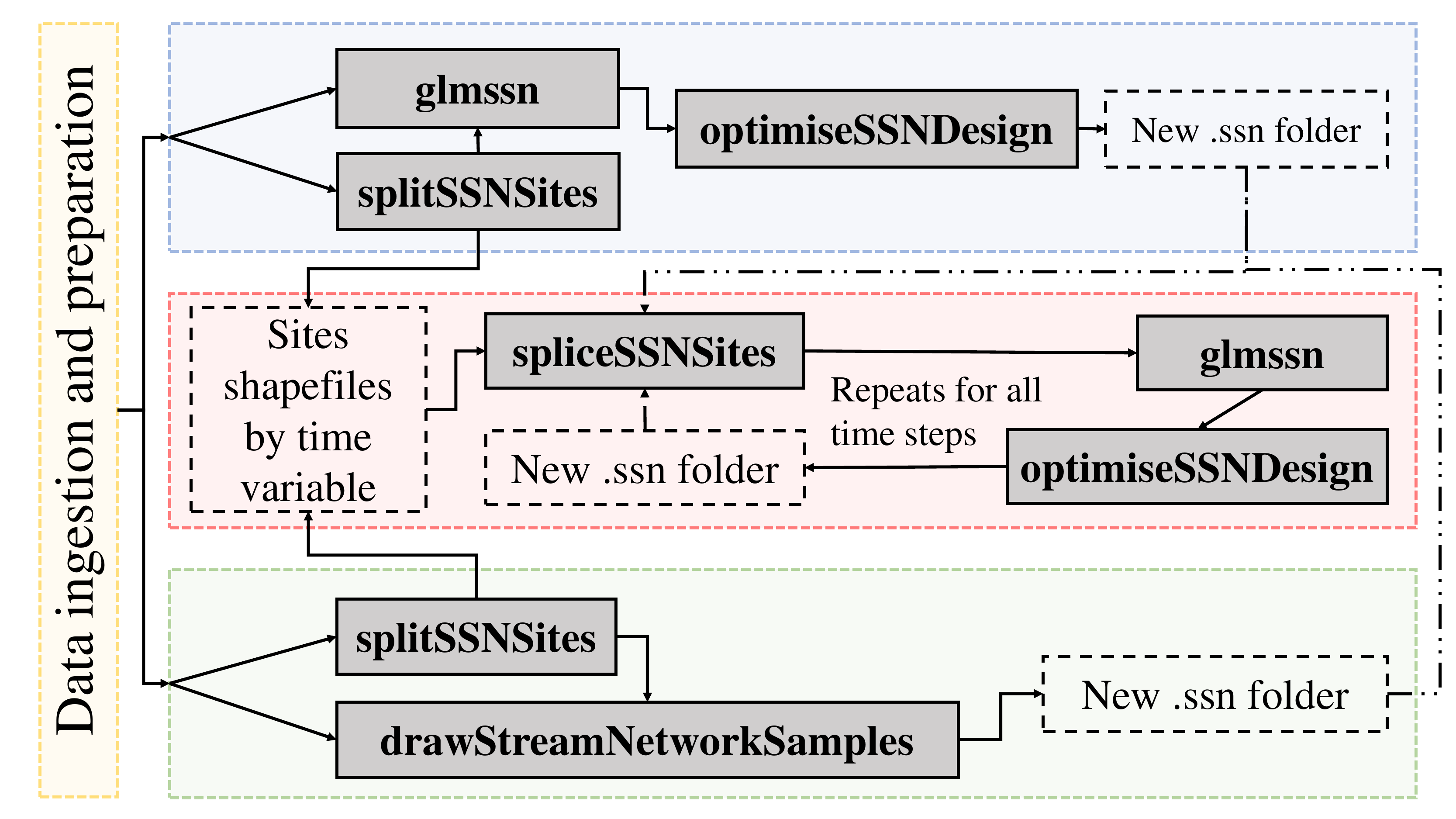}
    \captionof{figure}{A flow chart of the function calls used to construct optimal (blue), adaptive (red), and probability-based (green) designs for stream networks. The yellow box is discussed in more detail in Figure 1. Grey boxes with solid outlines represent a call to a function (N.B. \texttt{glmssn} belongs to \texttt{SSN}, not \texttt{SSNdesign}). Clear boxes with dashed outlines indicate a file, folder or R object that is created as a result of a function call. The dash-dot-dot lines represent optional steps or connections between function calls that do not always occur.}
    \label{Fig:flowchart_design}
\end{center}
    \end{landscape}
    \clearpage
}

\subsection{Data format, ingestion and
manipulation}\label{data-format-ingestion-and-manipulation}

The \texttt{SSNdesign} package builds on the functionality in the
package SSN (Ver Hoef et al. 2014), most notably the S4
\texttt{SpatialStreamNetwork} class for stream network datasets and the
function and class \texttt{glmssn} which fits and stores fitted spatial
stream network models. SpatialStreamNetwork objects are an extention of
the sp class, but are unique because they contain a set of stream lines
(i.e.~edges) and a set of observed sites. Prediction sites can also be
included, but are optional. These spatial data are imported from .ssn
folders, which are created using the Spatial Tools for the Analysis of
River Sytems (STARS) ArcGIS custom toolset (Peterson and Ver Hoef 2014).
The \texttt{importSSN} found in the SSN package is used to ingest data
contained in the .ssn folders, but this function will not work if there
are no observed sites in the .ssn folder. Therefore, the
\texttt{SSNdesign} package provides the function \texttt{importStreams}
for creating a \texttt{SpatialStreamNetwork} object with no observed or
predicted sites. Additional functions such as \texttt{generateSites} are
provided to add potential sampling sites to these empty networks. Four
distinct workflows for importing and preprocessing stream data are shown
in Fig. \ref{Fig:flowchart_import}, but the end result is a
SpatialStreamNetwork object that contains streams and observed sites
where data were collected or simulated that can be used for optimal or
adaptive designs.

The end result of the data processing (Fig. \ref{Fig:flowchart_import})
and design (Fig. \ref{Fig:flowchart_design}) workflows is an object of
class \texttt{ssndesign}. These objects are lists containing 1) the
information about the way the design optimisation function
(\texttt{optimiseSSNdesign}) was used, including the
\texttt{SpatialStreamNetwork} objects before and after an optimal design
is found, 2) data about the optimal or adaptive design, and 3)
diagnostic information about the optimisation procedure. This class of
objects also has a plot method, \texttt{plot.ssndesign}, which plots the
trace for the optimisation algorithm. The method
\texttt{plot.SpatialStreamNetwork} from the package \texttt{SSN} can be
used to visualise the locations of the selected sites. Further details
are provided in the package vignette (Supplementary Information B). The
vignette is not only intended as a practical guide that demonstrates the
functions and workflow of the package, but it can also serve as a
practical reference for managers.

\subsection{Expected utility estimation and
maximisation}\label{expected-utility-estimation-and-maximisation}

There are many ways to configure a fixed number of monitoring sites.
Each potential configuration represents a `design' \(d\), and the set of
all possible configurations is denoted as \(D\). The goal of optimal
design theory as applied to stream networks is to find which
configuration of sites is most suitable to achieve a purpose (e.g.,
precise parameter estimation in a statistical model). We refer to the
optimal configuration of sites as \(d^*\). Inside \texttt{SSNdesign},
the quality of a design and its suitability for a stated goal is
measured using a function called the expected utility \(U(d)\) (Mueller
1999). Larger values of \(U(d)\) indicate better designs and the
calculation of \(U(d)\) is linked to the utility function,
\(U(d, \theta, y)\). The utility function may depend on elements of the
geostatistical model fitted over the design, including its parameters
\(\theta\) and either observed or predicted data \(y\) (Mueller 1999).
However, in many cases of pseudo-Bayesian utility functions, the utility
function does not depend on \(y\) and can be written \(U(d, \theta)\)
(Falk, McGree, and Pettitt 2014). This function mathematically encodes
the criterion used to compare designs. Examples of utility functions can
be found in Section 2.1.2. The utility function, however, cannot be used
directly to assess the quality of designs. This is because
\(U(d, \theta, y)\) depends on specific values of \(\theta\) and \(y\)
and the relative rankings of designs may change, sometimes dramatically,
if there are small variations in these two quantities. Therefore, the
parameters \(\theta\) and the data \(y\) must be integrated out such
that the values used to rank designs depend only on the designs
themselves. We achieve this using Monte-Carlo integration (Mueller
1999). Further details are provided in Supplementary Info A.

The set of possible designs \(D\) is usually large and, due to time and
computational constraints, we cannot find \(d^*\) by evaluating \(U(d)\)
for every \(d \in D\). \texttt{SSNdesign} deals with this problem in two
ways. Firstly, we do not treat the design problem as a continuous one.
That is, we do not allow sites to shift to any place along the stream
edges during the search for the best design. The user must first create
a set of \(N\) candidate points, and a design containing \(n\) points is
chosen from among them. This ensures that \(D\) has a finite size.
Secondly, we use a coordinate exchange algorithm to reduce the
computational load of finding \(d^*\) (Royle 2002). The coordinate
exchange algorithm we use is called the Greedy Exchange Algorithm
(Supplementary Information A, Algorithm 1). This algorithm is known to
rapidly converge on highly efficient designs, although this efficient
design may not be the best design (Evangelou and Zhu 2012). Note that
the Greedy Exchange Algorithm has previously been used for optimal
designs on stream networks (Falk, McGree, and Pettitt 2014).

\subsection{Utility functions for optimal and adaptive experimental
designs}\label{utility-functions-for-optimal-and-adaptive-experimental-designs}

The utility functions implemented in the \texttt{SSNdesign} package can
be characterised as suitable for solving either static optimal design
problems or evolving, adaptive design problems. With regard to the
former category, there are six utility functions for common monitoring
objectives including parameter estimation and prediction (Table
1;Supplementary Information A). With regard to the latter category,
there are three utility functions for similar monitoring objectives that
are appropriate for adaptive decision-making. These are intended to be
used with the function \texttt{optimiseSSNDesign}. We also provide two
utility functions for finding space-filling designs (Table 1), intended
to be used with the optimisation function
\texttt{constructSpaceFillingDesigns}. Space-filling designs are designs
which, ideally, contain roughly equally spaced and unclustered sets of
monitoring sites along the stream network.

To solve adaptive design problems, we use a myopic design approach; that
is, when making an adaptive design decision at a given point in time, we
try to find the best decision for the next time step only (McGree et al.
2012). An alternative method is to use backward induction, which
involves enumerating all possible future decisions for a set number of
time steps and before choosing the best series of decisions from among
them (Mueller et al. 2007). However, this is often computationally
prohibitive. Here we used an algorithm that assumes an initial design
\(d_0\) that we seek to improve by adding sites or by rearranging the
existing ones. We know \textit{a priori} that, instead of solving this
problem once, we will have to make a further \(T\) decisions in the
future about the arrangement of the design points (Algorithm 2,
Supplementary Information A). At each step \(t = 1, 2, ..., T\) in this
iterative process, we fit a model to the existing design at \(d_{t-1}\)
and summarise the fit of the model using the summary statistic
\(O_t(\theta)\). The summary statistic can be arbitrarily defined,
though it should be low-dimensional and preferably fast to compute
(Mueller et al. 2007). In evaluating the designs at each timestep \(t\),
we incorporate the summary statistic \(O_{t-1}(\theta)\) in the
calculation of the expected utility given the previous design decisions
\(U(d|d_{0:t-1},y_{0:t-1})\). Data simulated from the likelihood at time
\(t\), \(p(y|\theta^t, d)\), can involve real data collection. However,
for most design studies, this step will simply involve generating
predictions, with error, from an assumed model. As with optimal design
problems, the best design \(d_t^{*}\) at step \(t\) is the one which
maximises \(U(d|d_{0:t-1},y_{0:t-1})\). We also update the priors on the
parameters for each new design. The process is repeated until the best
design has been found for each of \(T\) time steps. For the objectives
of covariance parameter estimation and fixed effects parameter
estimation, we have defined utility functions specifically. The main
difference between the adaptive and static utility functions is the use
of the summary statistic \(O_t(\theta)\) from the model fitted to the
existing sites (Table 1; Supplementary Information A). These utility
functions can be used with \texttt{optimiseSSNDesign}.

\newgeometry{a4paper,left=.5in,right=.5in,top=0.5in,bottom=0.5in}
    \begin{landscape}
        \begin{center}
        \begin{tabular}{p{4.3cm}p{4.5cm}p{1.7cm}p{1.7cm}p{5.25cm}p{3.20cm}}
          \multicolumn{6}{p{22.70cm}}{Table 1: Utility functions implemented in \texttt{SSNdesign}. Empirical utility functions are utility functions where the covariance parameters are estimated from data simulated using the prior draws.}\\
            \hline\\
            Name & Purpose & Application & Empirical & Definition of the expected utility & Reference\\
            \hline\\
            CP-optimality & Covariance parameters & OP & $\times$ & $\log\det\left[I\left(\theta\right)^{-1}\right]$ & Falk et al., 2014\\
            D-optimality & Fixed effects parameters & OP & $\times$ & $\log\det\left[Var(\hat{\beta}_{\text{gls}})^{-1}\right]$ & Falk et al., 2014; Som et al., 2014\\
            ED-optimality & Fixed effects parameters & OP & \checkmark & $\log\det\left[\hat{Var}(\hat{\beta}_{\text{gls}})^{-1}\right]$ & Falk et al., 2014; Som et al., 2014\\
            CPD-optimality & Fixed effects and covariance parameters, a mixture of CP- and D-optimality & OP & $\times$ & $\log\det\left[Var(\hat{\beta}_{\text{gls}})^{-1}\right] + \log\det\left[I\left(\theta\right)^{-1}\right]$ & \\
            K-optimality & Predictions & OP, AD & $\times$ & $\left(\sum_{s_z \in S} Var\left(\hat{y}(s_z)\right)\right)^{-1}$ & Falk et al., 2014; Som et al., 2014\\
            EK-optimality & Predictions & OP, AD & \checkmark & $\left(\sum_{s_z \in S} \hat{Var}\left(\hat{y}(s_z)\right)\right)^{-1}$ & Falk et al., 2014; Som et al., 2014\\
            Sequential CP-optimality & Covariance parameters & AD & $\times$ & $\log\det\left[\left(I\left(\theta\right)+O_t\left(\theta\right)\right)^{-1}\right]$ & Supplementary Info A\\
            Sequential D-optimality & Fixed effects parameters & AD & $\times$ & $\log\det\left[\left(Var\left(\hat{\beta}_{\text{gls}}\right) + O_t\left(\theta\right)\right)^{-1}\right]$ & Supplementary Info A\\
            Sequential ED-optimality & Fixed effects parameters & AD & \checkmark & $\log\det\left[\left(\hat{Var}\left(\hat{\beta}_{\text{gls}}\right) + O_t\left(\theta\right)\right)^{-1}\right]$ & Supplementary Info A\\
            Maximin & Space-filling, with an emphasis on increasing the minimum distance between pairs of points & OP & n/a & $\min_{i \neq j} D\left(x_i, x_j\right)$ & Pronzato and Muller, 2012\\
            Morris-Mitchell & Space-filling, with an emphasis on increasing separations larger than the minimum & OP & n/a & $-\left(\sum_{w = 1}^W\left(J_wD_w\right)^{p}\right)^{1/p}$ & Morris and Mitchell, 1995\\
            \hline\\
            \multicolumn{6}{p{22.70cm}}{Note: $\theta =$ a vector of covariance parameters from a geostatistical model; and $y =$ data that is either directly observed from a process or simulated from it. OP $=$ optimal design; AD $=$ adaptive design. n/a $=$ no covariance parameters involved. $I(\theta) =$ the expected Fisher Information Matrix; $\hat{\beta}_\text{gls} = $ the estimates of the fixed effects; $Var(\hat{\beta_{\text{gls}}}) = $ covariance matrix for the fixed effects. $s_z =$ a prediction site; $S=$ the set of all prediction sites. $\hat{y}(s_z) =$ the predicted value at a prediction site. $Var(\hat{y}(s_z)) = $ the kriging variance. $O_t(\theta) =$ a summary statistic from the existing design. $D(x_i, x_j) =$ the distance between two points $x_i$ and $x_j$. The distance can be measured as Euclidean distance or hydrological distance along the stream network (Ver Hoef and Peterson, 2010). $D =$ a sorted vector of non-zero distances in a distance matrix; $J =$ the number of times each distance occurs in one triangle of the matrix. The subscript $w = 1, 2, ..., W$ counts the $W$ unique non-zero entries in the distance matrix. $p = $ a weighting power, with $p \geq 1 $. In the Empirical column, $\times$ means No, $\checkmark$ means Yes.}
        \end{tabular}
\end{center}
    \end{landscape}
\restoregeometry

Users may also define their own utility functions using the
\texttt{optimiseSSNDesign} function, which has the flexibility to accept
utility functions as an argument. The utility function must be defined
in this format:
\texttt{utility.function(ssn,\ glmssn,\ design,\ prior.parameters,\ n.draws,\ extra.arguments)}.
The exact requirements in terms of input type and additional data
accessible within the function \texttt{optimiseSSNDesign} are described
in the function documentation. It is not necessary to use all the
arguments inside the function. Ultimately, the only requirement for a
working user-defined utility function is that the function returns a
single number, representing the expected utility.

\subsection{Other standard designs}\label{other-standard-designs}

The SSNdesign package focuses on optimal and adaptive design problems,
but we also include a number of standard designs such as simple random
sampling and GRTS (Stevens and Olsen 2004). We have also included
heuristic sampling schemes designed specifically for stream networks,
such as designs with sites allocated to headwater streams (i.e.~streams
at the top of the network), to outlets (i.e.~most downstream location on
the network), or in clusters around stream confluences (i.e. junctions
where stream segments converge; Som et al., 2014). These are all options
for the function \texttt{drawStreamNetworkSamples} (Table 2).

\begin{flushleft}
\begin{tabular}{p{0.8cm}p{5.75cm}p{9.45cm}}
 \multicolumn{3}{p{16cm}}{Table 2: Standard designs from Som et al. (2014). Note that the names in the second column are the string arguments which must be passed to the \texttt{drawStreamNetworkSamples} function to use the sampling scheme.}\\
 \hline
 ID & Name in package & Description \\
\hline
 SRS & SRS & Simple random sampling. An unstratified random sample of sites.\\
 G1 & GRTS & GRTS. Spatially balanced design (Stevens and Olsen, 2004). \\
 G2 & GRTSmouth & GRTS with one site always assigned to the stream outlet.\\
 G4 & GRTSclus & GRTS with clusters of sites around confluences.\\
 H1 & Headwater.Clusts.and.Singles & Headwater samples. One site allocated to the outlet and others preferentially allocated to tributaries.\\
 C3 & Trib.Sets.Head.Singles.sample & Clustered triplets. As many points as possible are allocated to triplets clustered where each segment meets at a confluence. All remaining points are assigned to tributaries. \\ 
 C4 & Trib.Sets.Head.Singles.Mouth.sample & C3 with a single point allocated to the outlet segment.\\
\hline
\end{tabular}\hfill
\end{flushleft}

\subsection{Computational performance}\label{computational-performance}

The optimisation of Bayesian and pseudo-Bayesian experimental designs
via simulation is notoriously slow (Royle 2002). Other experimental
design packages explicitly warn users to expect this (e.g.~acebayes
(Overstall, Woods, and Adamou 2017)). In our case, we have parallelised
the functions (compatible with any OS) to increase computational
efficiency for large problems. However, in many common situations users
can expect run-times of hours, to days, and even weeks. The expected
computation time in hours is given by
\(K \times L \times T \times n \times (N - n) / (3600 \times C)\), where
\(K\) is the number of random starts that are used to seed the
algorithm, \(L\) is the number of times the algorithm must iterate
before converging, \(T\) is the time in seconds that is required to
calculate \(U(d)\) for a single design, \(n\) is the number of desired
sampling locations, \(N\) is the number of potential sampling
locations,and \(C\) is the number of CPUs allocated to the task. The
parameter \(K\) is specified by the user in \texttt{optimiseSSNDesign},
but \(L\) is more difficult to constrain. The number of times the
algorithm must iterate until convergence is stochastic but, in our
experience, \(L = 2\) is most common; though we have observed
\(L \in \{3, 4, 5\}\). Unsurprisingly, the number of potential sampling
locations and the number of desired sampling locations strongly
influence computing time, and these make the largest contribution when
\(n \approx N/2\). Large problems \(N \geq 300\) generally take at least
a day, if not more, to compute. Our second case study below with
\(N = 900\) required approximately 4 days to complete using 20-32 CPUs
and \(n \ll N\). Using complicated or intensive utility functions will
also add significantly to computation time. The empirical prediction or
estimation utilities are particularly prone to computational
inefficiences because they rely on iteratively fitting and predicting
from geostatistical stream network models. This is compounded by the
fact that larger \(n\) is often associated with larger \(T\) values, due
to the increased size of the matrices (e.g.~covariance matrix, design
matrix) that must be created and/or inverted.

\section{Case studies}\label{case-studies}

We present two case studies where we illustrate how \texttt{SSNdesign}
can be used to address common monitoring challenges. Field data
collection can be expensive and so existing monitoring programs often
need to reduce sampling effort due to resource constraints. In the first
case study, we demonstrate how to ensure that the remaining survey sites
are optimally placed to minimize information loss when sampling effort
is reduced. In the second case study, we show how adaptive sampling can
be used to design a monitoring program in a previously unsampled area,
by gradually adding additional survey sites year-by-year. The R code
used to create these examples is provided in Supplementary Information B
so that readers can re-create them and also apply these methods to their
own data.

In a model-based design problem, a `true model' must be specified. Here,
a true model refers to the statistical model which most adequately
characterises the underlying spatial process given what we know about
the system. If a historical dataset from the study area exists, a
standard model-selection process can be used to determine which model
has the most support in the data using a range of approaches
(e.g.~information criteria or cross validation). If historical data are
not available, simulated data can also be used to implement the
model-based design. In this case, the general approach is to:

\begin{enumerate}
\item Specify the form of the statistical model;
\item identify which potential covariates should be related to the
response (e.g. temperature affects the solubility of dissolved oxygen
in water) based on prior knowledge of the system, or similar systems; and
\item set priors that specify what the relationship between the
covariates and the response is likely to be based on expert opinion
and/or a literature review. 
\end{enumerate}

The same process must be undertaken to specify the spatial covariance
structure and covariance parameters for the model. The
\texttt{SimulateOnSSN} function from \texttt{SSN} (Ver Hoef et al. 2014)
can then be used to simulate data that are subsequently used to
implement the model-based design. Needless to say, the quality of the
design will depend strongly based on the quality of the prior
information.

\subsection{Case study 1: Lake Eacham}\label{case-study-1-lake-eacham}

Water temperature samples were collected at 88 sites along a stream
network near Lake Eacham in north Queensland, Australia (Fig.
\ref{Fig:lakeeacham}). The dataset includes a shapefile of the streams,
the 88 observed sites with rainfall data and GIS-derived urban and
grazing land use percentages in the catchment, and 237 prediction sites
with the same covariates. Most survey sites were clustered at stream
confluences, with multiple sites located in the upstream and downstream
segments. Here we use optimal experimental design to reduce the number
of survey sites by half, with the least amount of information loss
possible. More specifically, we wanted to retain the ability to 1)
estimate the effect that land use and rainfall are having on water
temperature and 2) accurately predict temperature at unsampled
locations. In the former case, we optimised the design using
CPD-optimality, which is maximised when uncertainties in both the
fixed-effects and covariance parameters are minimised (Table 1). In the
latter case, we used K-optimality, which is maximised when the total
prediction uncertainty across all sites is minimised. Note that this
situation where two designs are built separately using different utility
functions is not ideal. There is no way to reconcile the two resulting
designs into a single one. Ideally, we would be able to use the
EK-optimality function. This utility function aims to maximise
prediction accuracy but also involves a parameter estimation step, and
therefore serves as a dual purpose utility function that yields designs
that are efficient for parameter estimation and prediction. However, it
was not practical to use the EK-optimality function because it is
extremely computationally expensive, even for this dataset with only 88
potential sites.

\begin{center}
    \includegraphics[width = \textwidth]{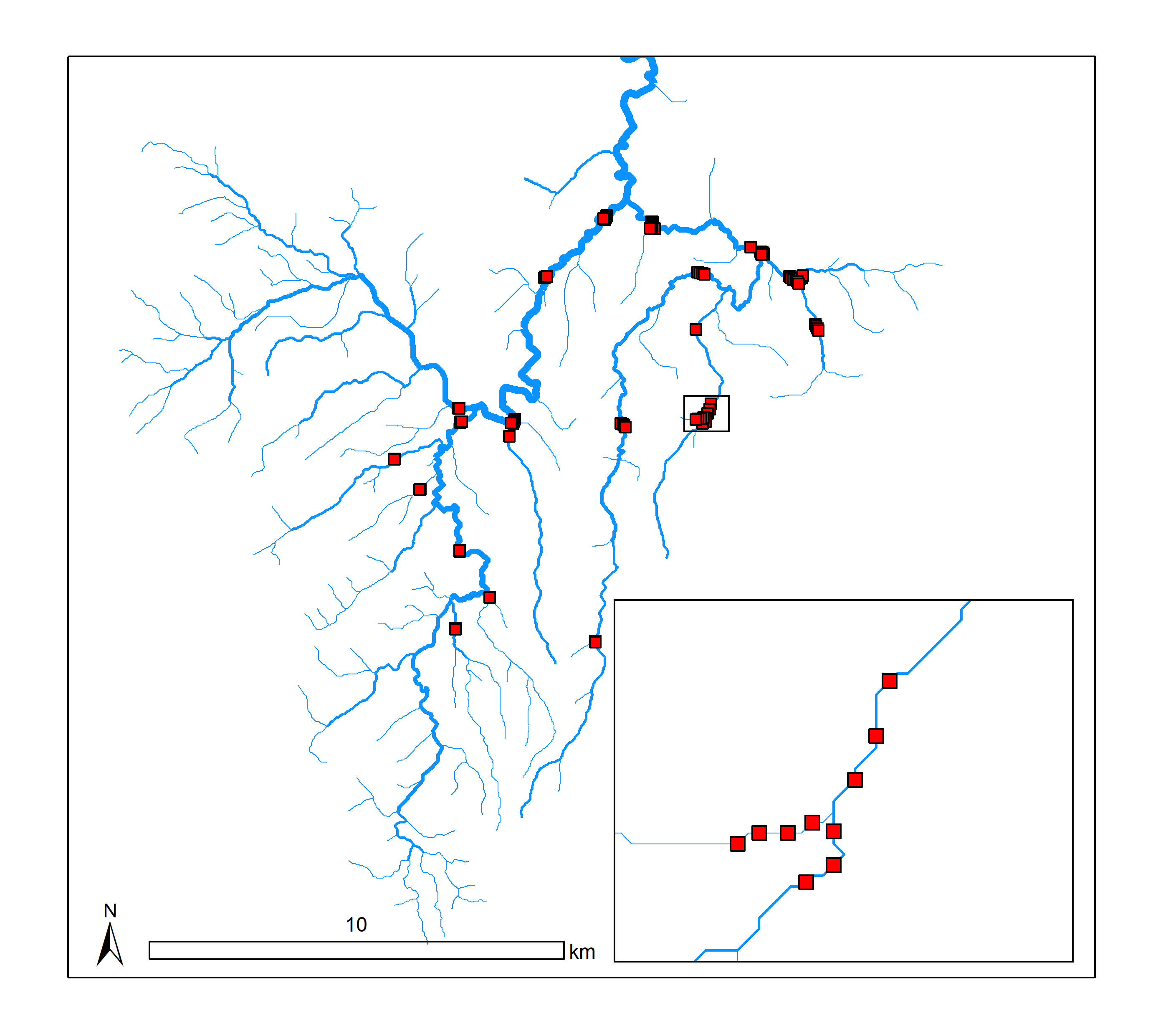}
    \captionof{figure}{The Lake Eacham stream network (blue lines) with all
    potential sampling sites (req squares). The width of the stream lines is proportional to the catchment area for each stream segment. The inset shows the spatial distribution. of sampling sites around a confluence in the stream network.}  
    \label{Fig:lakeeacham} \end{center}

We fit a spatial stream-network model to the temperature data, using
riparian land use (i.e.~percent grazing and urban area) and the total
rainfall recorded on the sampling date (mm) as covariates. The
covariance structure contained exponential tail-up and tail-down
components (Ver Hoef and Peterson 2010). Log-normal priors were set on
the covariance parameters using the log of the estimated covariance
parameters as the means and the log-scale standard errors of
\(\{0.35, 0.56, 0.63, 0.69, 0.68\}\), which were also estimated from the
existing data using restricted maximum likelihood (REML).

We set about finding CPD- and K-optimal designs with 44 sites by
removing one site at a time from the original 88 sites. An alternative
approach would be to optimise for a 44 site design with no intermediate
steps. However, we chose to remove one site at a time because it helps
reveal differences in the decision process between the CPD- and
K-optimality utility functions that would otherwise be difficult to
identify. Note that this is not an adaptive design because the model is
not refit and the priors on the covariance parameters are not updated at
each step. The expected CPD- and K-optimal utilities were calculated at
each step using 500 Monte Carlo draws.

The results revealed differences between the intermediate and final
designs discovered under CPD- and K-optimality. Both utility functions
preserved clusters in groups of at least three sites around confluences
(one on each of the upstream segments and one on the downstream
segment). However, CPD-optimality appeared to remove single sites that
were not part of clusters in preference to sites within clusters. By
comparison, K-optimality appeared to reduce clusters around confluences
down to three sites much more quickly, while preserving sites that were
located away from confluences. This reflects the previously observed
tendency of designs constructed for parameter estimation to favour
spatial clusters and the tendency of designs that optimise prediction to
contain sites spread out in space, with some clusters to characterise
spatial dependence at short distances (Falk, McGree, and Pettitt 2014;
Som et al. 2014). These results suggest that clusters located around
confluences provide valuable information about the covariance structure
that is needed to generate precise parameter estimates and accurate
predictions.

We tracked the information loss from the design process over time and
compared the performance of our final 44-site optimal designs against 20
random and GRTS designs of the same size (Fig. \ref{Fig:information}).
We compared them to 20 GRTS and 20 random designs because these designs
have multiple potential configurations and therefore we need to
characterise the range of their performances under the chosen expected
utilities. Information in this context is measured as relative design
efficiency; a ratio of the expected utility of a given design and the
expected utility of the `best' design, which in this case contained all
88 sites. There was a linear reduction of information available for
parameter estimation as sites were eliminated from the design (Fig.
\ref{Fig:information}a). While the optimal design containing 44 sites
provided only 20\% of the information gained from 88 sites, it provided
significantly more information than the random and GRTS designs. These
results suggest that reducing sampling effort by 50\% would signficantly
impact parameter estimation. However, the same was not true for
prediction. We observed only minor reductions in the efficiency of the
optimal 44-site design compared to the full 88-site design, while also
demonstrating considerable gains in efficiency over random and GRTS
designs (Fig. \ref{Fig:information}b). As such, a 50\% reduction in
sampling effort would have little impact on the predictive ability of
the models.

\begin{center}
\includegraphics[width = \textwidth]{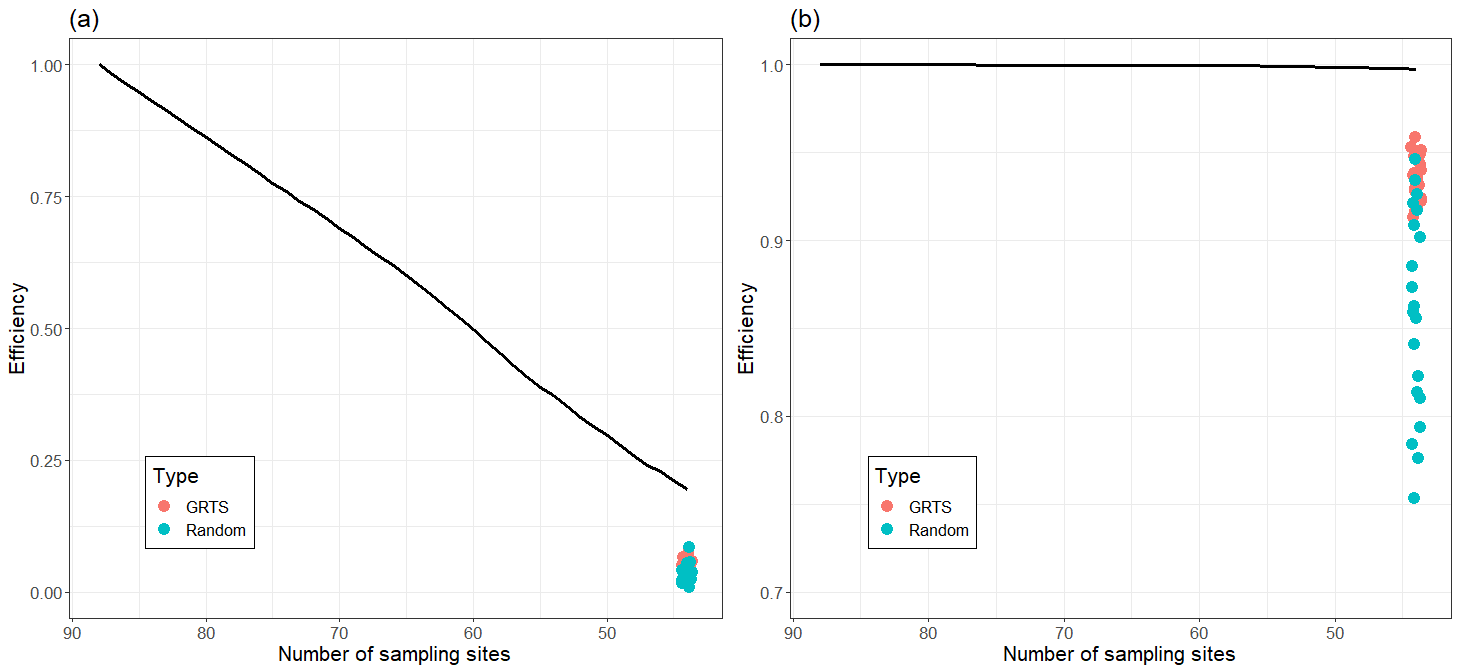}
\captionof{figure}{Information loss in our optimal designs as we
remove sites one-by-one using the (a) CPD-optimal and (b)  K-optimal
utility functions. Efficiency represents a ratio of the expected utilities of each design and the full 88 site design. The black line indicates the efficiency of the optimal design with a certain number of sampling sites. The 20 red dots and 20 blue dots in each panel represent the efficiencies of 44-site GRTS and random designs, respectively. These serve as a baseline measure of comparison for the optimal design, which should have higher relative efficiency. We only compare the efficiencies of the GRTS, random and optimal designs when there are 44 sampling sites in the design because the 44 site monitoring program is the final result.}
\label{Fig:information}
\end{center}

\subsection{Case study 2: Pine River}\label{case-study-2-pine-river}

In the second case study, we demonstrate how additional sites can be
selected to complement the information provided by a set of legacy sites
using simulated data. The objective of the adaptive design process is to
generate a design that can be used to accurately predict dissolved
oxygen over the stream network at unsampled locations.

We did not actually have data at legacy sites. Therefore, for this
example, we started by simulating four years of maximum daily dissovled
oxygen (DO, mg/L) data at 900 locations throughout the river network. We
then selected 200 of these sites using a GRTS design, which were treated
as legacy sites. The first two years of data from the legacy sites were
treated as historical data and used to estimate the fixed effects and
covariance parameters using a spatial statistical stream-network model
(Supplementary Information B). Five random starts were used to find an
adaptive design which maximises the K-optimality utility function (Table
1). We estimated \(U(d)\) using \(M = 500\) Monte-Carlo draws from our
independent log-normal priors on the covariance parameters. The results
were used to select an additional 50 sites in year 3; after which the
model was refit to the full dataset and an additional 50 sites were
selected in year 4. Thus, the final dataset included 300 sites, giving
rise to 950 observed DO measurements collected across four years (Fig.
\ref{Fig:the_plan}). The result is shown in Fig.
\ref{Fig:adaptive_design}.

\begin{center}
\includegraphics[width = \textwidth]{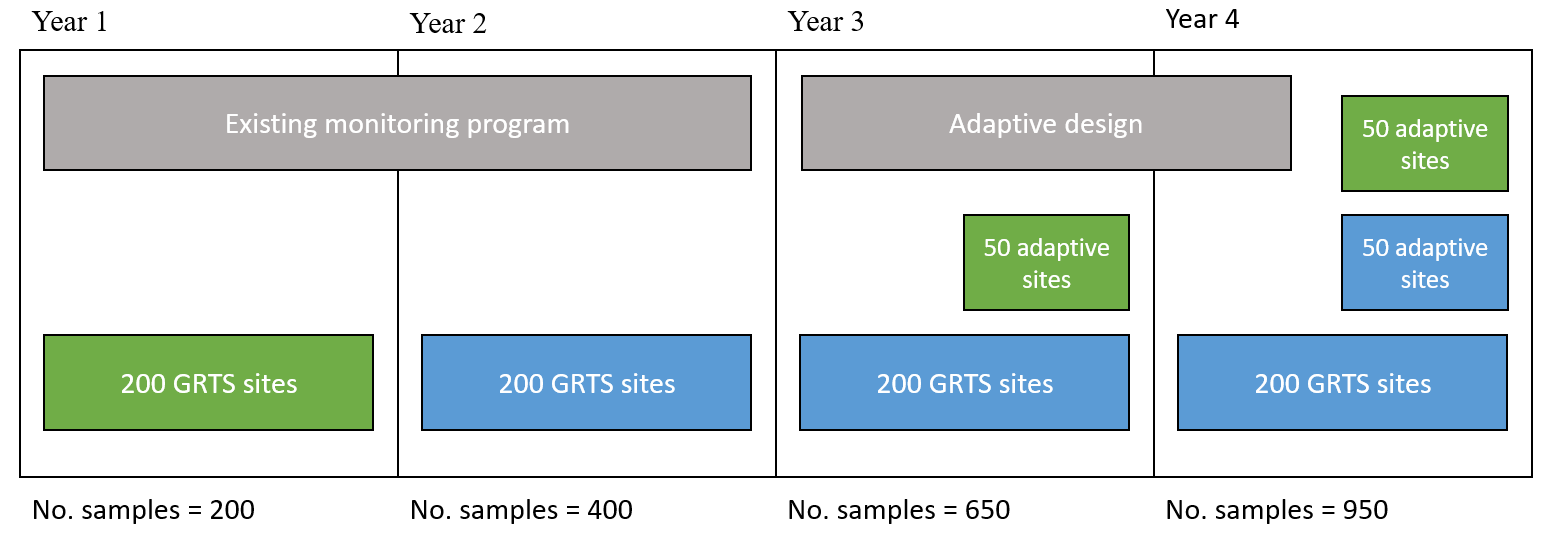}
\captionof{figure}{A schematic diagram showing the adaptive design
process over four years. Green rectangles indicate new sites
(i.e. sampling locations) that have
been added to the monitoring program, and blue rectangles indicate that sites have been retained from previous years.}
\label{Fig:the_plan}
\end{center}

\begin{center}
\includegraphics[width = \textwidth]{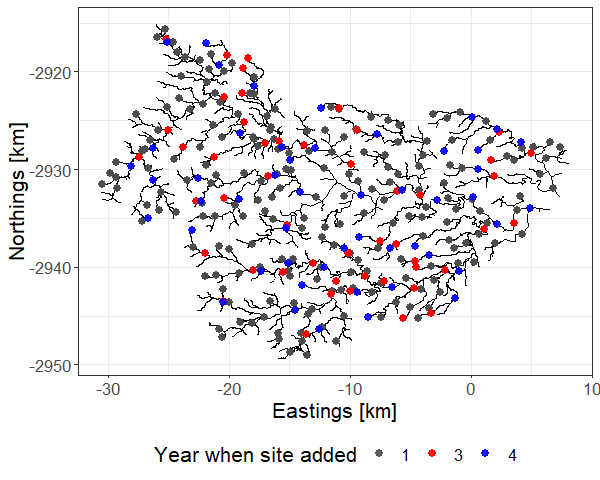}
\captionof{figure}{This is the Pine River stream network. The sites on the network represent the evolution of the adaptive design through time.}
\label{Fig:adaptive_design}
\end{center}

We validated the adaptive design by computing its relative efficiency
compared to 20 GRTS and random designs of the same size. The GRTS
designs were sequentially constructed using the `master sample' approach
(Larsen, Olsen, and Stevens 2008). We compared relative efficiency by
computing the sum of the kriging variances for the same 900 prediction
sites that were used when optimising the design. Note that the sum of
the kriging variances is simply the inverse of the expected utility. For
the purpose of validation, this expected utility was computed using 1000
prior draws to ensure our approximations to the expected utility were
accurate. We did not include the 200 fixed GRTS sites in the validation
procedure because we wanted to assess whether there were any additional
benefits gleaned from the adaptive design. Thus, the efficiencies for
each design (i.e.~adaptive, GRTS, and random) were based on the 150
measurements collected at 100 locations in years 3 and 4.

The results showed that the adaptive design was more efficient than both
the GRTS and random designs, which had 74.5-84\% relative efficiency. In
terms of variance units, using the adaptive design reduces the total
variance across 900 prediction sites by approximately 545-1004 variance
units compared to GRTS and random designs (Fig. \ref{Fig:validation}).
This demonstrates that adaptive design represents a far better
investment in terms of predictive ability than designs formed without
optimisation.

\begin{center}
\includegraphics[width = \textwidth]{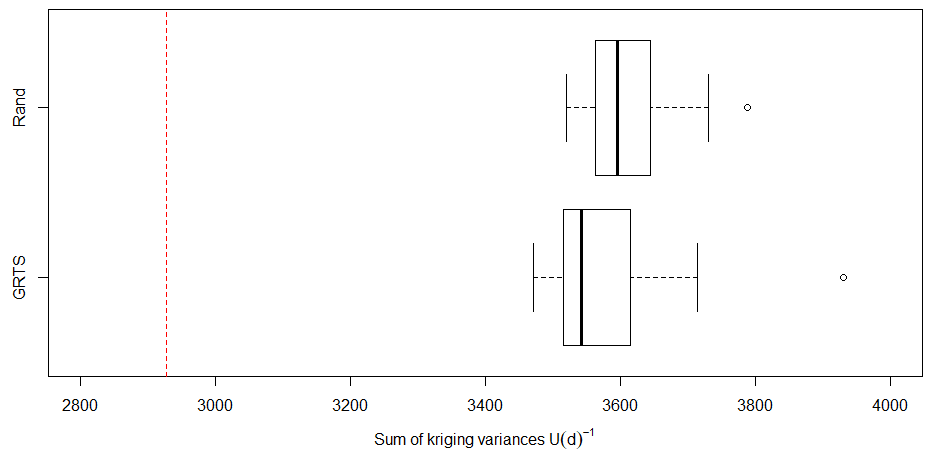}
\captionof{figure}{The sum of kriging variances for both
Generalised Random Tessellation Sampling (GRTS) and random designs
computed using 1000 draws from the priors set on the covariance
parameters. The dashed red line represents the sum of the kriging variances for the adaptive design. The performance of the adaptive design is plotted this way as opposed to as a boxplot because there is only one adaptive design.}  \label{Fig:validation}
\end{center}

\section{Conclusions}

The SSNdesign package brings together a large body of work on optimal
spatial sampling, pseudo-Bayesian experimental design, and the complex
spatial data processing and spatial statistical modelling of stream
network data. We demonstrate how the package can be used in two contexts
which should prove useful for scientists and managers working in stream
ecosystems; particularly where monitoring programs lack the resources to
comprehensively sample the network, but must nevertheless estimate
parameters for complicated spatial processes and accurately predict
in-stream conditions across broad areas.

Compared to other packages for spatial sampling, such as \texttt{geospt}
(C. Melo, Santacruz, and Melo 2012) and spsurvey (Kincaid and Olsen
2018), \texttt{SSNdesign} represents a significant advance in
functionality for stream network data. The \texttt{SSNdesign} package
integrates directly with the data structures, modelling and model
diagnostic functions from a well-established R package for streams,
\texttt{SSN} (Ver Hoef et al. 2014). As a result, the hydrological
distance measures and unique covariance functions for streams data can
also be used in the design process. This cannot be accomplished using
other packages for spatial design, which are restricted to using
Euclidean distance in the conventional 2-dimensional domain of
geostatistics. \texttt{SSNdesign} has been written specifically to deal
with problems of model-based design (i.e.~obtaining the best information
from a model given that a spatial statistical model for in-stream
processes can be specified). This distinguishes \texttt{SSNdesign} from
packages such as \texttt{spsurvey} (Kincaid and Olsen 2018), which
selects designs based on factors such as the inclusion probability for a
site given characteristics of the underlying stream segments. However,
note that we also include tailored functions to simplify the process of
generating designs for stream networks as per Som et al. (2014), which
include designs that can be formed using \texttt{spsurvey}.

In addition to the existing functionality in \texttt{SSNdesign}, there
are opportunities for further development. In particular, the authors of
the \texttt{SSN} package (Ver Hoef et al. 2014) are working on extending
its functionality to include computationally efficient models for big
data (Personal Comm., J. Ver Hoef) and we expect there will be major
performance boosts for the empirical utility functions where spatial
stream network models must be fit iteratively. Computationally efficient
spatial stream network models also open up possibilities to include new
functionality within the \texttt{SSNdesign} package. For example, users
must currently specify a single spatial stream network model as the
`true' model underpinning the design problem. This raises several
important questions about the possibility of true models, and the
uncertainty about which of several plausible, competing models ought to
be chosen as the true model. Increased computational efficiency would
allow us to implement a model-averaging approach (Clyde 1999) so that
users will be able to specify several plausible models as reasonable
approximations for the true model. Expected utilities for designs will
then be averages of the expected utilities of the design under each
plausible model. This approach would make optimal and adaptive designs
more robust, because the averaging mitigates the possibility that
designs are being chosen to obtain the best information about the wrong
model. Our hope is that managers of freshwater monitoring programs can
more efficiently allocate scarce resources using optimal and adaptive
designs for stream networks, which we have made accessible through the
\texttt{SSNdesign} package.

\section{Acknowledgements}\label{acknowledgements}

We thank the SEQ Healthy Land and Water Partnership for their financial
support and guidance. Computational resources and services used in this
work were provided by the eResearch Office, Queensland University of
Technology, Brisbane, Australia.

\section{Software and data
availability}\label{software-and-data-availability}

The \texttt{SSNdesign} package is publicly available at
\url{https://github.com/apear9/SSNdesign}, along with the data used in
this paper. This R package requires R version 3.5.0 or later, and
depends on the packages \texttt{SSN}, \texttt{doParallel},
\texttt{doRNG}, \texttt{spsurvey}, and \texttt{shp2graph}. These will be
downloaded from CRAN during the installation of the package.

\section*{References}\label{references}
\addcontentsline{toc}{section}{References}

\setlength{\parindent}{-0.5in} \setlength{\leftskip}{0.5in}
\setlength{\parskip}{8pt} \noindent

\hypertarget{refs}{}
\hypertarget{ref-Atkinson2007}{}
Atkinson, A. C., A. N. Donev, and R. D. Tobias. 2007. \emph{Optimum
Experimental Designs, with Sas}. Oxford University Press.

\hypertarget{ref-Clyde1999}{}
Clyde, Merlise A. 1999. ``Bayesian Model Averaging and Model Search
Strategies.'' \emph{Bayesian Statistics} 6: 157--85.

\hypertarget{ref-Cressie1993}{}
Cressie, Noel. 1993. \emph{Statistics for Spatial Data}. New York:
Wiley.

\hypertarget{ref-Dudgeon2006}{}
David, Dudgeon, Arthington Angela H., Gessner Mark O., Kawabata
Zen-Ichiro, Knowler Duncan J., Lévêque Christian, Naiman Robert J., et
al. 2006. ``Freshwater Biodiversity: Importance, Threats, Status and
Conservation Challenges.'' \emph{Biological Reviews} 81 (2): 163--82.
doi:\href{https://doi.org/10.1017/S1464793105006950}{10.1017/S1464793105006950}.

\hypertarget{ref-Evangelou2012}{}
Evangelou, E., and Z. Zhu. 2012. ``Optimal Predictive Design
Augmentation for Spatial Generalised Linear Mixed Models.''
\emph{Journal of Statistical Planning and Inference} 142 (12): 3242--53.

\hypertarget{ref-Falk2014}{}
Falk, Matthew G., James M. McGree, and Anthony N. Pettitt. 2014.
``Sampling Designs on Stream Networks Using the Pseudo-Bayesian
Approach.'' \emph{Environmental and Ecological Statistics} 21: 751--73.
doi:\href{https://doi.org/10.1007/s10651-014-0279-2}{10.1007/s10651-014-0279-2}.

\hypertarget{ref-Isaak2017}{}
Isaak, Daniel J., Seth J. Wenger, Erin E. Peterson, Jay M. Ver Hoef,
David E. Nagel, Charles H. Luce, Steven W. Hostetler, et al. 2017. ``The
Norwest Summer Stream Temperature Model and Scenarios for the Western
U.S.: A Crowd-Sourced Database and New Geospatial Tools Foster a User
Community and Predict Broad Climate Warming of Rivers and Streams.''
\emph{Water Resources Research} 53 (11): 9181--9205.
doi:\href{https://doi.org/10.1002/2017WR020969}{10.1002/2017WR020969}.

\hypertarget{ref-Kang2016}{}
Kang, Su Yun, James M. McGree, Christopher C. Drovandi, M. Julian Caley,
and Kerrie L. Mengersen. 2016. ``Bayesian Adaptive Design: Improving the
Effectiveness of Monitoring of the Great Barrier Reef.''
\emph{Ecological Applications} 26 (8): 2637--48.
doi:\href{https://doi.org/10.1002/eap.1409}{10.1002/eap.1409}.

\hypertarget{ref-spsurvey}{}
Kincaid, Thomas M., and Anthony R. Olsen. 2018. \emph{Spsurvey: Spatial
Survey Design and Analysis}.

\hypertarget{ref-Larsen2008}{}
Larsen, David P., Anthony R. Olsen, and Donald L. Stevens. 2008. ``Using
a Master Sample to Integrate Stream Monitoring Programs.'' \emph{Journal
of the Agricultural, Biological and Environmental Statistics} 13 (3):
243--54.
doi:\href{https://doi.org/10.1198/10857110X336593}{10.1198/10857110X336593}.

\hypertarget{ref-mcgree_et_al_adapt}{}
McGree, James M., Chris C. Drovandi, Mary H. Thompson, John A.
Eccleston, Steve B. Duffull, Kerrie Mengersen, Antony N. Pettitt, and
Tim Goggin. 2012. ``Adaptive Bayesian Compound Designs for Dose Finding
Studies.'' \emph{Journal of Statistical Planning and Inference} 142 (6):
1480--92.

\hypertarget{ref-geospt}{}
Melo, C., A. Santacruz, and O. Melo. 2012. \emph{Geospt: An R Package
for Spatial Statistics.} \url{geospt.r-forge.r-project.org/}.

\hypertarget{ref-Muller}{}
Mueller, Peter. 1999. ``Simulation-Based Optimal Design.''
\emph{Bayesian Statistics} 6: 459--74.

\hypertarget{ref-Muller2007}{}
Mueller, Peter, Don A. Berry, Andy P. Grieve, Michael Smith, and Michael
Krams. 2007. ``Simulation Based Sequential Bayesian Design.''
\emph{Journal of Statistical Planning and Inference} 137: 3140--50.
doi:\href{https://doi.org/10.1016/j.jspi.2006.05.021}{10.1016/j.jspi.2006.05.021}.

\hypertarget{ref-acebayes}{}
Overstall, Antony M., David C. Woods, and Maria Adamou. 2017.
\emph{Acebayes: Optimal Bayesian Experimental Design Using the Ace
Algorithm}. \url{https://CRAN.R-project.org/package=acebayes}.

\hypertarget{ref-STARS}{}
Peterson, Erin E., and Jay M. Ver Hoef. 2014. ``STARS: An Arcgis Toolset
Used to Calculate the Spatial Data Needed to Fit Spatial Statistical
Models to Stream Network Data.'' \emph{Journal of Statistical Software}
56 (2): 1--17.

\hypertarget{ref-Peterson2013}{}
Peterson, Erin E., Jay M. Ver Hoef, D. J. Isaak, J. A. Falke, M. J.
Fortin, C. E. Jordan, Kristina McNyset, et al. 2013. ``Modelling
Dendritic Ecological Networks in Space: An Integrated Network
Perspective.'' \emph{Ecology Letters} 16 (5).
doi:\href{https://doi.org/10.1111/ele.12084}{10.1111/ele.12084}.

\hypertarget{ref-Poff2012}{}
Poff, N. LeRoy, Julian D. Olden, and David L. Strayer. 2012. ``Climate
Change and Freshwater Fauna Extinction Risk.'' In \emph{Saving a Million
Species: Extinction Risk from Climate Change}, edited by Lee Hannah,
309--36. Washington, DC: Island Press/Center for Resource Economics.
doi:\href{https://doi.org/10.5822/978-1-61091-182-5_17}{10.5822/978-1-61091-182-5\_17}.

\hypertarget{ref-RCoreTeam}{}
R Core Team. 2018. \emph{R: A Language and Environment for Statistical
Computing}. Vienna, Austria: R Foundation for Statistical Computing.
\url{https://www.R-project.org/}.

\hypertarget{ref-Royle2002}{}
Royle, J. Andrew. 2002. ``Exchange Algorithms for Constructing Large
Spatial Designs.'' \emph{Journal of Statistical Planning and Inference}
100 (2): 121--34.

\hypertarget{ref-Som2014}{}
Som, Nicholas A., Pascal Monestiez, Jay M. Ver Hoef, Dale L. Zimmerman,
and Erin E. Peterson. 2014. ``Spatial Sampling on Streams: Principles
for Inference on Aquatic Networks.'' \emph{Environmetrics} 25 (5):
306--23. doi:\href{https://doi.org/10.1002/env.2284}{10.1002/env.2284}.

\hypertarget{ref-Stevens2004}{}
Stevens, Donald L., and Anthony R. Olsen. 2004. ``Spatially Balanced
Sampling of Natural Resources.'' \emph{Journal of the American
Statistical Association} 99 (465): 262--78.
doi:\href{https://doi.org/10.1198/016214504000000250}{10.1198/016214504000000250}.

\hypertarget{ref-UNWATER2016}{}
United Nations Water. 2016. ``The Right to Privacy in the Digital Age.''
Annual report of the United Nations High Commissioner for Human Rights
and reports of the Office of the High Commissioner and the
Secretary-General. Geneva, Switzerland.

\hypertarget{ref-VerHoef2018}{}
Ver Hoef, Jay M. 2018. ``Kriging Models for Linear Networks and
Non-Euclidean Distances: Cautions and Solutions.'' \emph{Methods in
Ecology and Evolution} 9: 1600--1613.
doi:\href{https://doi.org/10.1111/2041-210X.12979}{10.1111/2041-210X.12979}.

\hypertarget{ref-HoefPetersonMA}{}
Ver Hoef, Jay M., and Erin E. Peterson. 2010. ``A Moving Average
Approach for Spatial Statistical Models of Stream Networks.''
\emph{Journal of the American Statistical Association} 105 (489).
doi:\href{https://doi.org/10.1198/jasa.2009.ap09248}{10.1198/jasa.2009.ap09248}.

\hypertarget{ref-VerHoef2001}{}
Ver Hoef, Jay M., Noel Cressie, Robert N. Fisher, and Ted J. Case. 2001.
``Uncertainty and Spatial Linear Models for Ecological Data.'' In
\emph{Spatial Uncertainty in Ecology}, 214--37. New York, NY: Springer.

\hypertarget{ref-VerHoef2006}{}
Ver Hoef, Jay M., Erin E. Peterson, and David Theobald. 2006. ``Spatial
Statistical Models That Use Flow and Stream Distance.''
\emph{Environmental and Ecological Statistics} 13: 449--64.
doi:\href{https://doi.org/10.1007/s10651-006-0022-8}{10.1007/s10651-006-0022-8}.

\hypertarget{ref-SSNJOSS}{}
Ver Hoef, Jay M., Erin E. Peterson, David Clifford, and Rohan Shah.
2014. ``SSN: An R Package for Spatial Statistical Modelling on Stream
Networks.'' \emph{Journal of Statistical Software} 56 (3).
doi:\href{https://doi.org/10.18637/jss.v056.i03}{10.18637/jss.v056.i03}.

\hypertarget{ref-Vorosmarty2010}{}
Vorosmarty, C. J., P. B. McIntyre, M. O. Gessner, D. Dudgeon, A.
Prusevich, P. Green, S. Glidden, et al. 2010. ``Global Threats to Human
Water Security and River Biodiversity.'' \emph{Nature} 467: 555--61.
doi:\href{https://doi.org/10.1038/Nature09440}{10.1038/Nature09440}.

\end{document}

\maketitle

\section{Notation for geostatistical models and experimental
designs}\label{notation-for-geostatistical-models-and-experimental-designs}

Geostatistical models (Cressie 1993), specifically, spatial stream
network models (Ver Hoef and Peterson 2010), are a fundamental aspect of
our work in experimental design for stream networks. Our utility
functions which enable optimisation of experimental designs are derived
from various matrices and theoretical aspects of these models. Here, we
provide definitions of fundamental statistical elements of these models
(Table 1).

\begin{figure}
\captionof{table}{Definitions of statistical elements of spatial stream network models}
\centering
\includegraphics[width = \textwidth]{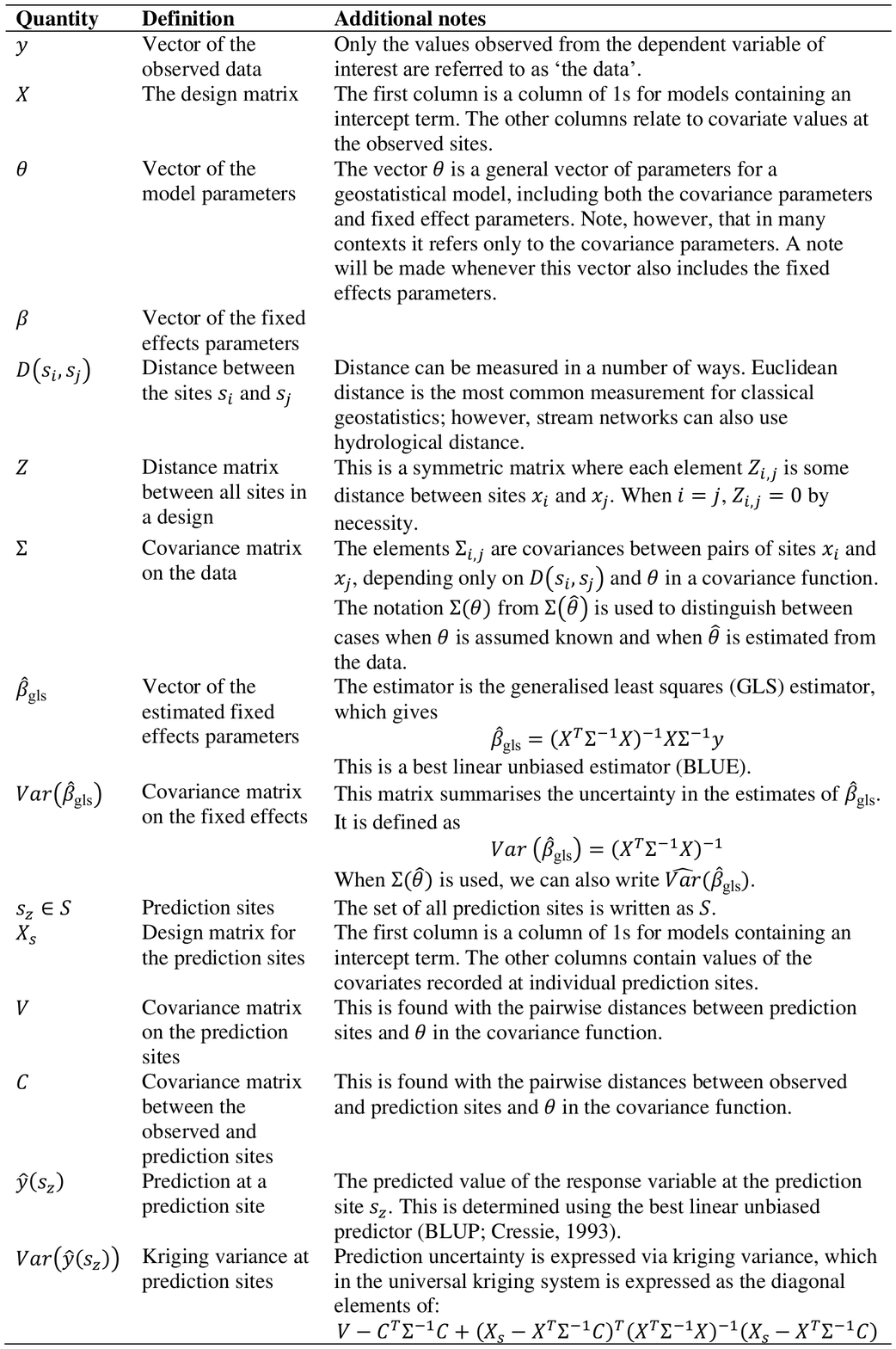}
\end{figure}

\section{Moving average autocovariance models for stream
networks}\label{moving-average-autocovariance-models-for-stream-networks}

As geostatistical domains, stream networks exhibit greater constraints
than the conventional domain of 2D Euclidean space. The unique branching
nature of stream networks and the unidirectional flow of water within
these branches necessitates the development and use of special spatial
covariance functions. This is what distinguishes stream network models
and the design of experiments for stream network models from similar
problems in geostatistics.

Consider a stream network as a set of lines (stream segments) that
branch upstream from the most downstream segment on the network (outlet
segment) to the most upstream segments on the network (headwaters). We
assume that the branching is binary (i.e., three or more segments never
branch upstream from the same confluence). Observations are represented
by points on the network, which have two coordinate systems (Peterson et
al. 2013); one is the usual two-dimensional coordinate system, and the
other is based on the network topology (i.e., branching structure and
connectivity of segments). Note that separation distance between two
locations along the network (e.g., stream distance) is the shortest
distance between them when movement is confined to the network (Dent and
Grimm 1999). If water flows from an upstream location to a downstream
location, we refer to these locations as flow-connected, and we refer to
two locations within the same network not connected by flowing water as
flow-unconnected.

Models for stream networks, based on moving average constructions, were
initially described by Ver Hoef, Peterson, and Theobald (2006) and
Cressie et al. (2006). The models summarized in Ver Hoef and Peterson
(2010) extend this work and use a spatial moving-average approach to
construct Gaussian random fields based on the network topology, rather
than the usual two-dimensional coordinate system commonly used in
geostatistics. These approaches yield random processes that are similar
to typical geostatistical models; they can be described by a mean
function that depends on the location within the network, and a
second-order, stationary covariance function that depends on the
separation distance between two locations.

Using the moving average constructions, if a moving average function
starts at some location and is non-zero only upstream of that location,
it is called a ``tail-up'' model. The function must split at confluences
as it goes upstream to maintain stationarity of variances, so some
weighting of segments must occur. If a moving average function starts at
some location and is non-zero only downstream of that location, it is
called a ``tail-down'' model. Consider two pairs of sites that have the
same stream distance between them, but one pair is flow-connected, and
the other pair is flow-unconnected; in general the amount of
autocorrelation will be different between them. Let \(r_i\) and \(s_j\)
denote two locations on a stream network, and let \(h\) be the stream
distance between them. Then the following models have been developed to
describe different forms of covariance of the response at locations
\(r_i\) and \(s_j\).

The moving average construction for tail-up models, as described by Ver
Hoef, Peterson, and Theobald (2006), is

\begin{equation}
\label{eq:moveaveup}
C_u(r_i,s_j|\theta_u)=  \left\{ \begin{array}{ll} \pi_{i,j}C_t(h|\theta_u) &
      \textrm{if $r_i$ and $s_j$ are flow-connected,} \\
      0 & \textrm{if $r_i$ and $s_j$ are flow-unconnected,}
    \end{array} \right.
\end{equation}

where \(C_u(r_i,s_j|\theta_u)\) is the spatial autocovariance between
\(r_i\) and \(s_j\), \(u\) denotes a tail-up model, \(\theta_u\) is the
set of covariance function parameters, \(C_t(h|\theta_u)\) is the value
of a covariance function based on \(h\), \(\theta_u\), and a selected
covariance model (e.g., exponential), and \(\pi_{i,j}\) are weights to
account for the branching characteristics of the stream and maintain
variance stationarity. The weights reflect the relative shared flow
among locations, and more details can be found in Ver Hoef and Peterson
(2010), including ways to create an additive function from values
associated with stream segments, such as flow volume, a proxy for flow
volume (e.g., basin area), or any other ecologically relevant variable.

For this introduction we focus on the exponential stream-network
covariance function because its geostatistical counterpart is frequently
applied by practitioners, but there are many other useful covariance
functions, and we encourage interested readers to explore them among the
stream-network covariance model citations. For the exponential
stream-network covariance function, \(C_t(h|\theta_u)\) has the
following form (Ver Hoef, Peterson, and Theobald 2006):

\begin{equation}
\label{eq:tailup}
  C_t(h|\theta_u)= \sigma^2_u\exp(-3h/\alpha_u),
\end{equation}

where \(\sigma^2_u > 0\) is an overall variance parameter (also known as
the partial sill), \(\alpha_u > 0\) is the range parameter, and
\(\theta_u = (\sigma^2_u,\alpha_u)^{\prime}\). Via equation
(\ref{eq:moveaveup}), spatial autocorrelation is only permitted between
flow-connected locations in the tail-up model.

For tail-down models, spatial autocorrelation is permitted between both
flow-connected and flow-unconnected locations, but we generally
distinguish between the two cases. When two sites are flow-unconnected,
there will always be at least one common confluence (i.e., a downstream
confluence that receives water from each of the two upstream sites). Let
\(b\) denote the longer of the two distances to the closest common
downstream confluence, and \(a\) denote the shorter of the two
distances. If two sites are flow-connected, again use \(h\) to denote
their stream distance. Again, the only tail-down model we consider is
the exponential, defined as follows:

\begin{equation} 
\label{eq:taildown}
    C_d(a,b,h|\theta_d)= \left\{ \begin{array}{ll}
      \sigma^2_d\exp(-3h/\alpha_d) &
      \textrm{if flow-connected,}\\
      \sigma^2_d\exp(-3(a+b)/\alpha_d) &
      \textrm{if flow-unconnected,}
    \end{array} \right.
\end{equation}

where \(C_d(a,b,h|\theta_d)\) is the spatial autocovariance between
\(r_i\) and \(s_j\), \(\sigma^2_d > 0\) is an overall variance
parameter, \(\alpha_d > 0\) is the range parameter,
\(\theta_d = (\sigma^2_d,\alpha_d)^{\prime}\), and \(d\) denotes a
tail-down model. We note, for the exponential model, that when
\(a + b = h\) the flow-connected and flow-unconnected models are
equivalent, and stress this is a unique property of the exponential form
of tail-down covariance models (Garreta, Monestiez, and Ver Hoef 2010).
A full development and more detail regarding the suite of stream-network
moving-average models can be found in Ver Hoef and Peterson (2010).

A mixed linear model combining tail-up and tail-down components is

\begin{equation} 
\label{eq:spLinearModel}
\bf{Y} = X\beta + z_u + z_d + \epsilon,
\end{equation}

where \(Y\) is the vector of random variables for an observable stream
attribute at sampled locations, \(X\) is a design matrix of fixed
effects, \(\beta\) contains fixed effects parameters, \(z_u\) contains
spatially-autocorrelated random variables with a tail-up autocovariance
(e.g., equation (\ref{eq:tailup})), with
\(\textrm{var}(z_u) = \sigma_u^2R(\alpha_u)\) and \(R(\alpha_u)\) is a
correlation matrix that depends on the range parameter \(\alpha_u\);
\(z_d\) contains spatially-autocorrelated random variables with a
tail-down autocovariance (e.g., equation (\ref{eq:taildown})) such that
\(\textrm{var}(z_d) =\sigma_d^2R(\alpha_d)\); and \(\epsilon\) contains
independent random variables with
\(\textrm{var}(\epsilon)=\sigma^2_0I\). When used for spatial
prediction, this model is referred to as ``universal'' kriging (Le and
Zidek 2006), with ``ordinary''" kriging being the special case where the
design matrix \(X\) is a single column of ones (Cressie 1993). This
yields a covariance matrix of the form

\begin{equation} 
\label{eq:covStructure}
\textrm{var}(Y)=\Sigma=\sigma_u^2R(\alpha_u) + \sigma_d^2R(\alpha_d) +
     \sigma_0^2I.
\end{equation}

\section{The expected utility}\label{the-expected-utility}

In Bayesian and pseudo-Bayesian experimental design, an optimal design
\(d^*\) is found by maximising an expected utility function \(U(d)\)
through the choice of design \(d\) from a set of possible designs \(D\).
The expected utility function is specified to capture the goal of data
collection, such as precise estimation of model parameters and accurate
prediction of a response at unobserved locations.

To define the expected utility, we first define the utility function
denoted as \(U(d,\theta,y)\) which depends on a vector of parameters
from a geostatistical model \(\theta \sim p(\theta)\) and the data we
expect to observe under that model \(y \sim p(y|\theta,d)\). Note,
however, that many pseudo-Bayesian utility functions do not depend on
\(y\) and so in many cases the utility function can simply be written
\(U(d, \theta)\) (Falk, McGree, and Pettitt 2014). The utility function
then quantifies the aim of data collection. For example, one may be
interested in the precise estimation of the parameters. In this case, we
could define our utility function as the negative sum of the variances
for each parameter estimate. We take the negative sum because we
generally define utility functions such that they should be maximised.
As the notation suggests, such a quantity depends on the design \(d\),
the data \(y\) and on \(\theta\) through the likelihood of \(y\).
However, in reality, we do not know what data will be observed, and
hence we cannot evaluate \(U(d,\theta,y)\) directly to design
experiments. Instead, we use prior information about \(\theta\) and
\(y\) to capture their join distribution, and integrate
\(U(d,\theta,y)\) over this uncertainty. This leads to the following
definition of the expected utility (Chaloner and Verdinelli 1995):

\begin{equation}
    \begin{aligned}
        U(d) = \int_\theta\int_y U(d, \theta, y)p(y|\theta, d)p(\theta) ~dy~d\theta. \label{eqn:integrate_out_y_theta}
    \end{aligned}
\end{equation}

This integral is slightly modified for pseudo-Bayesian utility functions
\(U(d, \theta)\) that do not depend on \(y\). For these utility
functions, the integral which gives the expected utility is simplified
to an integral over the parameters \(\theta\) such that

\begin{equation}
    \begin{aligned}
        U(d) = \int_\theta U(d, \theta)p(\theta)~d\theta. \label{eqn:integrate_out_theta}
    \end{aligned}
\end{equation}

Unfortunately, the above integrals are analytically intractable for most
applications, meaning that they have no closed form solution. In
practice, this is inconvenient but we can still approximate the expected
utility. Monte Carlo integration (Algorithm \ref{Alg:exp_uti}) is
commonly used for this purpose (Mueller 1999). For a utility function
\(U(d, \theta, y)\), Monte Carlo integration requires taking \(M\) draws
from the prior \(\theta^{(m)} \sim p(\theta)\) and then the likelihood
\(y^{(m)} \sim p(y|\theta^{(m)}, d)\). For each \(m\), the utility
function is evaluated for \(\theta^{(m)}, y^{(m)}\) to give
\(U(d,\theta^{(m)},y^{(m)})\). For a utility function \(U(d, \theta)\),
the process is the same but we ignore
\(y^{(m)} \sim p(y|\theta^{(m)}, d)\). The values of the utilities are
then averaged such that
\(U(d) \approx \sum_{m=1}^MU(d,\theta^{(m)},y^{(m)})/M\) or
\(U(d) \approx \sum_{m=1}^MU(d,\theta^{(m)})/M\) depending on whether
the utility funciton depends on the parameters and the data or the
parameters only. In order to accurately estimate \(U(d)\), we need large
\(M\), which usually means \(M \geq 500\).

\begin{algorithm}
\caption{Algorithm for estimating the expected utility $U(d)$ by Monte Carlo integration}
\label{Alg:exp_uti}
\begin{algorithmic}[1]
\STATE Specify $U(d, \theta, y)$ or $U(d, \theta)$ as appropriate.
\STATE Specify a prior $p(\theta)$ and, if necessary, the likelihood $p(y | \theta, d)$.
\STATE Set $M$ to be the total number of Monte Carlo draws to be used for approximating $U(d)$.
\FOR {$m = 1:M$ \# each Monte Carlo draw \#}
\STATE Take $\theta^{(m)} \sim p(\theta)$.
\STATE Take $y^{(m)} \sim p(y|\theta^{(m)}, d)$ if required.
\STATE Evaluate $U(d, \theta^{(m)}, y^{(m)})$ or $U(d, \theta^{(m)})$.
\ENDFOR
\STATE Evaluate $U(d) \approx \sum_{m = 1}^M U(d, \theta^{(m)}, y^{(m)}) / M$ or $U(d) \approx \sum_{m = 1}^M U(d, \theta^{(m)}) / M$ as appropriate.
\end{algorithmic}
\end{algorithm}

\section{Searching for an optimal
design}\label{searching-for-an-optimal-design}

In this section, we outline how we maximise the expected utility
function over a set of possible designs (Algorithm 2). The use of
optimisation algorithms such as exchange algorithms (Royle 2002) is
necessary because many design problems are impossible to solve
analytically and are too large to efficiently solve numerically with a
computer under a brute-force search scheme. If one wants to find an
optimal design of size \(n\) and there are \(N\) sites to choose from,
then the optimal design \(d^*\) will exist somewhere among \(N\) choose
\(n\) potential solutions. This number is exceedingly large for all but
the smallest values of \(N\). Therefore, an optimisation algorithm is
used to greatly reduce the costs associated with the search for \(d^*\)
(Royle 2002).

In this work, we use a greedy exchange algorithm (Algorithm
\ref{Alg:gea}) to locate optimal designs (Evangelou and Zhu 2012; Falk,
McGree, and Pettitt 2014). The greedy exchange algorithm works by
optimising the choice of each of \(n\) sites one-by-one. A random design
with \(n\) sites is proposed and becomes \(d_0\) (the initial design)
and \(d^*\) (the design which currently has the highest value of
\(U(d)\)). From this point, we begin the coordinate exchange. Note that
there are \(N - n\) candidate points not currently in \(d^*\). The first
of the \(n\) sites in \(d_0\) is then swapped out for each of the
\(N - n\) candidate sites. The expected utilities of the resulting
designs are recorded, and, if any design has an expected utility larger
than \(U(d^*)\), this design replaces \(d^*\), we update our pool of
candidate sites, and we begin to exchange the next site. Otherwise, the
design reverts to \(d^*\) and the next site in the design is exchanged
for each candidate site. This process continues until we have exchanged
all \(n\) sites. If \(d^*\) changed at any point in this process
(i.e.~if we see any improvement in the design), we repeat the sequence
of exchanges. This continues until we finally observe no improvement in
the expected utility of \(d^*\). We then exit the algorithm. The
algorithm is called the greedy exchange algorithm because it only
accepts improvements in the design and stops when no further improvement
can be achieved.

\begin{algorithm}
\caption{The greedy exchange algorithm}
\label{Alg:gea}
\begin{algorithmic}[1]
\STATE Set $K$, the number of searches from random starts. This is to mitigate against becoming trapped in local maxima.
\FOR {$k = 1:K$}
\STATE Initialise $d_0$ as a randomly selected design with $n$ of $N$ points.
\STATE Set $d^* = d_0$
\STATE Evaluate $U_0 = U(d_0)$ to initialise the search for designs
\STATE Store $U^{*} = U_0$; the expected utility of the global `best design'.
\STATE Temporarily store $U_{ij}^* = U^{*} + \epsilon$ (the expected utility of the `best-within-search' design) for small $\epsilon$. (This is simply to force the while loop to iterate at least once.)
\STATE Initialise $U_{ij} = \emptyset$. $U_{ij}$ will be used to store the expected utilities for designs evaluated during the following search.
\WHILE {$U_{ij}^* > U^{*}$ \# until there is no improvement \#}
\STATE Update $U^{*} = U_{ij}^*$ 
\FOR {$i = 1:n$ \# each point currently in the design \#}
\STATE Find the $N - n$ points not in $d^*$
\FOR {$j = 1:(N - n)$ \# each point not in the design \#}
\STATE Form $d_{ij}$ by swapping out the $i^{th}$ point in the design for the $j^{th}$ point not in the design 
\STATE Evaluate $U = U(d_{ij})$
\STATE Define $U_{ij} = U_{ij} \cup U$
\ENDFOR
\IF {$\max(U_{ij}) > U_{ij}^{*}$}
  \STATE Update $d^* = \text{argmax}_{d\in D} U(d_{ij})$
  \STATE Update $U_{ij}^* = U(d^*)$
\ELSE
  \STATE Keep the previous $d^*$ and $U_{ij}^*$.
\ENDIF
\ENDFOR
\ENDWHILE
\ENDFOR
\end{algorithmic}
\end{algorithm}

The greedy exchange algorithm can be analysed to yield an approximate
expression for the run-time. We first assume that estimating \(U(d)\) is
the most time-consuming part of the algorithm and that any intermediate
storage operations and data manipulation between evaluations of \(U(d)\)
are inconsequential. Let us assume that it takes \(S\) seconds to
evaluate \(U(d)\). The number of expected utilities that are calculated
in the greedy exchange algorithm each time an optimal design is found
(for each of \(K\) iterations, in Algorithm 2), is stochastic. However,
the stochasticity is due to only a single step (the condition at Line 9,
Algorithm 2) and the number of times the expected utility must be
estimated is otherwise well-constrained. There are \(L\) iterations, and
\(L\) is random. In every iteration of the greedy exchange, there are
\(n \times (N - n)\) designs to be evaluated. This process is repeated
\(K\) times from \(K\) random starts, so the number of times the
expected utility is estimated is
\(K \times L \times n \times (N - n) + K\). The additional \(K\)
evaluations are from the random starts, which must be evaluated.
However, \(K\) extra evaluations of \(U(d)\) are unlikely to be of any
consequence for calculating expected run-time, so we discard them.
Altogether, the expected run-time for the greedy exchange algorithm is
\(K \times L \times n \times (N - n) \times S\) seconds. However, some
utility functions have large \(S\) and therefore expected runtimes may
still be large. Therefore, we use parallel computing to further reduce
the total runtime. Let \(C\) be the number of CPUs across which the
greedy exchange algorithm is to be run. Then the expected runtime
reduces to approximately
\((K \times L \times n \times (N - n) \times S) / C\) seconds.

\section{Utility functions for static optimal
design}\label{utility-functions-for-static-optimal-design}

In this section, we set out detailed notation and explanations of our
utility functions. In our package, we have provided an off-the-shelf
utility function for

\begin{itemize}
\tightlist
\item
  Precision of covariance parameter estimation
\item
  Precision of fixed effect parameter estimation
\item
  Precision of estimation of both covariance and fixed effect parameters
\item
  Precision of predictions, and
\item
  Approximately evenly-spaced sites in the stream network.
\end{itemize}

Our covariance parameter estimation utility is called CP-optimality and
was used in both Falk, McGree, and Pettitt (2014) and N. A. Som et al.
(2014). It is given by

\begin{equation}
    \begin{aligned}
        U(d,\theta) &= \log\det\left[I(d, \theta)\right],\label{eqn:CP}
    \end{aligned}
\end{equation}

where \(I(d, \theta)\) is the expected Fisher information for the
covariance parameters. To compute the expected Fisher information, we
use the restricted error maximum likelihood (REML) estimator (N. A. Som
et al. 2014; Falk, McGree, and Pettitt 2014). This means each element
\([I_{i,j}(\theta, d)]\) is defined by

\begin{equation}
    \begin{aligned}
        I_{i,j}(\theta, d) &= \frac{1}{2}\text{tr}\left(\frac{\partial\Sigma}{\partial\theta_i}P\frac{\partial\Sigma}{\partial\theta_j}P\right),\label{eqn:FI}
    \end{aligned}
\end{equation}

where the matrix \(P\) in Eq. \ref{eqn:FI} is defined as
\(P = \Sigma^{-1} - \Sigma^{-1}X\left(X^T\Sigma^{-1}X\right)^{-1}X^T\Sigma^{-1}\).
This utility function works because larger values of
\(\det[I(d, \theta)]\) correspond to lower uncertainties on \(\theta\),
as given by the elements of \(I(d, \theta)\).

Our fixed effects estimation utility is called D-optimality (Falk,
McGree, and Pettitt 2014; N. A. Som et al. 2014). This utility works on
the same principle as CP-optimality, though it minimises the uncertainty
in a different set of parameters. Formally, the utility function is

\begin{equation}
    \begin{aligned}
        U(d,\theta) &= \log\det\left[I(d, \beta_{\text{gls}})\right],\label{eqn:D}
    \end{aligned}
\end{equation}

where \(I(d, \beta_{\text{gls}}) = Var(\beta_{\text{gls}}, d)^{-1}\) is
the Fisher information for the fixed effects parameters. Note that this
assumes the covariance parameters are known up to a prior. For cases
when there is little information about the covariance parameters and it
is advantageous to estimate them from the data, we use empirical
D-optimality (ED-optimality, after N. A. Som et al. (2014)). In this
case, the criterion is modified from Eq. \ref{eqn:D} such that

\begin{equation}
    \begin{aligned}
        U(d,\theta, y) &= \log\det\left[I(d, \hat{\beta}_{\text{gls}})\right],\label{eqn:ED}
    \end{aligned}
\end{equation}

where \(I(d, \hat{\beta})\) is the observed Fisher information for the
fixed effects parameters. For the empirical D-optimality utility
function, the vector \(\theta\) includes the fixed effects \(\beta\).
These are needed to generate the data \(y\) from which
\(\hat{\beta_{\text{gls}}}\) are estimated. Though Som et al. (2014)
adjust the utility function with the addition of another quantity
derived from the inverse Fisher information, we do not. Their reasoning
for making this adjustment was to account for changes to the sampling
distributions of the fixed effects when the covariance parameters are
estimated from the data. However, since we are averaging over a set of
prior draws for the covariance parameters, we are in effect constructing
the sampling distribution of the fixed effects through simulation.

A dual purpose utility function is also defined for improving the
precision of both fixed effects and covariance parameter estimates at
the same time. We call this CPD-optimality. Instead of considering the
information matrices for the fixed effects and covariance parameters
separately, we consider a combination of the two as a block diagonal
matrix such that

\begin{equation}
F = \begin{pmatrix} D & 0 \\ 0 & C \end{pmatrix},
\end{equation}

where \(D = I(d, \beta_{\text{gls}})\) and \(C = I(d, \theta)\). Again
we define our utility function as the log-determinant of this matrix,
which reduces to

\begin{equation}
  U(d, \theta) = \log\left[\det\left(D\right)\det\left(C\right)\right] = \log\left[\det\left(D\right)\right] + \log\left[\det\left(C\right)\right].
\end{equation}

This is simply the sum of the two utility functions D- and
CP-optimality.

Our prediction utility is called K-optimality, where K is for kriging.
It is the inverse sum of the kriging variances defined at a set of
prediction sites \(s_z \in S\) for \(z = 1, ..., Z\) where \(Z\) is the
number of prediction sites. This utility function favours designs where
the total uncertainty is small. When covariance parameters are known (Z.
Zhu and Stein 2006), this is

\begin{equation}
    \begin{aligned}
        U(d,\theta) &= \left(\sum_{s_z \in S}\text{var}\left(\hat{y}(s_z, \theta), d\right)\right)^{-1}.\label{eqn:K}
    \end{aligned}
\end{equation}

We use the universal kriging system to estimate the kriging variances
(Cressie 1993). When we need to empirically estimate the covariance
parameters due to a lack of strong beliefs about them, we can use
empirical kriging variances. In this situation, we get the empirical
K-optimality function (EK-optimality), which is

\begin{equation}
    \begin{aligned}
        U(d,\theta, y) &= \left(\sum_{s_z \in S}\hat{\text{var}}\left(\hat{y}(s_z, \hat{\theta}), d\right)\right)^{-1}.\label{eqn:EK}
    \end{aligned}
\end{equation}

The vector \(\theta\) includes the fixed effects \(\beta\) because they
are needed to generate \(y^{(m)} \sim p(y | \theta^{(m)}, d)\) in Alg.
\ref{Alg:exp_uti}. Note there is a parameter estimation step in this
empirical utility, so it serves the dual purpose of prediction accuracy
and parameter estimation (Falk, McGree, and Pettitt 2014).

Two space-filling utilities are also provided in the package.
Space-filling designs are used to construct designs with roughly equally
spaced and unclustered sets of monitoring sites in space. The first
space-filling utility function is the maximin utility function (Pronzato
and Muller 2012), which is

\begin{equation}
U(d) = \min_{i\neq j} D(s_i, s_j) \label{eqn:maximin},
\end{equation}

where the distance \(D(s_i, s_j)\) (Table 1) can be either Euclidean or
hydrological distance (Ver Hoef and Peterson 2010). This utility
function unsurprisingly favours configurations of sites that maximise
the minimum distance among any two sites. The second is the modified
maximin design criterion proposed by Morris and Mitchell (1995). This is

\begin{equation}
  U(d) = -\left(\sum_{i=1}^w(J_iZ_i)^{p}\right)^{1/p} \label{eqn:MorrisMitchell}.
\end{equation}

In this utility function, \(w\) is the number of unique non-zero
distances between pairs of points in a design. The vector \(Z\) contains
\(w\) distance elements sorted from smallest to largest. The vector
\(J\) contains the number of times each of these distances occur in one
triangle (upper or lower) of the distance matrix for a given design. The
parameter \(p\) is a weighting power which determines the weighting to
be given to smaller distances. As \(p \rightarrow \infty\), the
contribution of the smallest non-zero distance \(Z_1\) to the utility
will far outweigh the contribution of any other term in the sum and this
utility will reduce to the maximin utility described earlier. Note that
the value of \(p\) is arbitrary and user defined. Morris and Mitchell
(1995) recommend that \(p\) be set between \(p \in [20, 40]\) but any
\(p > 1\) is viable. Compared to the maximin utility function, this
utility function has the advantage of being able to incorporate
information about the distances between pairs of points in the design
which are larger than the minimum distance, with a view to providing a
more spatially balanced design where not only is the minimum distance
between points large but that larger distances also increase
accordingly. As a final note on these two utility functions, it can be
seen that neither depend on \(\theta\) or \(y\). Therefore, no
integration is needed to obtain the expected utility.

\section{Utility functions for adaptive
design}\label{utility-functions-for-adaptive-design}

Adaptive designs differ from optimal designs because, instead of making
a single decision about where to sample within a stream network,
adaptive designs involve a series of decisions about where to sample
that evolve over time as new data becomes available. We use a myopic
design approach in \texttt{SSNdesign}, which means that we only look one
step forward in the series of design decisions we have to make and try
to make the best decision for the next timestep only. This is in
contrast to backward induction, which involves enumerating every
possible decision we could make in the future and selecting the series
of decisions that, retrospectively, should lead to the best result
(Mueller et al. 2007).

Adaptive designs account for the designs used and data collected at
previous timesteps by modifying the expected utility function (Algorithm
\ref{Alg:seq_design}). Let \(t\) be a timestep with
\(t = 0, 1, 2, ..., T\) for some total number of time periods \(T\). At
time period \(t\), a total of \(t-1\) design decisions and datasets have
been collected. In adaptive design, we leverage this information to
improve our design. Therefore, our expected utility function can be
written as \(U(d|d_{0:t-1}, y_{0:t-1})\). That is, the utility of any
design under consideration in the current time period depends on the
designs and data from all previous time periods. To avoid continually
refitting models to a potentially large number of data points (i.e.~data
from previous timesteps), we summarise the information obtained about
\(\theta\) from previous timesteps through a summary statistic
\(O_t(d_{0:t}, y_{0:t}, \theta)\). An example of such a summary
statistic that we frequently use is the observed Fisher information
about \(\theta\) from previous time steps. Expected utility functions
can then be interpreted as evaluating the information gain that is
additional to what has been previous observed. Then, within this
context, expected utility functions are optimised as described in
Algorithm \ref{Alg:gea} for time period \(t\).

\begin{algorithm}
\caption{Algorithm for finding adaptive designs}
\label{Alg:seq_design}
\begin{algorithmic}[1]
\STATE Initialise $d_0$ and $y_0$
\STATE Estimate $\theta$ given $y_0$ and $d_0$ to form $p(\theta|d_0,y_0)$
\STATE Obtain summary of model fit, e.g., $O_0(d_0, y_0, \theta)$
\FOR {$t = 1:T$}
\STATE Find $d_t = \max_{d\in D} U(d|d_{0:t-1},y_{0:t-1})$ where  $U(d|d_{0:t-1},y_{0:t-1})$ depends on all previous design decisions through the statistic $O_{t-1}(d_0, y_0, \theta)$
\STATE Collect new data $y_t$ in accordance with $d_t$. If no data collection can be performed, simulate data collection by generating $y_t$ from the data-generating model ($p(y|\theta,d_t)$), with assumed parameters $\theta$ and the design under consideration $d_t$. 
\STATE Estimate $\theta$ given $y_{0:t}$ and $d_{0:t}$ to form $p(\theta |d_{0:t},y_{0:t})$
\STATE Update the statistic $O_t(d_{0:t}, y_{0:t}, \theta)$
\ENDFOR
\end{algorithmic}
\end{algorithm}

We have included three utility functions for adaptive design in our
package:

\begin{itemize}
\tightlist
\item
  Sequential CP-optimality, for adaptive covariance parameter
  estimation.
\item
  Sequential D-optimality, for adaptive fixed effects estimation.
\item
  Sequential ED-optimality, for adaptive fixed effects estimation with
  empirically estimated covariance parameters.
\end{itemize}

In sequential CP-optimality, we define \(O_t(d_{0:t}, y_{0:t}, \theta)\)
to be the observed Fisher information matrix for the covariance
parameters from a spatial stream network model fitted over the existing
design. This leads to the following definition of sequential
CP-optimality:

\begin{equation}
  \begin{aligned}
    U(d, \theta|d_{0:t-1}, y_{0:t-1}) &= \log\det\left[I(d, \theta) + O_{t-1}(d_{0:t-1}, y_{0:t-1}, \theta)\right]\label{eqn:seqC}
  \end{aligned}
\end{equation}

Note that, in practice, we cannot guarantee that it will always be
possible to run this utility function (\texttt{sequentialCPOptimality})
because the observed Fisher information matrix for the covariance
parameters is not always returned in objects of class \texttt{glmssn}.

In sequential D-optimality and ED-optimality, we define
\(O_t(d_{0:t}, y_{0:t}, \theta)\) to be the observed Fisher information
matrix for the fixed effects. We obtain this by fitting a stream network
model over the data that have been collected using the existing design.
The sequential D-optimality function is effectively the same as Eq.
\ref{eqn:seqC} where \(\beta\) is substituted in for \(\theta\). The
sequential ED-optimality function is similar, except that it uses the
observed Fisher information matrix for the fixed effects
\(I(d, \hat{\beta})\) instead of the expected Fisher information matrix
\(I(d, \beta)\) like sequential D-optimality. Therefore, the utility
function is written as

\begin{equation}
  \begin{aligned}
    U(d, \theta, y_t|d_{0:t-1}, y_{0:t-1}) &= \log\det\left[I(d, \hat{\beta}) + O_{t-1}(d_{0:t-1}, y_{0:t-1}, \theta)\right].
  \end{aligned}
\end{equation}

No special functions are defined as adaptive equivalents of K and
EK-optimality. This is because the only appropriate quantity that might
be used as \(O_t(\theta)\) for K and EK-optimality is the sum or inverse
sum of the kriging variances defined at the prediction sites. However,
simply adding this quantity in the utility function would have no impact
on the results because it would offset every calculation by the same
amount. Instead, K and EK-optimality can both be used `as-is' for
adaptive designs. Each optimisation will still be conditioned on
previous designs and observed data through any legacy sites in the
design, as well as through the updated estimates and priors of
parameters in the spatial stream network model.

\section*{References}\label{references}
\addcontentsline{toc}{section}{References}

\setlength{\parindent}{-0.5in} \setlength{\leftskip}{0.5in}
\setlength{\parskip}{8pt} \noindent

\hypertarget{refs}{}
\hypertarget{ref-Chaloner}{}
Chaloner, Kathryn, and Isabella Verdinelli. 1995. ``Bayesian
Experimental Design: A Review.'' \emph{Statistical Science} 10 (3):
273--304.

\hypertarget{ref-Cressie1993}{}
Cressie, Noel. 1993. \emph{Statistics for Spatial Data}. Wiley, New
York.

\hypertarget{ref-Cressie2006}{}
Cressie, Noel, J. Frey, Bronwyn Harch, and M. Smith. 2006. ``Spatial
Prediction on a River Network.'' \emph{Journal of Agricultural,
Biological, and Environmental Statistics} 11: 127--50.
doi:\href{https://doi.org/10.1198/108571106X110649}{10.1198/108571106X110649}.

\hypertarget{ref-DentGrimm1999}{}
Dent, C. Lisa, and Nancy B. Grimm. 1999. ``Spatial Heterogeneity of
Stream Water Nutrient Concentrations over Successional Time.''
\emph{Ecology} 80 (7). Ecological Society of America: 2283--98.
\url{http://www.jstor.org/stable/176910}.

\hypertarget{ref-Evangelou2012}{}
Evangelou, E., and Z. Zhu. 2012. ``Optimal Predictive Design
Augmentation for Spatial Generalised Linear Mixed Models.''
\emph{Journal of Statistical Planning and Inference} 142 (12): 3242--53.

\hypertarget{ref-Falk2014}{}
Falk, Matthew G., James M. McGree, and Anthony N. Pettitt. 2014.
``Sampling Designs on Stream Networks Using the Pseudo-Bayesian
Approach.'' \emph{Environmental and Ecological Statistics} 21: 751--73.
doi:\href{https://doi.org/10.1007/s10651-014-0279-2}{10.1007/s10651-014-0279-2}.

\hypertarget{ref-GarretaMAupdown2010}{}
Garreta, Vincent, Pascal Monestiez, and Jay M. Ver Hoef. 2010. ``Spatial
Modelling and Prediction on River Networks: Up Model, down Model or
Hybrid?'' \emph{Environmetrics} 21 (5). John Wiley \& Sons, Ltd.:
439--56. doi:\href{https://doi.org/10.1002/env.995}{10.1002/env.995}.

\hypertarget{ref-LeZidek2006}{}
Le, Nhu D., and James V. Zidek. 2006. \emph{Statistical Analysis of
Environmental Space-Time Processes}. Springer Series in Statistics. New
York, NY: Springer.

\hypertarget{ref-MorrisMitchell1995}{}
Morris, Max D., and Toby J. Mitchell. 1995. ``Exploratory Designs for
Computational Experiments.'' \emph{Journal of Statistical Planning and
Inference} 43 (3): 381--402.
doi:\href{https://doi.org/10.1016/0378-3758(4)00035-T}{10.1016/0378-3758(4)00035-T}.

\hypertarget{ref-Muller}{}
Mueller, Peter. 1999. ``Simulation-Based Optimal Design.''
\emph{Bayesian Statistics} 6: 459--74.

\hypertarget{ref-Muller2007}{}
Mueller, Peter, Don A. Berry, Andy P. Grieve, Michael Smith, and Michael
Krams. 2007. ``Simulation-Based Sequential Bayesian Design.''
\emph{Journal of Statistical Planning and Inference} 137: 3140--50.
doi:\href{https://doi.org/10.1016/j.jspi.2006.05.021}{10.1016/j.jspi.2006.05.021}.

\hypertarget{ref-NCEASEcoLet}{}
Peterson, Erin E., Jay M. Ver Hoef, Dan J. Isaak, Jeffrey A. Falke,
Marie-Josée Fortin, Chris E. Jordan, Kristina McNyset, et al. 2013.
``Modelling Dendritic Ecological Networks in Space: An Integrated
Network Perspective.'' \emph{Ecology Letters} 16 (5): 707--19.
doi:\href{https://doi.org/10.1111/ele.12084}{10.1111/ele.12084}.

\hypertarget{ref-Pronzato_Muller_2012}{}
Pronzato, Luc, and Werner G. Muller. 2012. ``Design of Computer
Experiments: Space Filling and Beyond.'' \emph{Statistics and Computing}
22.
doi:\href{https://doi.org/10.1007/s11222-011-9242-3}{10.1007/s11222-011-9242-3}.

\hypertarget{ref-Royle2002}{}
Royle, J. Andrew. 2002. ``Exchange Algorithms for Constructing Large
Spatial Designs.'' \emph{Journal of Statistical Planning and Inference}
100 (2): 121--34.

\hypertarget{ref-Som2014}{}
Som, Nicholas A., Pascal Monestiez, Jay M. Ver Hoef, Dale L. Zimmerman,
and Erin E. Peterson. 2014. ``Spatial Sampling on Streams: Principles
for Inference on Aquatic Networks.'' \emph{Environmetrics} 25 (5):
306--23. doi:\href{https://doi.org/10.1002/env.2284}{10.1002/env.2284}.

\hypertarget{ref-HoefPetersonMA}{}
Ver Hoef, Jay M., and Erin E. Peterson. 2010. ``A Moving Average
Approach for Spatial Statistical Models of Stream Networks.''
\emph{Journal of the American Statistical Association} 105 (489).
doi:\href{https://doi.org/10.1198/jasa.2009.ap09248}{10.1198/jasa.2009.ap09248}.

\hypertarget{ref-VerHoef2006}{}
Ver Hoef, Jay M., Erin E. Peterson, and David Theobald. 2006. ``Spatial
Statistical Models That Use Flow and Stream Distance.''
\emph{Environmental and Ecological Statistics} 13: 449--64.
doi:\href{https://doi.org/10.1007/s10651-006-0022-8}{10.1007/s10651-006-0022-8}.

\hypertarget{ref-Zhu2006}{}
Zhu, Zhengyuan, and Michael L. Stein. 2006. ``Spatial Sampling Design
for Prediction with Estimated Parameters.'' \emph{Journal of
Agricultural, Biological, and Environmental Statistics} 11 (1): 24--44.
doi:\href{https://doi.org/10.1198/108571106X99751}{10.1198/108571106X99751}.

\maketitle

\section{The SSNdesign package}\label{the-ssndesign-package}

The \texttt{SSNdesign} package provides functions to find
pseudo-Bayesian optimal and adaptive designs for spatial stream network
models and streams data. Given a set of potential sampling locations on
a stream network and a utility function (i.e.~mathematical statement
about the sampling objective), SSNdesign will find the best subset of
sites to meet that objective. This vignette steps through two case
studies of stream sampling problems that can be solved using
\texttt{SSNdesign}. These involve:

\begin{enumerate}
\def\labelenumi{\arabic{enumi}.}
\tightlist
\item
  Using optimal design to reduce the number of sites in a monitoring
  program by half, using data collected near Lake Eacham, Queensland;
  and
\item
  Augmenting an existing monitoring program with new sites within the
  Pine River catchment, Queensland, using adaptive design.
\end{enumerate}

The data required for this tutorial are downloaded with the package.
Instructions for extracting these files are provided below.

\section{Installing and loading the
package}\label{installing-and-loading-the-package}

The package can be installed by running
\texttt{devtools::install\_github("apear9/SSNdesign")}. If the package
files have already been downloaded from GitHub and are stored in a local
directory, the package can also be installed by running
\texttt{devtools::install\_local(path)} where \texttt{path} is a string
specifying the location of the directory containing the package files.
For Windows users, \texttt{devtools::install\_local} requires
\href{https://cran.r-project.org/bin/windows/Rtools/}{Rtools}. Once
installed, the package can be loaded by running

\begin{Shaded}
\begin{Highlighting}[]
\KeywordTok{library}\NormalTok{(SSNdesign)}
\end{Highlighting}
\end{Shaded}

\section{Extracting the example data from the
package}\label{extracting-the-example-data-from-the-package}

Once installed and loaded, the example data required for this tutorial
can be extracted from the package by running the following code.

\begin{Shaded}
\begin{Highlighting}[]
\CommentTok{# The ... should be replaced with the path to an empty, }
\CommentTok{# new folder you have set up to receive the package data.}
\KeywordTok{setwd}\NormalTok{(}\StringTok{"..."}\NormalTok{)}
\KeywordTok{unpackExampleData}\NormalTok{(}\StringTok{"."}\NormalTok{) }\CommentTok{# this should return TRUE}
\end{Highlighting}
\end{Shaded}

\section{Case study 1: Lake Eacham}\label{case-study-1-lake-eacham}

The Lake Eacham dataset contains stream temperature measurements
collected at 88 sites throughout a stream network in northern
Queensland, Australia. Each site is also associated with a set of
GIS-derived covariates such as total rainfall (mm) on the day of
sampling, percent urban, grazing and agricultural land-use within the
watershed, stream slope, and the distance upstream of the outlet
(i.e.~the most downstream location on the stream network). In this case
study, the goal is to reduce the number of sampling sites from 88 to 44,
while retaining the sites that yield the most information. This is a
classic use-case for optimal experimental design because there is one
decision to be made at a single point in time (i.e.~which 44 sites
should be dropped from the monitoring program after this year). We can
optimise this design for a number of objectives, but two common
objectives that concern managers are to (1) characterise in-stream
processes using a statistical model and (2) accurately predict in-stream
variables throughout the stream network at unobserved locations. For the
first objective, we optimise the design using CPD-optimality, which is a
utility function that aims to minimise uncertainty in the fixed effect
and covariance parameters produced by the geostatistical model. For the
second objective, we optimise the design using K-optimality. This
utility function aims to maximise prediction accuracy at unobserved
locations by reducing the average uncertainty across prediction sites in
the stream network. In this section, we demonstrate how to solve this
problem using \texttt{SSNdesign}.

The first step in any design procedure using \texttt{SSNdesign} is to
import a \texttt{SpatialStreamNetwork} object. This object already
contains streams, observed sites with measurements. Thus, we simply use
\texttt{importSSN} from the package \texttt{SSN} to load a
SpatialStreamNetwork object into R. A dataset of prediction sites is
also imported, which will be needed to construct a design that focuses
on maintaining prediction accuracy at unsampled sites.

\begin{Shaded}
\begin{Highlighting}[]
\CommentTok{# import spatial stream network}
\NormalTok{lake.eacham <-}\StringTok{ }\KeywordTok{importSSN}\NormalTok{(}\StringTok{"lake-eacham-full.ssn"}\NormalTok{, }\DataTypeTok{predpts=}\StringTok{"preds"}\NormalTok{)}

\CommentTok{# create distance matrices for obs and preds sites}
\KeywordTok{createDistMat}\NormalTok{(lake.eacham, }\DataTypeTok{predpts=}\StringTok{"preds"}\NormalTok{, }\DataTypeTok{o.write=}\OtherTok{TRUE}\NormalTok{, }\DataTypeTok{amongpreds=}\OtherTok{TRUE}\NormalTok{)}
\end{Highlighting}
\end{Shaded}

We can check what this stream network looks like using the plot method
for \texttt{SpatialStreamNetwork} objects.

\begin{Shaded}
\begin{Highlighting}[]
\CommentTok{# plot the network}
\KeywordTok{plot}\NormalTok{(lake.eacham, }\StringTok{"Temp"}\NormalTok{)}
\end{Highlighting}
\end{Shaded}

\begin{figure}
\centering
\includegraphics{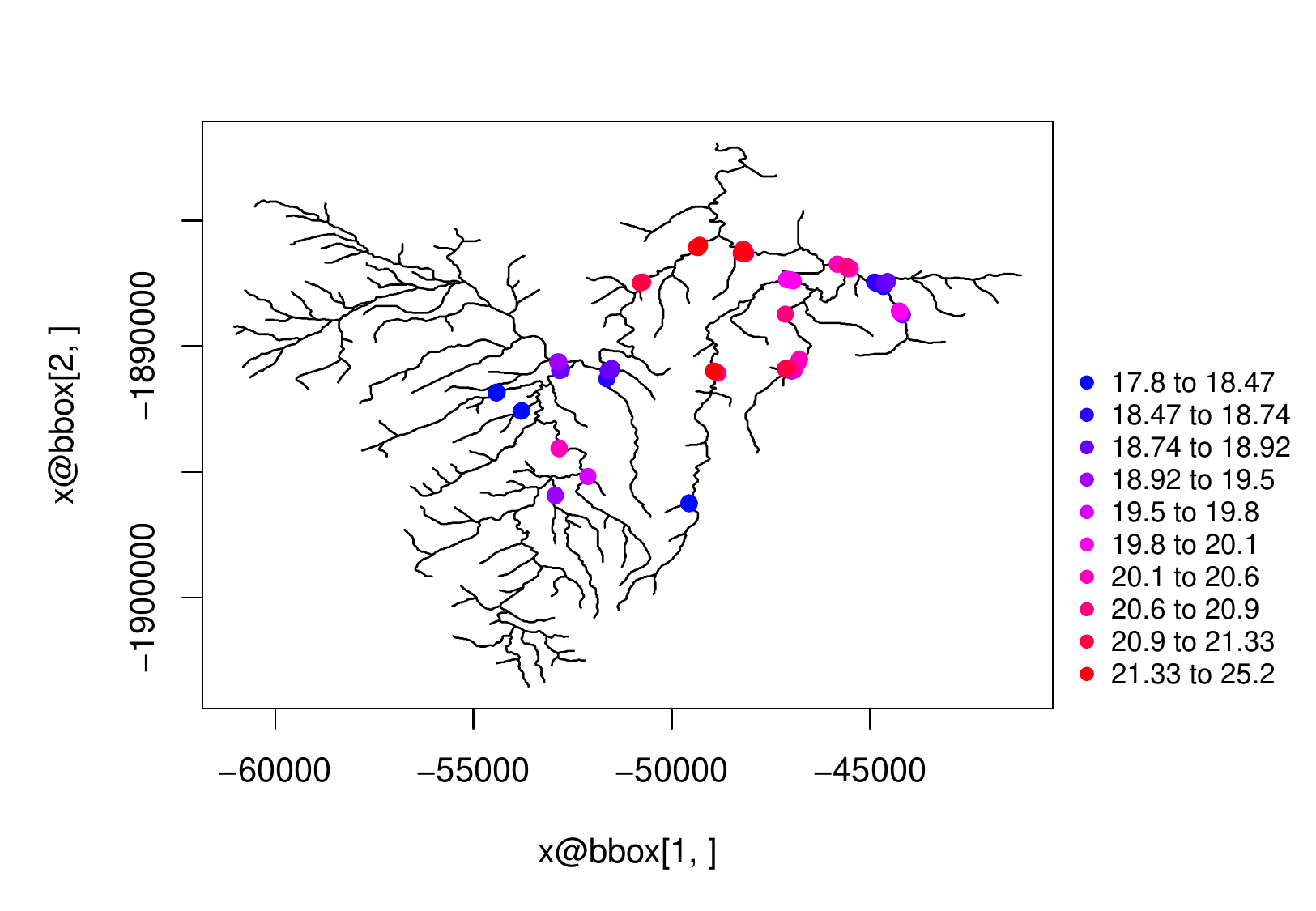}
\caption{The Lake Eacham stream network with sampling points.}
\end{figure}

A prerequisite of optimal design is that we are able to define a `true'
spatial statistical model. The notion of a `true' model refers to a
model that we believe adequately describes the process responsible for
generating the data. Designs are then optimised in relation to some
element of this true model (e.g.~the parameters and/or predictions).
Note that we only need to know the structure of the model, which, for
spatial stream-network models, includes the mean structure for the fixed
effects and the covariance function for the covariance parameters. In
our case, we assume that the mean structure for temperature is a
function of total rainfall (mm) and the percent of urban and grazing
land use in the riparian zone. We also include the exponential tail up
and tail down components in the covariance mixture. We then fit this
model to our data. Note that fitting the model does not necessarily mean
the estimates from the model are used when optimising designs using
\texttt{optimiseSSNDesign}. The main reason for fitting this model is to
give \texttt{optimiseSSNDesign} a template for its own operations, such
as constructing design and covariance matrices when evaluating a utility
function.

\begin{Shaded}
\begin{Highlighting}[]
\CommentTok{# Fit the 'true' model to the data}
\NormalTok{TC.model <-}\StringTok{ }\KeywordTok{glmssn}\NormalTok{(}
\NormalTok{  Temp }\OperatorTok{~}\StringTok{ }\NormalTok{rainfall }\OperatorTok{+}\StringTok{ }\NormalTok{ripURBAN }\OperatorTok{+}\StringTok{ }\NormalTok{ripGRAZE,}
\NormalTok{  lake.eacham,}
  \DataTypeTok{CorModels =} \KeywordTok{c}\NormalTok{(}\StringTok{"Exponential.tailup"}\NormalTok{, }\StringTok{"Exponential.taildown"}\NormalTok{),}
  \DataTypeTok{addfunccol =} \StringTok{"afvArea"}
\NormalTok{)}
\end{Highlighting}
\end{Shaded}

We define log-normal priors on the covariance parameters based on the
fitted model.

\begin{Shaded}
\begin{Highlighting}[]
\CommentTok{# Create a list of functions used to define log-normal priors based on the}
\CommentTok{# estimates of the covariance parameters and their standard errors.}
\NormalTok{priors.TC <-}\StringTok{ }\KeywordTok{constructLogNormalPriors}\NormalTok{(TC.model)}
\end{Highlighting}
\end{Shaded}

The next step is to optimise the design. We show the code for the
CPD-optimal design first. Note that, the \texttt{optimiseSSNDesign}
function could be used to find the optimal 44-site designs from a few (5
- 10) random starts. However, in this case study the aim is to remove
sites from an existing monitoring program one-by-one, so that we can
quantify how much information is lost as sites are removed.

Finding an optimal design using the functions \texttt{CPDOptimality} or
\texttt{KOptimality} is often computationally expensive. We have
provided the R code below so that users can recreate the examples using
their own data. However, we also provide the results as saved Rdata
workspaces, which can be found in the Supplementary Information and
loaded into R to save time.

\begin{Shaded}
\begin{Highlighting}[]
\CommentTok{# Find the CPD-optimal design for 44 of the 88 sites}
\CommentTok{# }\AlertTok{WARNING}\CommentTok{: this code takes approximately 5 hours to run.}

\CommentTok{# # Set random seed for reproducibility}
\CommentTok{# set.seed(987654321)}

\CommentTok{# # Initialise loop by dropping the first site}
\CommentTok{# step88to87 <- optimiseSSNDesign(}
\CommentTok{#   ssn = lake.eacham,}
\CommentTok{#   new.ssn.path = "./to87CPD.ssn",}
\CommentTok{#   glmssn = TC.model,}
\CommentTok{#   n.points = 87,}
\CommentTok{#   utility.function = CPDOptimality,}
\CommentTok{#   prior.parameters = priors.TC,}
\CommentTok{#   n.cores = 1,}
\CommentTok{#   parallelism = "none",}
\CommentTok{#   n.optim = 1,}
\CommentTok{#   n.draws = 500}
\CommentTok{# )}
\CommentTok{# createDistMat(step88to87$ssn.new)}

\CommentTok{# # Loop through the remaining steps dropping sites one-by-one}
\CommentTok{# indices <- 86:44}
\CommentTok{# counter <- 1}
\CommentTok{# n.indices <- length(indices)}
\CommentTok{# cpd.designs <- vector("list", n.indices + 1)}
\CommentTok{# cpd.designs[[1]] <- step88to77$final.points}

\CommentTok{# for(i in indices)\{}
\CommentTok{#   current.pth <- paste0("./to", i+1,"CPD.ssn")}
\CommentTok{#   ifuture.pth <- paste0("./to",i,"CPD.ssn")}
\CommentTok{#   current.ssn <- importSSN(current.pth)}
\CommentTok{#   current.ssd <- optimiseSSNDesign(}
\CommentTok{#     ssn = current.ssn,}
\CommentTok{#     new.ssn.path = ifuture.pth,}
\CommentTok{#     glmssn = TC.model,}
\CommentTok{#     n.points = i,}
\CommentTok{#     utility.function = CPDOptimality,}
\CommentTok{#     prior.parameters = priors.TC,}
\CommentTok{#     n.cores = 1,}
\CommentTok{#     parallelism = "none",}
\CommentTok{#     n.optim = 1,}
\CommentTok{#     n.draws = 500}
\CommentTok{#   )}

\CommentTok{#   createDistMat(current.ssd$ssn.new)}
\CommentTok{#   cpd.designs[[counter + 1]] <- current.ssd$final.points}
\CommentTok{#   counter <- counter + 1}
\CommentTok{# \}}
\CommentTok{# save.image("CPD-OPTIMAL-RESULTS.Rdata")}

\CommentTok{# Load the results from the CPD-optimal design process}
\KeywordTok{load}\NormalTok{(}\StringTok{"CPD-OPTIMAL-RESULTS.Rdata"}\NormalTok{)}
\end{Highlighting}
\end{Shaded}

\bigskip
The process for finding the K-optimal design is very similar. The
differences are that

\begin{itemize}
\tightlist
\item
  We need to use the \texttt{KOptimality} utility function instead of
  \texttt{CPDOptimality}.
\item
  The \texttt{SpatialStreamNetwork} object must contain prediction
  sites.
\item
  Distance matrices must be generated for both the observed and
  prediction sites when the \texttt{SpatialStreamNetwork} object is
  imported. \bigskip
\end{itemize}

\begin{Shaded}
\begin{Highlighting}[]
\CommentTok{# Find a K-optimal design for 44 of the 88 sites}
\CommentTok{# Warning, this code may take up to 5 hours to run}

\CommentTok{# # Set random seed for reproducibility}
\CommentTok{# set.seed(123456789)}
\CommentTok{#}
\CommentTok{# step88to87 <- optimiseSSNDesign(}
\CommentTok{#   ssn = lake.eacham,}
\CommentTok{#   new.ssn.path = "./to87K.ssn",}
\CommentTok{#   glmssn = TC.model,}
\CommentTok{#   n.points = 87,}
\CommentTok{#   utility.function = KOptimality,}
\CommentTok{#   prior.parameters = priors.TC,}
\CommentTok{#   n.cores = 1,}
\CommentTok{#   parallelism = "none",}
\CommentTok{#   n.optim = 1,}
\CommentTok{#   n.draws = 500}
\CommentTok{# )}
\CommentTok{# createDistMat(step88to87$ssn.new, "preds", T, T)}
\CommentTok{#}
\CommentTok{# # Loop to find optimal designs using one-dimensional optimisation}
\CommentTok{# indices <- 86:44}
\CommentTok{# counter <- 1}
\CommentTok{# n.indices <- length(indices)}
\CommentTok{# k.designs <- vector("list", n.indices + 1)}
\CommentTok{# k.designs[[1]] <- step88to77$final.points}

\CommentTok{# for(i in indices)\{}
\CommentTok{#   current.pth <- paste0("./to", i+1,"K.ssn")}
\CommentTok{#   ifuture.pth <- paste0("./to",i,"K.ssn")}
\CommentTok{#   current.ssn <- importSSN(current.pth, "preds")}
\CommentTok{#   createDistMat(current.ssn, "preds", T, T)}
\CommentTok{#   current.ssd <- optimiseSSNDesign(}
\CommentTok{#     ssn = current.ssn,}
\CommentTok{#     new.ssn.path = ifuture.pth,}
\CommentTok{#     glmssn = TC.model,}
\CommentTok{#     n.points = i,}
\CommentTok{#     utility.function = KOptimality,}
\CommentTok{#     prior.parameters = priors.TC,}
\CommentTok{#     n.cores = 1,}
\CommentTok{#     parallelism = "none",}
\CommentTok{#     n.optim = 1,}
\CommentTok{#     n.draws = 500}
\CommentTok{#   )}
\CommentTok{#   createDistMat(current.ssd$ssn.new, "preds", TRUE, TRUE)}
\CommentTok{#   k.designs[[counter + 1]] <- current.ssd$final.points}
\CommentTok{#   counter <- counter + 1}
\CommentTok{# \}}
\CommentTok{# save.image("K-OPTIMAL-RESULTS.Rdata")}

\CommentTok{# Load the results from the K-optimal design process}
\KeywordTok{load}\NormalTok{(}\StringTok{"K-OPTIMAL-RESULTS.Rdata"}\NormalTok{)}
\end{Highlighting}
\end{Shaded}

The output of \texttt{optimiseSSNDesign} is an S3 object of class
\texttt{ssndesign}. This is a list of 14 elements that contains the
\texttt{SpatialStreamNetwork} object passed to it, a modified version
containing only the observed sites associated with the optimal or
adaptive design, and diagnostic information associated with the Greedy
Exchange Algorithm it uses to find optimal and adaptive designs.
Additionally, objects of class \texttt{ssndesign} contain information
about the user's call to \texttt{optimiseSSNDesign}, including, for
example, the values of key parameters and also the prior draws that were
used in the Monte Carlo integration.

Having computed an optimal design, the next step is to check that it
performs well compared to random and Generalised Random Tessellation
Sampling (GRTS) designs. We chose to benchmark our solution against GRTS
designs because they are a powerful tool for constructing spatially
balanced designs. However, any standard design (e.g.~random sampling)
can be chosen as a benchmark.

In practical terms, validating a design using the functions in
\texttt{SSNdesign} requires the following:

\begin{enumerate}
\def\labelenumi{\arabic{enumi}.}
\tightlist
\item
  Finding one or more optimal design(s) and recording the sites included
  in each design.
\item
  Identify designs of the same size as the optimal design to benchmark
  against. Here, size refers to either the number of sites or number of
  samples collected per sampling period. If it makes sense to compare
  the optimal design against other designs of different sizes then this
  is also possible.
\item
  Evaluating the expected utility of the benchmarking designs with a
  large number of Monte-Carlo draws and then computing the designs'
  relative efficiency compared to the optimal design.
\end{enumerate}

The first step was completed in part when we found a series of optimal
designs. We must now record the sites included in those designs, which
we can do with the following code:

\begin{Shaded}
\begin{Highlighting}[]
\CommentTok{# Record designs for CPD-optimality}

\CommentTok{# Set up an empty vector to store the designs}
\CommentTok{# Designs are stored as a list of vectors containing pid or locID values}
\NormalTok{opt.cpd.reference <-}\StringTok{ }\KeywordTok{append}\NormalTok{(}
  \KeywordTok{list}\NormalTok{(}\DecValTok{1}\OperatorTok{:}\DecValTok{88}\NormalTok{), }\CommentTok{# the full desig with all 88 sites}
\NormalTok{  cpd.designs}
\NormalTok{)}

\CommentTok{# Record designs for K-optimality, repeating the same process}
\NormalTok{opt.k.reference <-}\StringTok{ }\KeywordTok{append}\NormalTok{(}
  \KeywordTok{list}\NormalTok{(}\DecValTok{1}\OperatorTok{:}\DecValTok{88}\NormalTok{), }\CommentTok{# the full design}
\NormalTok{  k.designs}
\NormalTok{)}

\CommentTok{# Save the output as a .Rdata file}
\CommentTok{# save(opt.cpd.reference, opt.k.reference, file = "optimal-designs.Rdata")}
\end{Highlighting}
\end{Shaded}

The second step is to set up the GRTS and random benchmarking designs.

\begin{Shaded}
\begin{Highlighting}[]
\CommentTok{# The following code takes a few minutes to run.}
\CommentTok{# If you do not run the following code, you must run}
\CommentTok{# load("reference-designs.Rdata")}
\CommentTok{# before continuing to the next code block}

\CommentTok{# Need to reset the path for lake eacham data}
\NormalTok{lake.eacham <-}\StringTok{ }\KeywordTok{updatePath}\NormalTok{(}
\NormalTok{  lake.eacham, }
  \CommentTok{# We need the full path }
  \KeywordTok{paste}\NormalTok{(}\KeywordTok{getwd}\NormalTok{(), }\StringTok{"lake-eacham-full.ssn"}\NormalTok{, }\DataTypeTok{sep =} \StringTok{"/"}\NormalTok{)}
\NormalTok{)}

\CommentTok{# Seed for reproducibility}
\KeywordTok{set.seed}\NormalTok{(}\DecValTok{1}\NormalTok{)}

\CommentTok{# Set up an empty vector to store the GRTS designs}
\NormalTok{grts.reference <-}\StringTok{ }\KeywordTok{vector}\NormalTok{(}\StringTok{"list"}\NormalTok{, }\DecValTok{20}\NormalTok{)}
\CommentTok{# Note: the following loop takes approximately}
\CommentTok{# 10-15 minutes to run. }
\ControlFlowTok{for}\NormalTok{(i }\ControlFlowTok{in} \DecValTok{1}\OperatorTok{:}\DecValTok{20}\NormalTok{)\{}
  \CommentTok{# For each iteration, we}
  \CommentTok{# 1. Create a GRTS design using drawStreamNetworkSamples}
  \CommentTok{# 2. Record the pid values associated with the GRTS design}
\NormalTok{  g <-}\StringTok{ }\KeywordTok{drawStreamNetworkSamples}\NormalTok{(}
\NormalTok{    lake.eacham, }\KeywordTok{paste0}\NormalTok{(}\KeywordTok{tempdir}\NormalTok{(), }\StringTok{"/g"}\NormalTok{, i, }\StringTok{".ssn"}\NormalTok{), T, }\StringTok{"GRTS"}\NormalTok{, }\DecValTok{44}
\NormalTok{  )}
\NormalTok{  grts.reference[[i]] <-}\StringTok{ }\KeywordTok{getSSNdata.frame}\NormalTok{(g)}\OperatorTok{$}\NormalTok{pid}
\NormalTok{\}}

\CommentTok{# Set up an empty vector to store random designs}
\NormalTok{rand.reference <-}\StringTok{ }\KeywordTok{vector}\NormalTok{(}\StringTok{"list"}\NormalTok{, }\DecValTok{20}\NormalTok{)}
\ControlFlowTok{for}\NormalTok{(i }\ControlFlowTok{in} \DecValTok{1}\OperatorTok{:}\DecValTok{20}\NormalTok{)\{}
  \CommentTok{# Simply store a random sample (w/out replacement)}
  \CommentTok{# of 44 out of 88 possible pid values}
\NormalTok{  rand.reference[[i]] <-}\StringTok{ }\KeywordTok{sample}\NormalTok{(}\DecValTok{1}\OperatorTok{:}\DecValTok{88}\NormalTok{, }\DecValTok{44}\NormalTok{, }\OtherTok{FALSE}\NormalTok{)}
\NormalTok{\}}

\CommentTok{# Save info in .Rdata file}
\CommentTok{# save(rand.reference, grts.reference, file = "reference-designs.Rdata")}
\end{Highlighting}
\end{Shaded}

The final step is to evaluate the expected utility for each of these
designs using a large number of Monte Carlo draws.

\begin{Shaded}
\begin{Highlighting}[]
\CommentTok{# The following code may take an hour to run.}
\CommentTok{# If you do not run the following code, you must run}
\CommentTok{# load("optimal-info.Rdata")}
\CommentTok{# load("rand-grts-info.Rdata")}
\CommentTok{# before continuing to the next code block}

\CommentTok{# Reset path to lake.eacham in case overwritten by loaded .Rdata file}
\NormalTok{lake.eacham <-}\StringTok{ }\KeywordTok{updatePath}\NormalTok{(}
\NormalTok{  lake.eacham, }
  \CommentTok{# We need the full path }
  \KeywordTok{paste}\NormalTok{(}\KeywordTok{getwd}\NormalTok{(), }\StringTok{"lake-eacham-full.ssn"}\NormalTok{, }\DataTypeTok{sep =} \StringTok{"/"}\NormalTok{)}
\NormalTok{)}

\CommentTok{# Set seed for reproducibility}
\KeywordTok{set.seed}\NormalTok{(}\FloatTok{1e6} \OperatorTok{+}\StringTok{ }\DecValTok{1}\NormalTok{)}

\CommentTok{# Evaluate the expected utility of the designs discovered}
\NormalTok{CPD_info <-}\StringTok{ }\KeywordTok{evaluateFixedDesigns}\NormalTok{(}
\NormalTok{  lake.eacham, TC.model, opt.cpd.reference,}
  \StringTok{"pid"}\NormalTok{, CPDOptimality, priors.TC, }\DecValTok{1000}
\NormalTok{)}
\NormalTok{K_info <-}\StringTok{ }\KeywordTok{evaluateFixedDesigns}\NormalTok{(}
\NormalTok{  lake.eacham, TC.model, opt.k.reference,}
  \StringTok{"pid"}\NormalTok{, KOptimality, priors.TC, }\DecValTok{1000}
\NormalTok{)}
\CommentTok{# save(CPD_info, K_info, file = "optimal-info.Rdata")}

\CommentTok{# Evaluate the 44-site random and GRTS design}
\NormalTok{R_CPD_info <-}\StringTok{ }\KeywordTok{evaluateFixedDesigns}\NormalTok{(}
\NormalTok{  lake.eacham, TC.model, }\KeywordTok{append}\NormalTok{(grts.reference, rand.reference),}
  \StringTok{"pid"}\NormalTok{, CPDOptimality, priors.TC, }\DecValTok{1000}
\NormalTok{)}
\NormalTok{R_K_info <-}\StringTok{ }\KeywordTok{evaluateFixedDesigns}\NormalTok{(}
\NormalTok{  lake.eacham, TC.model, }\KeywordTok{append}\NormalTok{(grts.reference, rand.reference),}
  \StringTok{"pid"}\NormalTok{, KOptimality, priors.TC, }\DecValTok{1000}
\NormalTok{)}
\CommentTok{# save(R_CPD_info, R_K_info, file = "rand-grts-info.Rdata")}
\end{Highlighting}
\end{Shaded}

Note that \texttt{evaluateFixedDesigns} has a \texttt{data.frame}
output, which looks like this:

\begin{Shaded}
\begin{Highlighting}[]
\CommentTok{# for the reference designs under CPD optimality}
\KeywordTok{head}\NormalTok{(R_CPD_info) }
\CommentTok{#>   ID Size Expected utility Efficiency Efficiency_Unlogged}
\CommentTok{#> 1  1   44        -20.73655   1.025710           0.5946610}
\CommentTok{#> 2  2   44        -20.49263   1.013645           0.7589271}
\CommentTok{#> 3  3   44        -20.74988   1.026369           0.5867867}
\CommentTok{#> 4  4   44        -20.89017   1.033308           0.5099772}
\CommentTok{#> 5  5   44        -20.70846   1.024320           0.6116022}
\CommentTok{#> 6  6   44        -20.63997   1.020932           0.6549581}
\end{Highlighting}
\end{Shaded}

Notice here that there are columns giving the expected utility of each
design, and also the efficiency of each design \emph{relative to the
design with the highest expected utility in the \texttt{data.frame}}.
The problem with this is that we have computed the expected utilities of
the optimal and reference designs separately. Therefore, an additional
step is needed at this juncture to compute the efficiencies of the
reference designs compared to the full 88-site design.

\begin{Shaded}
\begin{Highlighting}[]
\CommentTok{# Evaluate relative efficiencies}
\NormalTok{R_CPD_info}\OperatorTok{$}\NormalTok{Type <-}\StringTok{ }\KeywordTok{rep}\NormalTok{(}\KeywordTok{c}\NormalTok{(}\StringTok{"GRTS"}\NormalTok{, }\StringTok{"Random"}\NormalTok{), }\DataTypeTok{each =} \DecValTok{20}\NormalTok{)}
\NormalTok{R_CPD_info}\OperatorTok{$}\NormalTok{E2 <-}\StringTok{ }\KeywordTok{exp}\NormalTok{(}
\NormalTok{  R_CPD_info}\OperatorTok{$}\StringTok{`}\DataTypeTok{Expected utility}\StringTok{`} \OperatorTok{-}\StringTok{ }\KeywordTok{max}\NormalTok{(CPD_info}\OperatorTok{$}\StringTok{`}\DataTypeTok{Expected utility}\StringTok{`}\NormalTok{)}
\NormalTok{) }\CommentTok{# Compute efficiency relative to full design}
\NormalTok{R_K_info}\OperatorTok{$}\NormalTok{Type <-}\StringTok{ }\KeywordTok{rep}\NormalTok{(}\KeywordTok{c}\NormalTok{(}\StringTok{"GRTS"}\NormalTok{, }\StringTok{"Random"}\NormalTok{), }\DataTypeTok{each =} \DecValTok{20}\NormalTok{)}
\CommentTok{# Compute efficiency relative to full design}
\NormalTok{R_K_info}\OperatorTok{$}\NormalTok{E2 <-}\StringTok{ }\NormalTok{R_K_info}\OperatorTok{$}\StringTok{`}\DataTypeTok{Expected utility}\StringTok{`}\OperatorTok{/}\KeywordTok{max}\NormalTok{(K_info}\OperatorTok{$}\StringTok{`}\DataTypeTok{Expected utility}\StringTok{`}\NormalTok{)}
\end{Highlighting}
\end{Shaded}

Now that we have computed the relative efficiencies, we can plot the
summaries. We have chosen to use \texttt{ggplot2}.

\begin{Shaded}
\begin{Highlighting}[]
\KeywordTok{library}\NormalTok{(ggplot2)}
\KeywordTok{library}\NormalTok{(gridExtra)}

\NormalTok{a <-}\StringTok{ }\KeywordTok{ggplot}\NormalTok{(}\DataTypeTok{data =}\NormalTok{ CPD_info, }\KeywordTok{aes}\NormalTok{(}\DataTypeTok{x =}\NormalTok{ Size, }\DataTypeTok{y =}\NormalTok{ Efficiency_Unlogged)) }\OperatorTok{+}
\StringTok{  }\KeywordTok{geom_path}\NormalTok{(}\DataTypeTok{size =} \FloatTok{1.15}\NormalTok{) }\OperatorTok{+}
\StringTok{  }\KeywordTok{geom_jitter}\NormalTok{(}\DataTypeTok{data =}\NormalTok{ R_CPD_info, }\KeywordTok{aes}\NormalTok{(}\DataTypeTok{x =} \DecValTok{44}\NormalTok{, }\DataTypeTok{y =}\NormalTok{ E2, }\DataTypeTok{col =}\NormalTok{ Type), }\DataTypeTok{size =} \DecValTok{4}\NormalTok{) }\OperatorTok{+}
\StringTok{  }\KeywordTok{ylim}\NormalTok{(}\KeywordTok{c}\NormalTok{(}\DecValTok{0}\NormalTok{, }\DecValTok{1}\NormalTok{)) }\OperatorTok{+}
\StringTok{  }\KeywordTok{labs}\NormalTok{(}\DataTypeTok{x =} \StringTok{"Number of sampling sites"}\NormalTok{, }\DataTypeTok{y =} \StringTok{"Efficiency"}\NormalTok{, }\DataTypeTok{title =} \StringTok{"(a)"}\NormalTok{) }\OperatorTok{+}
\StringTok{  }\KeywordTok{scale_x_reverse}\NormalTok{() }\OperatorTok{+}
\StringTok{  }\KeywordTok{theme_bw}\NormalTok{() }\OperatorTok{+}
\StringTok{  }\KeywordTok{theme}\NormalTok{(}\DataTypeTok{legend.position =} \KeywordTok{c}\NormalTok{(}\FloatTok{0.2}\NormalTok{, }\FloatTok{0.2}\NormalTok{),}
        \DataTypeTok{legend.background =} \KeywordTok{element_rect}\NormalTok{(}\DataTypeTok{colour =} \StringTok{"black"}\NormalTok{),}
        \DataTypeTok{text =} \KeywordTok{element_text}\NormalTok{(}\DataTypeTok{size =} \DecValTok{16}\NormalTok{))}
\NormalTok{b <-}\StringTok{ }\KeywordTok{ggplot}\NormalTok{(}\DataTypeTok{data =}\NormalTok{ K_info, }\KeywordTok{aes}\NormalTok{(}\DataTypeTok{x =}\NormalTok{ Size, }\DataTypeTok{y =}\NormalTok{ Efficiency)) }\OperatorTok{+}
\StringTok{  }\KeywordTok{geom_path}\NormalTok{(}\DataTypeTok{size =} \FloatTok{1.15}\NormalTok{) }\OperatorTok{+}
\StringTok{  }\KeywordTok{geom_jitter}\NormalTok{(}\DataTypeTok{data =}\NormalTok{ R_K_info, }\KeywordTok{aes}\NormalTok{(}\DataTypeTok{x =} \DecValTok{44}\NormalTok{, }\DataTypeTok{y =}\NormalTok{ E2, }\DataTypeTok{col =}\NormalTok{ Type), }\DataTypeTok{size =} \DecValTok{4}\NormalTok{) }\OperatorTok{+}
\StringTok{  }\KeywordTok{ylim}\NormalTok{(}\KeywordTok{c}\NormalTok{(}\FloatTok{0.7}\NormalTok{, }\DecValTok{1}\NormalTok{)) }\OperatorTok{+}
\StringTok{  }\KeywordTok{labs}\NormalTok{(}\DataTypeTok{x =} \StringTok{"Number of sampling sites"}\NormalTok{, }\DataTypeTok{y =} \StringTok{"Efficiency"}\NormalTok{, }\DataTypeTok{title =} \StringTok{"(b)"}\NormalTok{) }\OperatorTok{+}
\StringTok{  }\KeywordTok{scale_x_reverse}\NormalTok{() }\OperatorTok{+}
\StringTok{  }\KeywordTok{theme_bw}\NormalTok{() }\OperatorTok{+}
\StringTok{  }\KeywordTok{theme}\NormalTok{(}\DataTypeTok{legend.position =} \KeywordTok{c}\NormalTok{(}\FloatTok{0.2}\NormalTok{, }\FloatTok{0.2}\NormalTok{),}
        \DataTypeTok{legend.background =} \KeywordTok{element_rect}\NormalTok{(}\DataTypeTok{colour =} \StringTok{"black"}\NormalTok{),}
        \DataTypeTok{text =} \KeywordTok{element_text}\NormalTok{(}\DataTypeTok{size =} \DecValTok{16}\NormalTok{))}

\KeywordTok{grid.arrange}\NormalTok{(}\DataTypeTok{ncol =} \DecValTok{2}\NormalTok{, a, b)}
\end{Highlighting}
\end{Shaded}

This gives us the graph below. This graph shows the efficiency of our
optimal designs relative to the full design (i.e.~the current monitoring
program with 88 sites). The black line tracks how much information we
get from sampling \(n\) sites in the optimal designs for
\(n \in \{87, 86, ..., 44\}\) compared to sampling all the sites. The
blue and red dots on the right-hand side of each plot panel represent
the amount of information we get from sampling 44 randomly chosen or
GRTS sites compared to the full 88-site monitoring program. The CPD- and
K-optimal designs consistently outperform the random and GRTS designs.
This shows the optimal design is an effective way of choosing which 44
sites to keep and which to discard, compared to two common sampling
strategies.

\begin{figure}[!h]
\includegraphics[width=\textwidth]{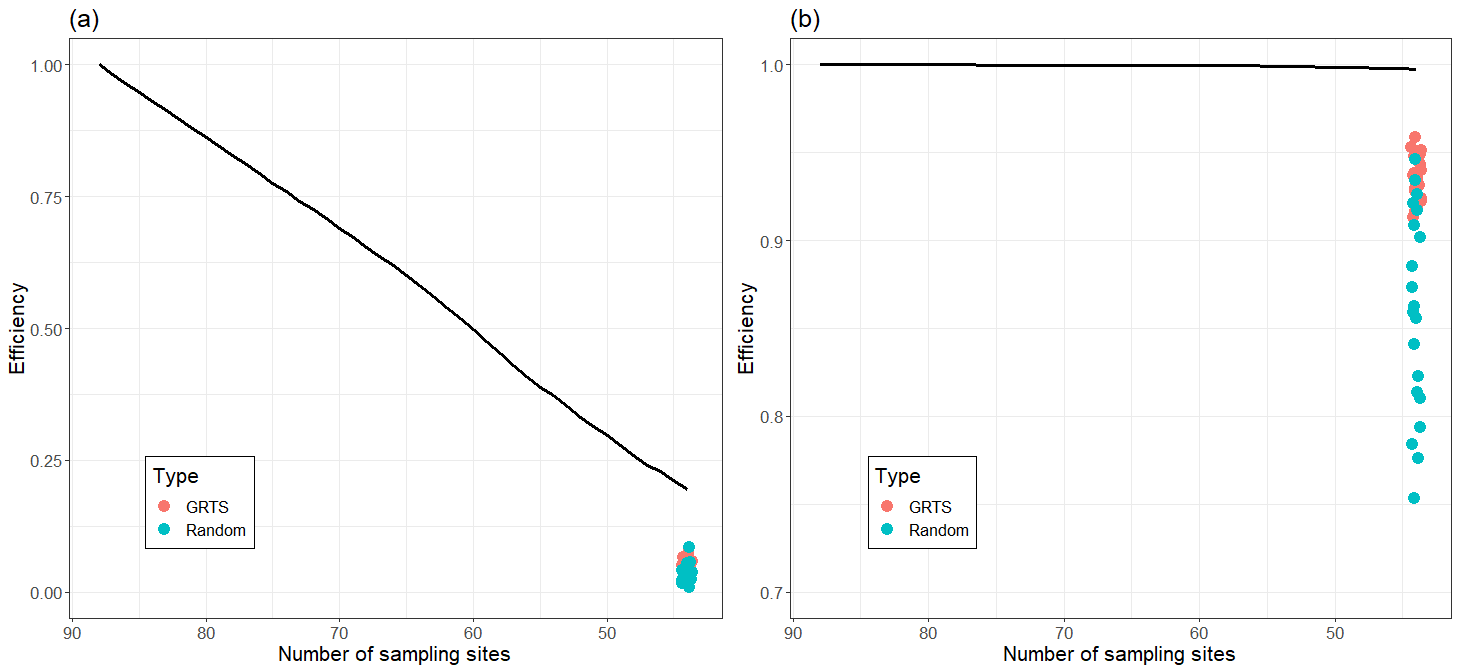}
\end{figure}

\section{Case study 2: Pine River}\label{case-study-2-pine-river}

Pine River is located in South East Queensland, Australia. Its catchment
is one of those monitored by the Ecosystem Health Monitoring Program
(EHMP) administered by Healthy Land and Water. In this synthetic
example, we demonstrate how to extend an existing monitoring program
using adaptive design using the Pine River catchment as an example
location. Currently, an extensive monitoring program does not exist
within Pine River alone. However, we assume for the sake of illustration
that a monitoring program has operated there for two years, using 200
sites, with the primary goal of measuring dissolved oxygen (DO) levels
(mg/L) and predicting them throughout the stream network. Our goal is to
extend this monitoring program optimally over two years by adding
another 100 sites to the program. We plan to add 50 sites per year for
the third and fourth year of the program. We will be using adaptive
design because there is an existing design which we expect will change
year-to-year depending on any new data we collect. The goal of the
monitoring program is to be able to accurately predict DO levels, so it
is appropriate to adaptively update our design using the K-optimality
utility function.

As before, we need to import our stream network into R. In this case, we
have a set of edges marking the locations of the streams but we do not
have any observed sites. In this situation, we must use
\texttt{importStreams} to import our .ssn folder into R and subsequently
use \texttt{generateSites} and \texttt{SimulateOnSSN} to create a set of
potential sampling sites and simulate observed data on them. The code
below is for the step where we generate the locations of the potential
sampling sites. We have the code commented out because it takes a few
minutes to run. We have provided the .ssn folder which is the output of
this code block in the data.

\begin{Shaded}
\begin{Highlighting}[]
\CommentTok{# NOTES:}

\CommentTok{# before running the following code blocks, you need to run}
\CommentTok{# setwd(...) where ... is the path to the folder containing the }
\CommentTok{# data for this tutorial}

\CommentTok{# in addition, the output of several code blocks is a .ssn folder.}
\CommentTok{# we have provided these same .ssn folders with the example data.}
\CommentTok{# the code blocks may result in errors if you attempt to run them}
\CommentTok{# while the existing .ssn folders of the same name are still saved}
\CommentTok{# in your data directory. a suggested strategy is to create a new}
\CommentTok{# folder on your computer and to move the example data to that folder,}
\CommentTok{# so the resulting .ssn folders can easily be written to your working }
\CommentTok{# directory.}

\CommentTok{# import the stream network with edges only as a}
\CommentTok{# SpatialStreamNetwork object}
\CommentTok{# pine_river <- importStreams(}
\CommentTok{#   "pine_river.ssn"}
\CommentTok{# )}

\CommentTok{# put sites on this network}
\CommentTok{# with_sites <- generateSites(}
\CommentTok{#   ssn = pine_river,}
\CommentTok{#   obsDesign = systematicDesign(}
\CommentTok{#     1500,}
\CommentTok{#     replications = 4,}
\CommentTok{#     rep.variable = "Year",}
\CommentTok{#     rep.values = 0:3}
\CommentTok{#   ),}
\CommentTok{#   predDesign = systematicDesign(1500),}
\CommentTok{#   o.write = TRUE}
\CommentTok{# )}
\CommentTok{# createDistMat(with_sites, "preds", TRUE, TRUE)}
\end{Highlighting}
\end{Shaded}

Now we simulate data on all 900 potential sampling sites for 4 years.
That is, we are simulating the data that we would observe at any site at
any point in time if we choose to sample it. We hope that this will
allow us to emulate a real data collection example. The process should
look like this:

\begin{enumerate}
\def\labelenumi{\arabic{enumi}.}
\tightlist
\item
  We simulate all the data that we could possibly observe at any given
  site for any given year.
\item
  We find a 200-site GRTS design and we form our initial dataset using
  the simulated samples observed from these sites in the first and
  second years. We did not need to know anything about the stream
  network to find the GRTS design so this represents a scenario where we
  are collecting an initial sample to form our first ideas about how
  in-stream processes in our study area operate.
\item
  We use the information from the two years of sampling at our 200 GRTS
  sites to choose our next 50 sites adaptively.
\item
  We add the simulated data at our GRTS sites and our new 50 adaptive
  sites to our sample.
\item
  We update our knowledge about in-stream processes based on these new
  data.
\item
  We use all of our current information to choose the next 50 sites
  adaptively.
\end{enumerate}

Note that, in a real situation, steps 1, 2, 4, and 6 would be slightly
different. We would not simulate any data at the first step. At the
second step, we would lay out the same GRTS design but instead of
forming a sample by using the simulated values at those sites, we would
go into the field and directly sample the GRTS sites. At the fourth
step, we would again sample the GRTS sites plus the 50 sites chosen
adaptively. At the sixth step, we would again sample all previously
chosen sites in addition to the extra 50 sites chosen adaptively. The
point here is that we only simulate data in lieu of being able to
perform real data collection. Therefore, it may not be necessary to
replicate this exact process of simulation for a user's own use-case.

The first step in the data simulation process is to import the .ssn
folder we created before, and to extract some covariates (in particular,
stream order and the additive function values) to the potential sampling
sites from the stream edges.

\begin{Shaded}
\begin{Highlighting}[]
\CommentTok{# import stream network}
\NormalTok{pine.river <-}\StringTok{ }\KeywordTok{importSSN}\NormalTok{(}\StringTok{"./pine_river.ssn"}\NormalTok{, }\StringTok{"preds"}\NormalTok{)}

\CommentTok{# Extract some covariates from the edges}
\CommentTok{# Start by computing Shreve Stream Orders for the edges}
\NormalTok{pine.river <-}\StringTok{ }\KeywordTok{calculateShreveStreamOrderAndAFVs}\NormalTok{(pine.river)}
\CommentTok{# We need this later }
\NormalTok{pine.river}\OperatorTok{@}\NormalTok{predpoints}\OperatorTok{@}\NormalTok{SSNPoints[[}\DecValTok{1}\NormalTok{]]}\OperatorTok{@}\NormalTok{point.data}\OperatorTok{$}\NormalTok{Year <-}\StringTok{ }\DecValTok{1}
\CommentTok{# Now extract the shreve stream order and edge AFV values to the points}
\NormalTok{pine.river <-}\StringTok{ }\KeywordTok{extractStreamEdgeCovariates}\NormalTok{(}
  \CommentTok{# The SpatialStreamNetwork object}
\NormalTok{  pine.river,}
  \CommentTok{# The columns to extract from the edges data}
  \KeywordTok{c}\NormalTok{(}\StringTok{"AreaAFV"}\NormalTok{, }\StringTok{"shreve"}\NormalTok{)}
\NormalTok{)}
\end{Highlighting}
\end{Shaded}

We then compute a new covariate, which is the standardised stream order
(stream order divided by maximum stream order).

\begin{Shaded}
\begin{Highlighting}[]
\CommentTok{# The following code takes approximately 2 minutes to run.}
\CommentTok{# The most computationally intensive part is creating the}
\CommentTok{# distance matrices.}
\NormalTok{pine.river <-}\StringTok{ }\KeywordTok{transformSSNVars}\NormalTok{(}
  \CommentTok{# First, give the SpatialStreamNetwork object}
\NormalTok{  pine.river, }
  \CommentTok{# Then give the name of a new output folder}
  \CommentTok{# Note that in the example data the output of this function call}
  \CommentTok{# is called pine_river_sim_.ssn; the name has been changed here to }
  \CommentTok{# avoid a name conflict when writing out the result}
  \StringTok{"pine_river_sim.ssn"}\NormalTok{,}
  \CommentTok{# This is the format for creating new columns in the }
  \CommentTok{# point.data slot of the obspoints@SSNPoints[[1]] of the}
  \CommentTok{# SpatialStreamNetwork. The left-hand-side is the new column name}
  \CommentTok{# and the right-hand side is the existing column name.}
  \KeywordTok{c}\NormalTok{(}\StringTok{"order"}\NormalTok{ =}\StringTok{ "shreve"}\NormalTok{),}
  \CommentTok{# The same as above but not for the prediction points}
  \KeywordTok{c}\NormalTok{(}\StringTok{"order"}\NormalTok{ =}\StringTok{ "shreve"}\NormalTok{),}
  \CommentTok{# The function which is applied to the RHS column to create}
  \CommentTok{# the new LHS column}
  \ControlFlowTok{function}\NormalTok{(x) x}\OperatorTok{/}\KeywordTok{max}\NormalTok{(x), }\OtherTok{TRUE}
\NormalTok{)}
\CommentTok{# We need to redo the distance matrix computations}
\CommentTok{# because we have created a new folder for this SSN}
\KeywordTok{createDistMat}\NormalTok{(pine.river, }\StringTok{"preds"}\NormalTok{, }\OtherTok{TRUE}\NormalTok{, }\OtherTok{TRUE}\NormalTok{)}

\CommentTok{# If you didn't run the above code block, you may want to}
\CommentTok{# run }
\CommentTok{# pine.river <- importSSN("pine_river_sim_.ssn", "preds")}
\CommentTok{# createDistMat(pine.river, "preds", TRUE, TRUE)}
\end{Highlighting}
\end{Shaded}

This is because we expect DO to decrease with increasing stream order.
Therefore, we model DO as a function of normalised stream order. We set
the regression parameter for normalised stream order to -5 mg/L per unit
DO. The covariance mixture included a Spherical tailup function and a
random effect for each site, to account for the temporal replication
which occurs as the adaptive design progresses. The partial sill, range,
random effect variance and nugget parameters were assumed to be 4,
20000, 1, and 1, respectively. This finally leads us to the point of
simulating the data from such a model:

\begin{Shaded}
\begin{Highlighting}[]
\CommentTok{# Set a random seed}
\KeywordTok{set.seed}\NormalTok{(}\DecValTok{123}\NormalTok{)}

\CommentTok{# Simulate data on SSN}
\CommentTok{# This takes ~ 4 - 5 minutes}
\NormalTok{pine.river <-}\StringTok{ }\KeywordTok{SimulateOnSSN2}\NormalTok{(}
  \CommentTok{# This function is used in exactly the same way as}
  \CommentTok{# SimulateOnSSN but all arguments must be matched}
  \CommentTok{# explicitly by name.}
  \DataTypeTok{ssn.object =}\NormalTok{ pine.river,}
  \DataTypeTok{ObsSimDF =} \KeywordTok{getSSNdata.frame}\NormalTok{(pine.river),}
  \DataTypeTok{PredSimDF =} \KeywordTok{getSSNdata.frame}\NormalTok{(pine.river, }\StringTok{"preds"}\NormalTok{),}
  \DataTypeTok{PredID =} \StringTok{"preds"}\NormalTok{,}
  \DataTypeTok{formula =} \OperatorTok{~}\StringTok{ }\NormalTok{order,}
  \CommentTok{# calculate stream order / max(stream order),}
  \CommentTok{# insert as variable with slight negative effect}
  \CommentTok{# without random errors, this means our nominal range of }
  \CommentTok{# Dissolved Oxygen (mg/L) is 6 - 11 mg/L. }
  \DataTypeTok{coefficients =} \KeywordTok{c}\NormalTok{(}\DecValTok{11}\NormalTok{, }\OperatorTok{-}\DecValTok{5}\NormalTok{),}
  \DataTypeTok{CorModels =} \KeywordTok{c}\NormalTok{(}\StringTok{"Spherical.tailup"}\NormalTok{, }\StringTok{"locID"}\NormalTok{),}
  \DataTypeTok{CorParms =} \KeywordTok{c}\NormalTok{(}\DecValTok{4}\NormalTok{, }\FloatTok{2e4}\NormalTok{, }\DecValTok{1}\NormalTok{, }\DecValTok{1}\NormalTok{),}
  \DataTypeTok{addfunccol =} \StringTok{"AreaAFV"}
\NormalTok{)}
\end{Highlighting}
\end{Shaded}

An additional and non-intuitive step is needed after simulating the
data. When designing a monitoring program over several years with the
intention of keeping the sites from earlier time periods of sampling as
legacy sites in later periods of sampling, we need to split up the
observed sites shapefile by levels of the time period variable. This can
be done with the following code:

\begin{Shaded}
\begin{Highlighting}[]
\CommentTok{# Split up sites by time period (year)}
\NormalTok{first.year <-}\StringTok{ }\KeywordTok{splitSSNSites}\NormalTok{(}
  \CommentTok{# We're splitting this object}
\NormalTok{  pine.river,}
  \CommentTok{# Name of new .ssn folder to create }
  \StringTok{"Year_1_.ssn"}\NormalTok{,}
  \CommentTok{# What column we are splitting by}
  \StringTok{"Year"}\NormalTok{,}
  \CommentTok{# Whether we split the predictions as well}
  \CommentTok{# We only have one year of prediction sites so }
  \CommentTok{# it does not make sense to do this.}
  \OtherTok{FALSE} 
\NormalTok{)}
\end{Highlighting}
\end{Shaded}

We can now continue with our example. In this case study, we start with
two years of data collected from an established monitoring program
containing 200 sites. To emulate this in our synthetic case study, we
set up a GRTS design to serve as the existing monitoring program. We
choose a GRTS design because the spatially balanced samples they provide
are known to be efficient for prediction. In our case study, where the
existing monitoring program was set up without any previously collected
data, using a GRTS design for the first phase of data collection is
reasonable. This is the code needed to establish the GRTS design for the
first two years of sampling:

\begin{Shaded}
\begin{Highlighting}[]
\CommentTok{# Select 200-site GRTS design}
\NormalTok{first.grts <-}\StringTok{ }\KeywordTok{drawStreamNetworkSamples}\NormalTok{(}
\NormalTok{  first.year,}
  \StringTok{"Year_1_GRTS.ssn"}\NormalTok{,}
  \OtherTok{TRUE}\NormalTok{,}
  \StringTok{"GRTS"}\NormalTok{,}
  \DecValTok{200}
\NormalTok{)}
\KeywordTok{createDistMat}\NormalTok{(first.grts, }\StringTok{"preds"}\NormalTok{, T, T)}
\CommentTok{# if you didn't run the above, then make sure to run}
\CommentTok{# first.grts <- importSSN("Year_1_GRTS.ssn", "preds")}
\CommentTok{# createDistMat(first.grts)}
\CommentTok{# before moving on. }

\CommentTok{# Record the locIDs of the chosen sites}
\NormalTok{first.fixed <-}\StringTok{ }\KeywordTok{as.character}\NormalTok{(}\KeywordTok{getSSNdata.frame}\NormalTok{(first.grts)}\OperatorTok{$}\NormalTok{locID)}

\CommentTok{# Splice second years' sites}
\NormalTok{second.year <-}\StringTok{ }\KeywordTok{spliceSSNSites}\NormalTok{(}
\NormalTok{  first.grts, }\StringTok{"Year_2_Potential.ssn"}\NormalTok{, }\StringTok{"sites2.shp"}
\NormalTok{) }\CommentTok{# note, this will print messages from readOGR(). }
\NormalTok{second.year <-}\StringTok{ }\KeywordTok{importSSN}\NormalTok{(}\StringTok{"Year_2_Potential.ssn"}\NormalTok{, }\StringTok{"preds"}\NormalTok{)}
\NormalTok{second.year <-}\StringTok{ }\KeywordTok{subsetSSN}\NormalTok{(}
\NormalTok{  second.year, }\StringTok{"Year_2_Selected.ssn"}\NormalTok{,}
\NormalTok{  locID }\OperatorTok{%in%}\StringTok{ }\NormalTok{first.fixed}
\NormalTok{)}
\KeywordTok{createDistMat}\NormalTok{(second.year, }\StringTok{"preds"}\NormalTok{, T, T)}
\CommentTok{# if you didn't run the above, then make sure to run}
\CommentTok{# second.year <- importSSN("Year_2_Selected.ssn", "preds")}
\CommentTok{# createDistMat(second.year, "preds", T, T)}
\end{Highlighting}
\end{Shaded}

We now have a starting design. By the end of the second year of
sampling, where we are now, we have collected 400 DO samples across 200
unique locations. We can use these data to fit a model and to form
priors about the covariance parameters in our `true' model based on
their estimates in that model.

\begin{Shaded}
\begin{Highlighting}[]
\CommentTok{# Fit model}
\NormalTok{first.model <-}\StringTok{ }\KeywordTok{glmssn}\NormalTok{(}
\NormalTok{  Sim_Values }\OperatorTok{~}\StringTok{ }\NormalTok{order,}
\NormalTok{  second.year,}
  \DataTypeTok{CorModels =} \KeywordTok{c}\NormalTok{(}\StringTok{"Spherical.tailup"}\NormalTok{, }\StringTok{"locID"}\NormalTok{),}
  \DataTypeTok{addfunccol =} \StringTok{"AreaAFV"}
\NormalTok{)}

\CommentTok{# Form priors}
\CommentTok{# This will give a warning that the observed information matrix}
\CommentTok{# does not exist. This is fine, and is the normal behaviour of this}
\CommentTok{# function. Instead, we will use an estimate }
\CommentTok{# of the expected fisher information. }
\NormalTok{first.priors <-}\StringTok{ }\KeywordTok{constructLogNormalPriors}\NormalTok{(first.model)}
\end{Highlighting}
\end{Shaded}

Again, there is a seemingly unnatural step here necessitated by the file
structure of \texttt{SpatialStreamNetwork} objects. We previously split
our sites by year. Now, we need to reintroduce the third year of
potential sampling sites back into the \texttt{SpatialStreamNetwork}. We
do this with the function \texttt{spliceSSNSites}, which requires the
following code:

\begin{Shaded}
\begin{Highlighting}[]
\CommentTok{# New sites for next year}
\NormalTok{second.year <-}\StringTok{ }\KeywordTok{spliceSSNSites}\NormalTok{(}
\NormalTok{  second.year, }\StringTok{"Year_3_Potential.ssn"}\NormalTok{, }\StringTok{"sites3.shp"}
\NormalTok{)}
\NormalTok{second.year <-}\StringTok{ }\KeywordTok{importSSN}\NormalTok{(}\StringTok{"Year_3_Potential.ssn"}\NormalTok{, }\StringTok{"preds"}\NormalTok{)}
\KeywordTok{createDistMat}\NormalTok{(second.year, }\StringTok{"preds"}\NormalTok{, T, T)}

\CommentTok{# Now save data for next run.}
\CommentTok{# You would uncomment and run the following line}
\CommentTok{# if you were planning on running optimiseSSNDesign}
\CommentTok{# over our results from before. Note, the save step is}
\CommentTok{# not necessary in many cases. We have done it here because}
\CommentTok{# we had to transfer the data from a local computer, where we}
\CommentTok{# did this preprocessing, to a high performance}
\CommentTok{# computing cluster.}
\CommentTok{# save.image("Sim_Data_Year_3.Rdata")}
\end{Highlighting}
\end{Shaded}

At this juncture, we can begin our adaptive design. We have two years of
data, a model, and priors. We do adaptive design by calling
\texttt{optimiseSSNDesign}, as shown below. Note that the code below is
set up to be used on the Queensland University of Technology's High
Performance Computing system, which runs a linux operating system with R
v. 3.5.1. We did this to save time. This code block takes between 50 and
55 hours to run on 32 CPUs.

Note that one feature of the code that is unusual for a local machine or
a smaller example is that we set \texttt{n.optim\ =\ 1}. Ordinarily,
this would be too low. Setting \texttt{n.optim\ =\ 5} would have been
reasonable. However, to save time, we set up five separate scripts with
\texttt{n.optim\ =\ 1} and set different random seeds (1, 1e6 + 1, 2e6 +
1, 3e6 + 1, and 4e6 + 1) to achieve the same result as
\texttt{n.optim\ =\ 5} but in a fifth of the time.

\begin{Shaded}
\begin{Highlighting}[]
\CommentTok{# }\AlertTok{WARNING}\CommentTok{: THIS CODE TAKES APPROXIMATELY 50 HOURS TO RUN}
\CommentTok{# USING 32 CPUs.}

\CommentTok{# Load preprocessed model and priors}
\CommentTok{# You MUST set your working directory to the folder on}
\CommentTok{# your computer containing the .Rdata file and .ssn folder}
\CommentTok{# load("Sim_Data_Year_3.Rdata")}

\CommentTok{# Import SpatialStreamNetwork object}
\CommentTok{# second.year <- importSSN(}
\CommentTok{#   "Year_3_Potential.ssn",}
\CommentTok{#   "preds"}
\CommentTok{# )}

\CommentTok{# Optimise a design}
\CommentTok{# O <- optimiseSSNDesign(}
\CommentTok{#   ssn = second.year,}
\CommentTok{#   new.ssn.path = "S_Up_1.ssn",}
\CommentTok{#   glmssn = first.model,}
\CommentTok{#   n.points = 250,}
\CommentTok{#   legacy.sites = first.fixed,}
\CommentTok{#   utility.function = KOptimality,}
\CommentTok{#   prior.parameters = first.priors,}
\CommentTok{#   n.cores = 32,}
\CommentTok{#   parallelism = "osx/linux",}
\CommentTok{#   parallelism.seed = 1,}
\CommentTok{#   # the above argument changes in increments}
\CommentTok{#   # of 1e6, being set to 1, 1e6 + 1, 2e6 + 1, etc.}
\CommentTok{#   # up to 4e6 + 1}
\CommentTok{#   # this allowed us to generate five adaptive designs}
\CommentTok{#   # from five random starts}
\CommentTok{#   n.optim = 1,}
\CommentTok{#   n.draws = 500}
\CommentTok{# )}

\CommentTok{# Save the result}
\CommentTok{# Note the number at the end of the file name changes}
\CommentTok{# according to the random start (1, 2, 3, 4, 5)}
\CommentTok{# save(O, first.priors, "S_Up_Optimal_1.Rdata")}
\end{Highlighting}
\end{Shaded}

Getting to this point is time-consuming; this is especially true at the
last step. To give some indication of the results, however, we load
pre-processed data. Note we are loading the file \texttt{S\_Up\_3.Rdata}
because the adaptive design found using
\texttt{parallelism.seed\ =\ 2e6\ +\ 1} produced the best result of our
five optimisation runs. We load the data with

\begin{Shaded}
\begin{Highlighting}[]
\KeywordTok{load}\NormalTok{(}\StringTok{"S_Up_Optimal_3.Rdata"}\NormalTok{)}
\end{Highlighting}
\end{Shaded}

We can plot diagnostics for the optimisation algorithm, such as the
trace plot of the maximum expected utility.

\begin{Shaded}
\begin{Highlighting}[]
\KeywordTok{plot}\NormalTok{(O)}
\end{Highlighting}
\end{Shaded}

\begin{figure}
\centering
\includegraphics{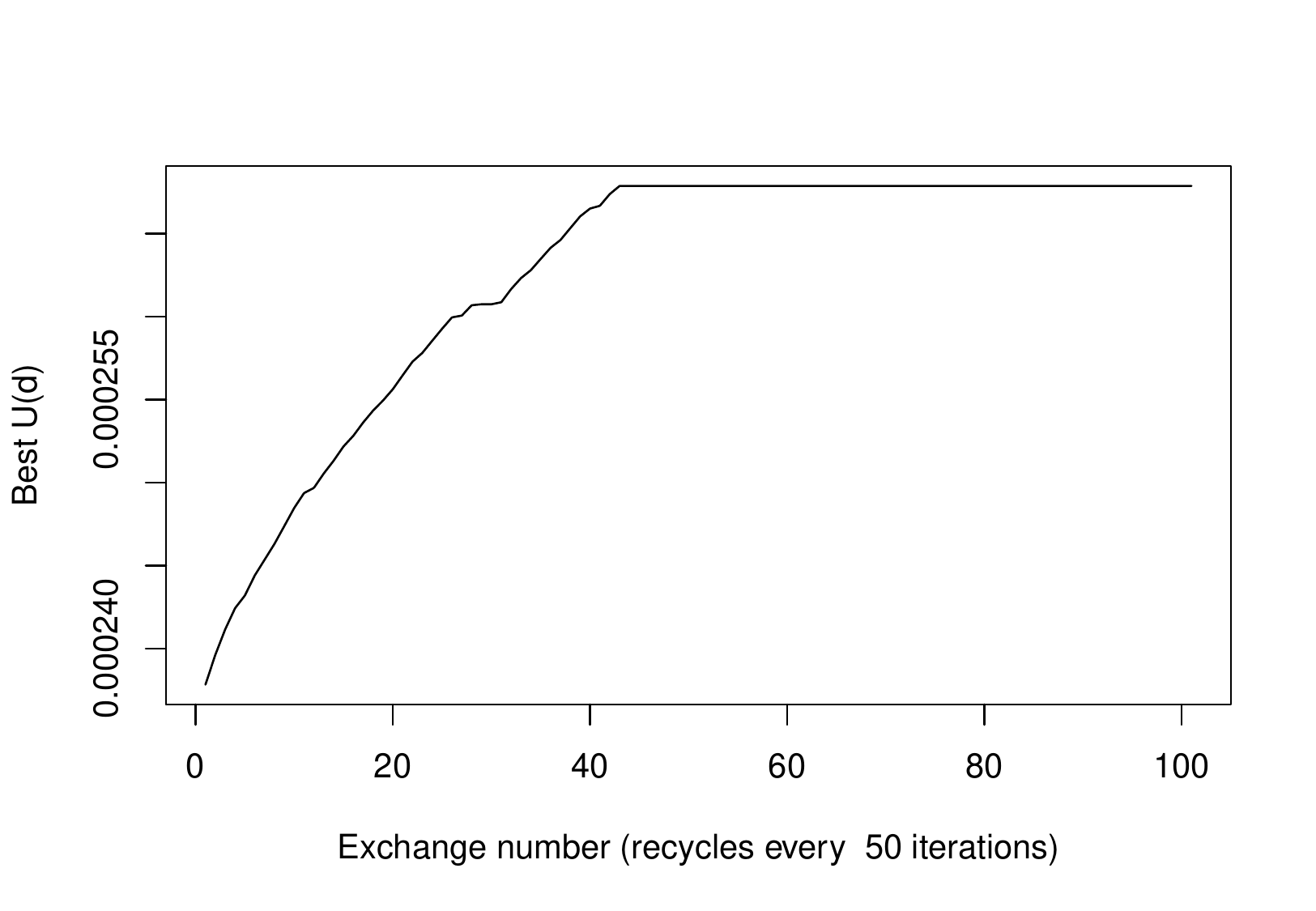}
\caption{The trace of the Greedy Exchange Algorithm. The y-axis
represents the maximum expected utility at each iteration in the
algorithm.}
\end{figure}

We can also plot the adaptive design, indicating which sites have been
added to the design.

\begin{Shaded}
\begin{Highlighting}[]
\CommentTok{# Data frame for the adaptive design observed sites}
\NormalTok{adaptive <-}\StringTok{ }\KeywordTok{getSSNdata.frame}\NormalTok{(O}\OperatorTok{$}\NormalTok{ssn.new)}
\CommentTok{# New variable to code for whether the sites are legacy sites}
\CommentTok{# or were freshly added}
\NormalTok{adaptive}\OperatorTok{$}\NormalTok{New <-}\StringTok{ }\KeywordTok{with}\NormalTok{(adaptive, }\DecValTok{1} \OperatorTok{*}\StringTok{ }\NormalTok{(}\OperatorTok{!}\NormalTok{locID }\OperatorTok{%in%}\StringTok{ }\NormalTok{O}\OperatorTok{$}\NormalTok{legacy.sites))}
\CommentTok{# Return data frame to the SpatialStreamNetwork object}
\NormalTok{O}\OperatorTok{$}\NormalTok{ssn.new <-}\StringTok{ }\KeywordTok{putSSNdata.frame}\NormalTok{(adaptive, O}\OperatorTok{$}\NormalTok{ssn.new)}
\CommentTok{# Plot}
\KeywordTok{plot}\NormalTok{(O}\OperatorTok{$}\NormalTok{ssn.new, }\StringTok{"New"}\NormalTok{, }\DataTypeTok{nclasses =} \DecValTok{2}\NormalTok{, }\DataTypeTok{breaktype =} \StringTok{"even"}\NormalTok{)}
\end{Highlighting}
\end{Shaded}

\begin{figure}
\centering
\includegraphics{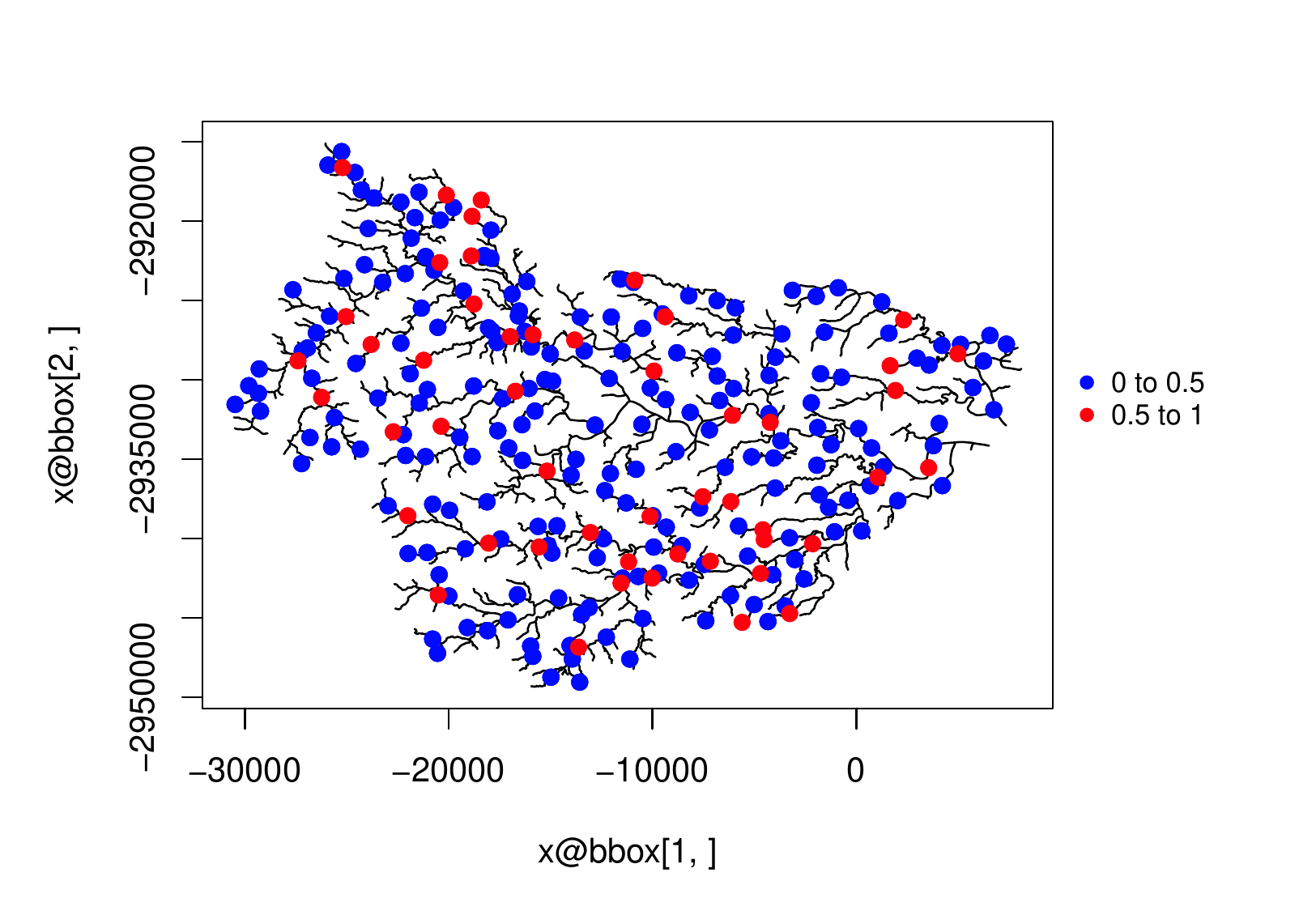}
\caption{The adaptive design at the second step.}
\end{figure}

At this stage, we have found the adaptive design for the third year of
sampling. The next task is to find the adaptive design in the fourth
year of sampling. However, before this can occur, we need to

\begin{enumerate}
\def\labelenumi{\arabic{enumi}.}
\tightlist
\item
  Refit our true model to the updated dataset (which represents the
  situation where we find the adaptive design, and sample from it).
\item
  Update our priors based on our new estimates from the updated model.
\item
  Update our list of legacy sites.
\end{enumerate}

We do this using the following code:

\begin{Shaded}
\begin{Highlighting}[]
\CommentTok{# Read in the best adaptive design out of 5}
\NormalTok{ssn <-}\StringTok{ }\KeywordTok{importSSN}\NormalTok{(}\StringTok{"./S_Up_3.ssn"}\NormalTok{, }\StringTok{"preds"}\NormalTok{)}
\KeywordTok{createDistMat}\NormalTok{(ssn, }\StringTok{"preds"}\NormalTok{, T, T)}

\CommentTok{# fit model to SSN}
\NormalTok{third.model <-}\StringTok{ }\KeywordTok{glmssn}\NormalTok{(}
\NormalTok{  Sim_Values }\OperatorTok{~}\StringTok{ }\NormalTok{order, ssn,}
  \DataTypeTok{CorModels =} \KeywordTok{c}\NormalTok{(}\StringTok{"Spherical.tailup"}\NormalTok{, }\StringTok{"locID"}\NormalTok{),}
  \DataTypeTok{addfunccol =} \StringTok{"AreaAFV"}
\NormalTok{)}

\CommentTok{# 'update' priors}
\CommentTok{# This will throw another warning about the observed information}
\CommentTok{# matrix, but this is okay.}
\NormalTok{third.priors <-}\StringTok{ }\KeywordTok{constructLogNormalPriors}\NormalTok{(third.model)}

\CommentTok{# Update fixed sites}
\NormalTok{third.fixed <-}\StringTok{ }\KeywordTok{unique}\NormalTok{(}\KeywordTok{as.character}\NormalTok{(}\KeywordTok{getSSNdata.frame}\NormalTok{(ssn)}\OperatorTok{$}\NormalTok{locID))}

\CommentTok{# now splice in fourth year of sites}
\NormalTok{ssn <-}\StringTok{ }\KeywordTok{spliceSSNSites}\NormalTok{(ssn, }\StringTok{"Year_4_Potential.ssn"}\NormalTok{, }\StringTok{"sites4.shp"}\NormalTok{)}
\NormalTok{ssn <-}\StringTok{ }\KeywordTok{importSSN}\NormalTok{(}\StringTok{"Year_4_Potential.ssn"}\NormalTok{, }\StringTok{"preds"}\NormalTok{)}

\CommentTok{# Save image}
\CommentTok{# Run the following line of code if you intend to run the next }
\CommentTok{# codeblock down. Note, the save step is not necessary in many}
\CommentTok{# cases. We have done it here because we had to transfer the }
\CommentTok{# data from a local computer, where we did this preprocessing,}
\CommentTok{# to a high performance computing cluster.}
\CommentTok{# save.image("Sim_Data_Year_4.Rdata")}
\end{Highlighting}
\end{Shaded}

The only thing that remains is to find the adaptive design for the
fourth year of sampling.

\begin{Shaded}
\begin{Highlighting}[]
\CommentTok{# Load the data we saved before}
\CommentTok{# Again, you MUST set your working directory to}
\CommentTok{# the folder containing these files (the .Rdata file}
\CommentTok{# and the .ssn folder).}
\CommentTok{# load("Sim_Data_Year_4.Rdata")}

\CommentTok{# Import the fourth year of data}
\CommentTok{# with potential sites}
\CommentTok{# fourth.year <- importSSN(}
\CommentTok{#   "Year_4_Potential.ssn",}
\CommentTok{#   "preds"}
\CommentTok{# )}

\CommentTok{# Find adaptive design}
\CommentTok{# O <- optimiseSSNDesign(}
\CommentTok{#   ssn = fourth.year,}
\CommentTok{#   new.ssn.path = "S_Up_1_1.ssn",}
\CommentTok{#   glmssn = third.model,}
\CommentTok{#   n.points = 300,}
\CommentTok{#   legacy.sites = third.fixed,}
\CommentTok{#   utility.function = KOptimality,}
\CommentTok{#   prior.parameters = third.priors,}
\CommentTok{#   n.cores = 20,}
\CommentTok{#   parallelism = "osx/linux",}
\CommentTok{#   parallelism.seed = 1,}
\CommentTok{#   ## this arguments increases in increments of}
\CommentTok{#   ## 1e6 as before}
\CommentTok{#   n.optim = 1,}
\CommentTok{#   n.draws = 500}
\CommentTok{# )}

\CommentTok{# Modify the contents of the glmssn object}
\CommentTok{# to save space in .Rdata file}
\CommentTok{# third.model$ssn.object <- NULL}
\CommentTok{# third.model$estimates$V <- NULL}
\CommentTok{# third.model$estimates$Vi <- NULL}
\CommentTok{# third.model$sampinfo$REs <- NULL}
\CommentTok{# O$ssn.old <- NULL}
\CommentTok{# O$glmssn <- NULL}

\CommentTok{# Save the workspace image}
\CommentTok{# Both numbers at the end of the filename increment}
\CommentTok{# with the random start (1, 2, 3, 4, 5)}
\CommentTok{# save(O, third.fixed, third.priors, file = "S_Up_Optimal_1_1.Rdata")}
\end{Highlighting}
\end{Shaded}

This gives us the final state of our adaptive design. If we want to
explore this further, we can import a pre-processed dataset:

\begin{Shaded}
\begin{Highlighting}[]
\KeywordTok{load}\NormalTok{(}\StringTok{"S_Up_Optimal_3_3.Rdata"}\NormalTok{)}
\end{Highlighting}
\end{Shaded}

We use \texttt{S\_Up\_Optimal\_3\_3.Rdata} because the optimisation run
with \texttt{parallelism.seed\ =\ 2e6\ +\ 1} again produced the adaptive
design with the highest expected utility.

\begin{Shaded}
\begin{Highlighting}[]
\CommentTok{# Same plot as before}
\NormalTok{adaptive <-}\StringTok{ }\KeywordTok{getSSNdata.frame}\NormalTok{(O}\OperatorTok{$}\NormalTok{ssn.new)}
\NormalTok{adaptive}\OperatorTok{$}\NormalTok{New <-}\StringTok{ }\KeywordTok{with}\NormalTok{(adaptive, }\DecValTok{1} \OperatorTok{*}\StringTok{ }\NormalTok{(}\OperatorTok{!}\NormalTok{locID }\OperatorTok{%in%}\StringTok{ }\NormalTok{O}\OperatorTok{$}\NormalTok{legacy.sites))}
\NormalTok{O}\OperatorTok{$}\NormalTok{ssn.new <-}\StringTok{ }\KeywordTok{putSSNdata.frame}\NormalTok{(adaptive, O}\OperatorTok{$}\NormalTok{ssn.new)}
\KeywordTok{plot}\NormalTok{(O}\OperatorTok{$}\NormalTok{ssn.new, }\StringTok{"New"}\NormalTok{, }\DataTypeTok{nclasses =} \DecValTok{2}\NormalTok{, }\DataTypeTok{breaktype =} \StringTok{"even"}\NormalTok{)}
\end{Highlighting}
\end{Shaded}

\begin{figure}
\centering
\includegraphics{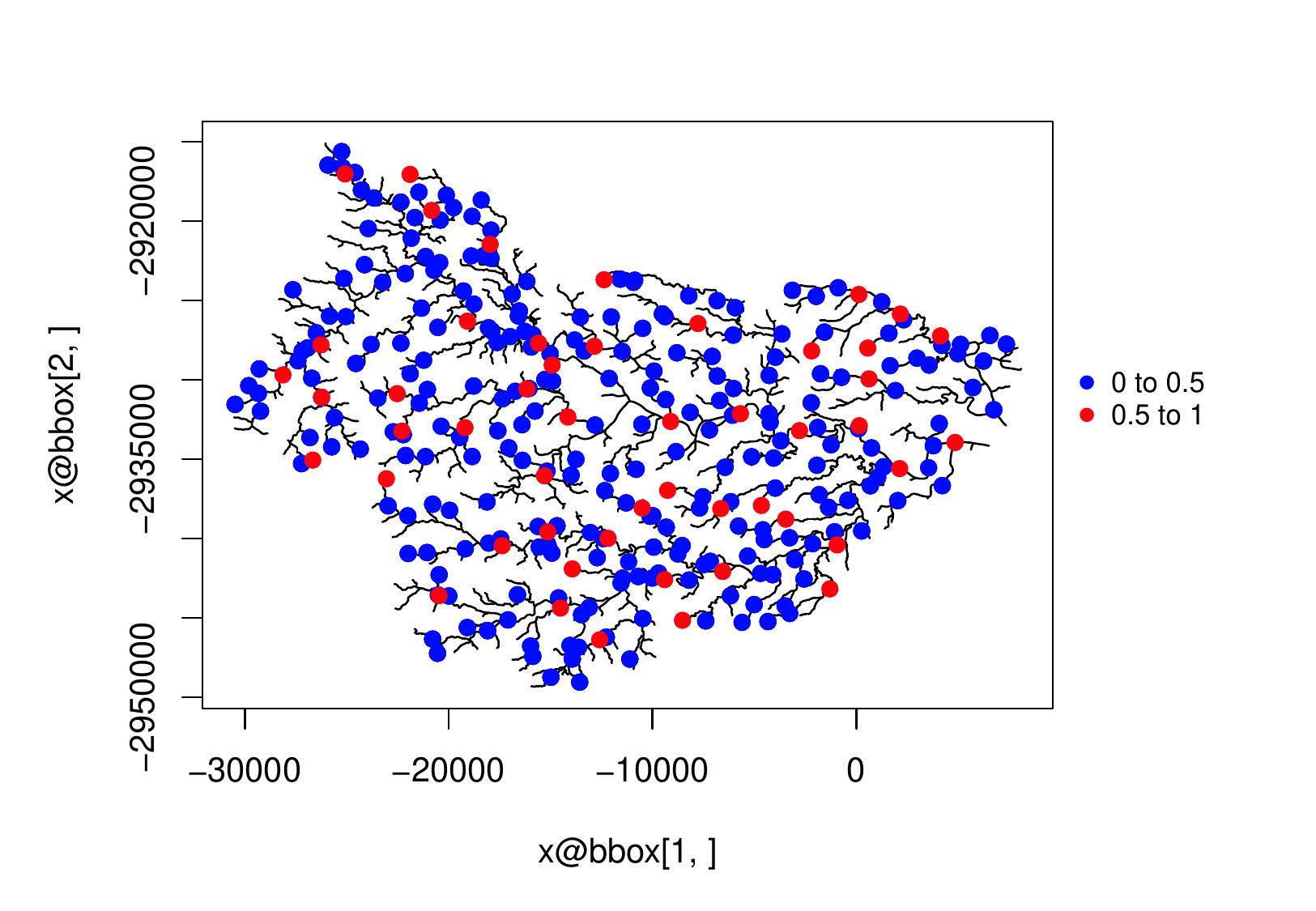}
\caption{The adaptive design at the final step.}
\end{figure}

All that remains is to validate our final design. This proceeds in a few
steps. Firstly, we need to import a \texttt{SpatialStreamNetwork} object
containing all potential sampling sites across all years of sampling.

\begin{Shaded}
\begin{Highlighting}[]
\CommentTok{# we need the ssn containing all POTENTIAL sampling sites for}
\CommentTok{# ALL sampling years. This will be used in evaluateFixedDesigns}
\NormalTok{total.ssn <-}\StringTok{ }\KeywordTok{importSSN}\NormalTok{(}\StringTok{"./pine_river_sim_.ssn"}\NormalTok{, }\StringTok{"preds"}\NormalTok{)}
\end{Highlighting}
\end{Shaded}

Secondly, we need to remove sites that were included in the GRTS designs
we used for the first two years of sampling. This is not a step that is
required in general. However, we do it because we want to compare the
efficiencies of our adaptive design against some standard designs, and
the adaptive design only includes the 100 sites we added between years
three and four.

\begin{Shaded}
\begin{Highlighting}[]
\CommentTok{# import adaptive design}
\NormalTok{fourth <-}\StringTok{ }\KeywordTok{importSSN}\NormalTok{(}\StringTok{"./S_Up_3_3.ssn"}\NormalTok{, }\StringTok{"preds"}\NormalTok{)}
\CommentTok{# import legacy sites and record them}
\NormalTok{first.two <-}\StringTok{ }\KeywordTok{importSSN}\NormalTok{(}\StringTok{"Year_2_Selected.ssn"}\NormalTok{)}
\NormalTok{first.fixed <-}\StringTok{ }\KeywordTok{getSSNdata.frame}\NormalTok{(first.two)}\OperatorTok{$}\NormalTok{locID}
\KeywordTok{rm}\NormalTok{(first.two)}
\CommentTok{# record the design as a vector of pids}
\CommentTok{# but first remove the 200 fixed GRTS sites we started with}
\NormalTok{fourth.design <-}\StringTok{ }\KeywordTok{with}\NormalTok{(}\KeywordTok{getSSNdata.frame}\NormalTok{(fourth), pid[}\OperatorTok{!}\NormalTok{locID }\OperatorTok{%in%}\StringTok{ }\NormalTok{first.fixed])}
\NormalTok{opt.designs <-}\StringTok{ }\KeywordTok{list}\NormalTok{(}
  \DataTypeTok{fourth =}\NormalTok{ fourth.design}
\NormalTok{)}
\end{Highlighting}
\end{Shaded}

Thirdly, we set up a range of standard designs we want to validate our
adaptive design against. For us, these are random and GRTS designs. We
generate 20 of each to account for the range of performance we might
expect to see because random and GRTS designs are stochastic.

\begin{Shaded}
\begin{Highlighting}[]
\CommentTok{# Set a seed for reproducibility}
\KeywordTok{set.seed}\NormalTok{(}\DecValTok{123456789}\NormalTok{)}

\CommentTok{# Construct some GRTS designs}
\CommentTok{# Set up an empty vector to store designs}
\NormalTok{grts.designs <-}\StringTok{ }\KeywordTok{vector}\NormalTok{(}\StringTok{"list"}\NormalTok{, }\DecValTok{20}\NormalTok{)}
\ControlFlowTok{for}\NormalTok{(i }\ControlFlowTok{in} \DecValTok{1}\OperatorTok{:}\DecValTok{20}\NormalTok{)\{}
  \CommentTok{# Use this function to create temporally evolving}
  \CommentTok{# GRTS designs, using the 'master-sample' approach}
\NormalTok{  grts.designs[[i]] <-}\StringTok{ }\KeywordTok{evolveGRTSOverTime}\NormalTok{(}
\NormalTok{    total.ssn,}
    \KeywordTok{c}\NormalTok{(}\DecValTok{0}\NormalTok{, }\DecValTok{0}\NormalTok{, }\DecValTok{50}\NormalTok{, }\DecValTok{50}\NormalTok{),}
    \StringTok{"Year"}
\NormalTok{  )}\OperatorTok{$}\NormalTok{Period_}\DecValTok{3}\OperatorTok{$}\NormalTok{by.pid}
\NormalTok{\}}
\KeywordTok{names}\NormalTok{(grts.designs) <-}\StringTok{ }\KeywordTok{rep}\NormalTok{(}\StringTok{"GRTS"}\NormalTok{, }\DecValTok{20}\NormalTok{)}

\CommentTok{# Random designs}
\CommentTok{# Set up an empty vector as before}
\NormalTok{rand.designs <-}\StringTok{ }\KeywordTok{vector}\NormalTok{(}\StringTok{"list"}\NormalTok{, }\DecValTok{20}\NormalTok{)}
\ControlFlowTok{for}\NormalTok{(i }\ControlFlowTok{in} \DecValTok{1}\OperatorTok{:}\DecValTok{20}\NormalTok{)\{}
  \CommentTok{# Use this function to build up a random design}
  \CommentTok{# over time.}
\NormalTok{  rand.designs[[i]] <-}\StringTok{ }\KeywordTok{evolveRandOverTime}\NormalTok{(}
\NormalTok{    total.ssn,}
    \KeywordTok{c}\NormalTok{(}\DecValTok{0}\NormalTok{, }\DecValTok{0}\NormalTok{, }\DecValTok{50}\NormalTok{, }\DecValTok{50}\NormalTok{),}
    \StringTok{"Year"}
\NormalTok{  )}\OperatorTok{$}\NormalTok{Period_}\DecValTok{3}\OperatorTok{$}\NormalTok{by.pid}
\NormalTok{\}}
\KeywordTok{names}\NormalTok{(rand.designs) <-}\StringTok{ }\KeywordTok{rep}\NormalTok{(}\StringTok{"Rand"}\NormalTok{, }\DecValTok{20}\NormalTok{)}
\CommentTok{# Combine our lists of data}
\NormalTok{designs <-}\StringTok{ }\KeywordTok{append}\NormalTok{(opt.designs, grts.designs)}
\NormalTok{designs <-}\StringTok{ }\KeywordTok{append}\NormalTok{(designs, rand.designs)}
\end{Highlighting}
\end{Shaded}

Finally, we benchmark our adaptive design against the GRTS and random
designs. We do two sets of benchmarking. The first set of benchmarking
compares the efficiencies of the designs under the true values of the
covariance parameters. THe second set of benchmarking compares the
efficiencies of the designs under the last formed set of priors for the
covariance parameters. In both cases, we expect the adaptive design to
outperform the random and GRTS designs.

\begin{Shaded}
\begin{Highlighting}[]
\CommentTok{# Benchmark our adpative designs aginst the GRTS and}
\CommentTok{# random designs}

\CommentTok{# compare designs under last best estimate of parameters}
\CommentTok{# warning: this can take an hour}
\NormalTok{efficiencies.emp <-}\StringTok{ }\KeywordTok{evaluateFixedDesigns}\NormalTok{(}
\NormalTok{  total.ssn,}
\NormalTok{  third.model,}
\NormalTok{  designs,}
  \StringTok{"pid"}\NormalTok{,}
\NormalTok{  KOptimality,}
\NormalTok{  third.priors,}
  \DecValTok{1000}
\NormalTok{)}

\CommentTok{# save designs and results in .Rdata file}
\CommentTok{# save(designs, efficiencies.emp, file = "benchmarked.Rdata")}
\end{Highlighting}
\end{Shaded}

Running the above code can take 40-60 minutes, so we will load
pre-processed data.

\begin{Shaded}
\begin{Highlighting}[]
\KeywordTok{load}\NormalTok{(}\StringTok{"benchmarked.Rdata"}\NormalTok{)}
\end{Highlighting}
\end{Shaded}

The first six rows of the \texttt{data.frame} object called
\texttt{efficiencies.emp} look like this:

\begin{Shaded}
\begin{Highlighting}[]
\KeywordTok{head}\NormalTok{(efficiencies.emp)}
\CommentTok{#>       ID Size Expected utility Efficiency Efficiency_Unlogged}
\CommentTok{#> 1 fourth  150     0.0003416538  1.0000000           1.0000000}
\CommentTok{#> 2   GRTS  150     0.0002837895  0.8306348           0.9999421}
\CommentTok{#> 3   GRTS  150     0.0002543829  0.7445634           0.9999127}
\CommentTok{#> 4   GRTS  150     0.0002849684  0.8340854           0.9999433}
\CommentTok{#> 5   GRTS  150     0.0002863867  0.8382367           0.9999447}
\CommentTok{#> 6   GRTS  150     0.0002786942  0.8157211           0.9999370}
\end{Highlighting}
\end{Shaded}

The last six rows look like this:

\begin{Shaded}
\begin{Highlighting}[]
\KeywordTok{tail}\NormalTok{(efficiencies.emp)}
\CommentTok{#>      ID Size Expected utility Efficiency Efficiency_Unlogged}
\CommentTok{#> 36 Rand  150     0.0002773608  0.8118184           0.9999357}
\CommentTok{#> 37 Rand  150     0.0002817689  0.8247207           0.9999401}
\CommentTok{#> 38 Rand  150     0.0002781254  0.8140564           0.9999365}
\CommentTok{#> 39 Rand  150     0.0002750161  0.8049556           0.9999334}
\CommentTok{#> 40 Rand  150     0.0002840155  0.8312962           0.9999424}
\CommentTok{#> 41 Rand  150     0.0002639778  0.7726472           0.9999223}
\end{Highlighting}
\end{Shaded}

The column \texttt{ID} represents an identifier given to the design.
These are the names of the list of designs provided to
\texttt{evaluateFixedDesigns} by the user but if these are null then the
\texttt{ID} column will simply contain unique numerical identifiers. The
\texttt{Size} column is the number of samples in the design. If the
argument \texttt{list.of\ =\ "pid"} then \texttt{Size} is the number of
samples across all sites across all years. If
\texttt{list.of\ =\ "locID"} then \texttt{Size} is the number of unique
sampling locations. The \texttt{Expected\ utility} column contains the
expected utility for each design, and the two \texttt{Efficiency}
columns are the efficiencies of the designs relative to the design with
the highest expected utility. Note that \texttt{Efficiency} is
calculated as a direct ratio, i.e. \(U(d_i)/U(d^*)\) where \(d^*\) is
the best design. In contrast, \texttt{Efficiency\_Unlogged} is
calculated as a ratio on the log-scale. That is, it is calculated as
\(\exp{\left\{U(d_i) - U(d^*)\right\}}\). This method is correct when
\(U(d_i)\) and \(U(d^*)\) are log-scale expected utilities.

We can plot the results in whatever way seems appropriate. In this
instance we used a boxplot of the performances of the GRTS and random
designs, with a dashed line to indicate the performance of the adaptive
design. The expected utility is the inverse sum of the kriging variances
across our 900 prediction sites, so we made boxplots of the sum of the
kriging variances for each of the design types.

\begin{Shaded}
\begin{Highlighting}[]
\KeywordTok{par}\NormalTok{(}\DataTypeTok{mai =} \KeywordTok{c}\NormalTok{(}\DecValTok{1}\NormalTok{, }\FloatTok{0.5}\NormalTok{, }\FloatTok{0.1}\NormalTok{, }\FloatTok{0.1}\NormalTok{))}
\KeywordTok{boxplot}\NormalTok{(}
  \CommentTok{# Note that 1/Expected utility is the sum of the}
  \CommentTok{# kriging variances}
  \DecValTok{1}\OperatorTok{/}\StringTok{`}\DataTypeTok{Expected utility}\StringTok{`} \OperatorTok{~}\StringTok{ }\NormalTok{ID,}
  \CommentTok{# Drop first row because this is the adaptive design}
\NormalTok{  efficiencies.emp[}\OperatorTok{-}\DecValTok{1}\NormalTok{, ],}
  \DataTypeTok{ylim =} \KeywordTok{c}\NormalTok{(}\DecValTok{2800}\NormalTok{, }\DecValTok{4000}\NormalTok{),}
  \DataTypeTok{xlab =} \KeywordTok{expression}\NormalTok{(Sum}\OperatorTok{~}\NormalTok{of}\OperatorTok{~}\NormalTok{kriging}\OperatorTok{~}\NormalTok{variances}\OperatorTok{~}\KeywordTok{U}\NormalTok{(d)}\OperatorTok{^-}\DecValTok{1}\NormalTok{),}
  \DataTypeTok{horizontal =}\NormalTok{ T}
\NormalTok{)}
\KeywordTok{abline}\NormalTok{(}
  \DataTypeTok{v =} \DecValTok{1}\OperatorTok{/}\NormalTok{efficiencies.emp}\OperatorTok{$}\StringTok{`}\DataTypeTok{Expected utility}\StringTok{`}\NormalTok{[}\DecValTok{1}\NormalTok{],}
  \DataTypeTok{col =} \StringTok{"red"}\NormalTok{,}
  \DataTypeTok{lty =} \DecValTok{2}
\NormalTok{)}
\end{Highlighting}
\end{Shaded}

\begin{figure}
\centering
\includegraphics{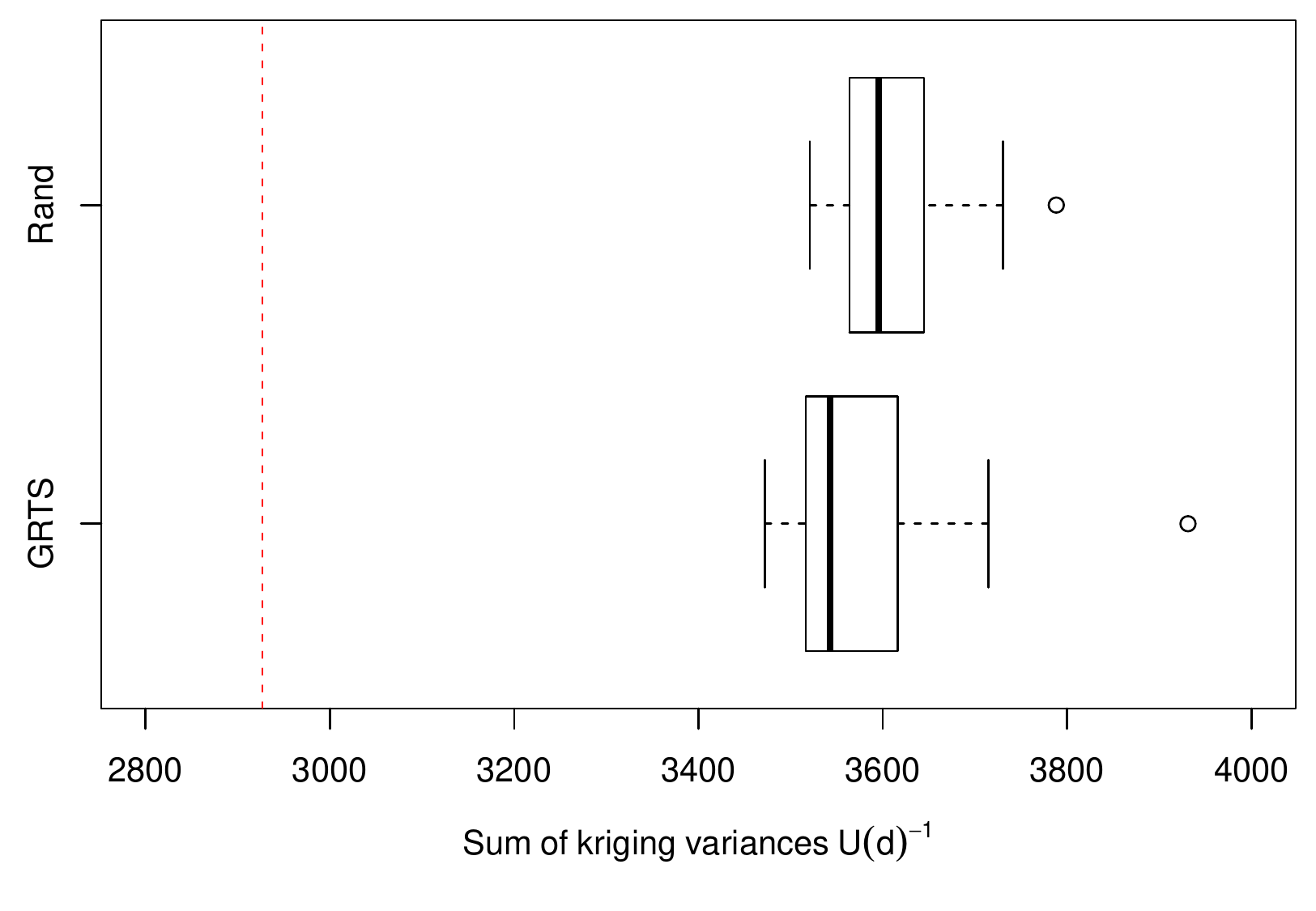}
\caption{Sum of the kriging variances from random and GRTS designs
(plotted as boxplots) compared to the sum of the kriging variances from
the adaptive design (plotted as the dashed red line). A smaller sum of
kriging variances indicates less uncertainty in the predictions from a
model.}
\end{figure}

These results indicate that random and GRTS designs perform to a similar
level but that the adaptive design is more efficient. That is, we can
make more accurate predictions of DO across the stream network using
fewer observations when we use the adaptive design.

\section{Summary}\label{summary}

The \texttt{SSNdesign} package is designed to solve optimal and adaptive
design problems on stream networks. The package contains functions for
preprocessing stream network data and functions for finding designs for
stream network data. The key functions are
\texttt{drawStreamNetworkSamples}, which allows users to construct
stream network designs based on standard spatial sampling schemes such
as GRTS, and \texttt{optimiseSSNdesign}, which is the main workhorse
function that can be used to find optimal and adaptive designs. We hope
\texttt{SSNdesign} will prove a useful tool for aquatic scientists and
managers.

\maketitle

\section{Glossary}\label{glossary}

\begin{description}
\item[\textbf{Adaptive design}]
A framework for sequentially modifying a monitoring program over time.
This framework uses data that have already been collected in an existing
monitoring program to inform the next decision(s) about new sites to
sample in the next sampling period.
\item[\textbf{Bayesian statistics}]
A framework for statistical inference characterised by the use of prior
information combined with information from the data (through the
likelihood function) to construct a posterior distribution of the
parameters. It is this distribution on which all inferences are based.
See also \textbf{pseudo-Bayesian}.
\item[\textbf{Covariance}]
A measure representing the strength of spatial autocorrelation between
measurements, given the distance separating them. Covariances are
unnormalised \textbf{Correlations}. That is, they do not necessarily
range between -1 and 1; instead, they can range between negative and
positive infinity.
\item[\textbf{Covariance function}]
A mathematical function used to estimate the covariance between any two
sets of locations in a study area. Covariance functions tend to have
three parameters: the \textbf{Nugget effect}, the \textbf{Range}, and
the \textbf{Sill}.
\item[\textbf{Covariate}]
A variable that is thought to affect the outcome of an experiment. These
variables are also measured or recorded alongside the
\textbf{Response variable} during the experiment so that the
relationship between them can be quantified. Often referred to as
predictors or independent variables.
\item[\textbf{Design}]
A particular configuration of an experiment. In the context of
\textbf{Stream networks} and \textbf{Spatial statistics}, a design is a
set of \textbf{Sites} that are sampled.
\item[\textbf{Design criterion}]
See \textbf{Utility function} and \textbf{Expected utility}.
\item[\textbf{Efficiency}]
The amount of information gained from unit effort in an experiment. More
efficient experiments need fewer samples to gain the same amount of
information as less efficient experiments. See also
\textbf{Relative efficiency}.
\item[\textbf{Expected utility}]
Qualitatively, this is a measure of how fit a \textbf{Design} is for a
user-specified objective. Higher values of the expected utility indicate
higher suitability. Mathematically, it is the average value of the
\textbf{Utility function} once integrating out uncertainty due to, for
example, the parameter values and the observed data.
\item[\textbf{Experimental design}]
The deliberate construction of experiments, usually controlled
experiments but also relates to how observational data are collected, in
a way that makes the observed data easier to statistically analyse and
may also strengthen any statistical conclusions drawn from those data.
\item[\textbf{Fisher information (matrix)}]
For a single \textbf{Parameter}, the Fisher information is a single
number that describes how much information there to estimate the value
of that parameter. For multiple parameters, the Fisher information is
instead a matrix where the elements on the diagonal describe the
information in the parameters and the off-diagonal elements describe how
related the parameter estimates are to each other.
\item[\textbf{Frequentist statistics}]
A framework for statistics where model parameters are estimated from
observed data only and no prior beliefs about the parameters are
formally incorporated into the estimation process.
\item[\textbf{Geostatistics}]
A branch of statistics focussed on the analysis of spatial datasets
i.e.~data sets that exhibit \textbf{spatial autocorrelation}.
\item[\textbf{Kriging}]
A spatial interpolation method used to generate a surface of predictions
based on a geostatistical model fit to empirical data.
\item[\textbf{Kriging variance}]
The uncertainty of a \emph{Prediction} made using \emph{Kriging}.
\item[\textbf{Likelihood}]
A mathematical function that describes the distribution of the data
given a value for the parameter.
\item[\textbf{Loss function}]
A function to be minimised. Loss functions typically describe error; for
example, absolute discrepancy between a modelled value and an observed
value.
\item[\textbf{Myopic design approach}]
In the context of \textbf{Adaptive design}, this is a framework for
building an experimental design in a series of steps where one tries to
optimise decisions about the placement of sites for a single step in the
future.
\item[\textbf{Nugget effect}]
The nugget effect is the natural variation between measurements at
locations as the distance between then approaches zero.
\item[\textbf{Optimal design}]
An \textbf{Experimental design} that has the largest value of the
\textbf{Expected utility} among a set of other potential designs.
\item[\textbf{Optimality}]
The state of having achieved the best possible value of a function. In
the context of \textbf{Experimental design}, this means having maximised
the \textbf{Expected utility}.
\item[\textbf{Parameter}]
An unobserved numerical value in a statistical model. These are usually
unknown \textit{a priori} and need to be inferred from the data.
\item[\textbf{Posterior}]
In Bayesian statistics, the posterior is a distribution that represents
one's knowledge about some parameter(s) of a system as a combination of
objective prior belief and observed data.
\item[\textbf{Prediction}]
The modelled value of a variable at an unsampled location.
\item[\textbf{Prior}]
In \textbf{Bayesian statistics}, a prior is typically `set' on a
\textbf{Parameter} in a \textbf{Model} to summarise objective beliefs or
knowledge about the distribution of that parameter. A prior is usually
expressed as a distribution but it can also be a single value (this is
called a point prior, but it is generally avoided because this is a
statement made with extreme certainty that a parameter has a specific
value). The prior is updated by the data one collects as part of one's
experiment(s) to yield the \textbf{Posterior}.
\item[\textbf{Pseudo-Bayesian statistics}]
A framework for designing experiments where the utility function is
based on the results of a frequentist analysis, and the expected utility
is formed by integating out uncertainty due to, for example, the
parameter values.
\item[\textbf{Random design}]
In the context of \textbf{Stream networks} and \textbf{Spatial
statistics}, this refers to a collection of randomly chosen locations on
a stream that are sampled for data. These designs are a helpful
benchmark because optimal and adaptive designs, which are very
deliberately constructed, should outperform randomly chosen designs
given that they are optimised for a specific objective.
\item[\textbf{Range}]
The range parameter describes how quickly covariance decays with
distance in an covariance function.
\item[\textbf{Relative efficiency}]
The ratio of the information gained from two different experimental
configurations (typically including the same number of observations).
This efficiency indicates how many times more sites are needed in the
less efficient design to yield the same information as the more
efficient design.
\item[\textbf{Response variable}]
The outcome or dependent variable in an experiment or model.
\item[\textbf{Sequential design}]
See \textbf{Adaptive design}. This terminology is avoided as much as
possible in this work, but is common in the broader experimental design
literature.
\item[\textbf{Sill}]
The sill is the variance found among uncorrelated (i.e.~spatially
independent) data.
\item[\textbf{Sites}]
A location on a stream network where stream processes are, or will be
measured. These may include water quality, biodiversity, or water flow,
height and/or volume.
\item[\textbf{Spatial autocorrelation}]
The tendency for measurements to show a pattern of similarity (positive
spatial autocorrelation) or dissimilarity (negative spatial
autocorrelation) relative to the distance separating them.
\item[\textbf{Spatial statistics}]
A family of statistical methods specifically designed to analyse data
that exhibit spatial autocorrelation. A spatial statistical model uses
the spatial location of data in the probabilistic model component
(i.e.~spatial dependence in the residual errors is modelled as a
function of space).
\item[\textbf{Stream network}]
A stream network is a connected system of rivers, streams, channels
and/or creeks that form a branching network and ultimately converge into
a single stream segment.
\item[\textbf{Stream network distance}]
Distance between two locations when movement is restricted to the
\textbf{stream network}.
\item[\textbf{Space-filling design}]
A configuration of sampling sites that ensures the maximum and most
uniform spatial coverage from a fixed number of sites.
\item[\textbf{Static design}]
See \textbf{Optimal design}.
\item[\textbf{True model}]
A \textbf{Model} that is assumed to adequately describe a process that
one is sampling and observing during an experiment. Note that,
philosophically, the notion of a `true model' is tenuous; it is rare
that the underlying process is fully described by a mathematical
function.
\item[\textbf{Utility function}]
A function to be maximised. Contrast to \textbf{Loss function}.
\item[\textbf{Variable}]
Any process or thing that can be measured. Examples of variables
commonly measured in stream surveys include dissolved oxygen, water
temperature, canopy cover, or width of the stream at the sampling site.
\item[\textbf{Variance}]
The uncertainty inherent in a variable or observation. Higher variances
indicate higher uncertainties.
\end{description}